\documentclass[aps,prc,twocolumn,groupaddress,showpacs,nofootinbib,floatfix]{revtex4}

\usepackage{bm}
\usepackage{epsfig}
\usepackage{longtable}
\usepackage{dcolumn}

\def\bea {\begin{eqnarray}}
\def\eea {\end{eqnarray}}
\def\be {\begin{equation}}
\def\ee {\end{equation}}
\def\ben{\begin{enumerate}}
\def\een{\end{enumerate}}
\def\bi{\begin{itemize}}
\def\ei{\end{itemize}}

\def\etal{{\it et al.}\ }

\def\F{{\cal F}}
\def\prl {Phys. Rev. Lett.\ }

\def\pr {Phys. Rev.\ }
\def\np {Nucl. Phys.\ }

\def\GV{G_{\mbox{\tiny V}}}
\def\GF{G_{\mbox{\tiny F}}}
\def\DRV{\Delta_{\mbox{\tiny R}}^{\mbox{\tiny V}}}

\newcommand{\nuc}[2]{$^{#1}$#2}

\begin{document} 

\title{Superallowed $0^+$$\rightarrow 0^+$ nuclear $\beta$ decays:
2014 critical survey, with precise results for $V_{ud}$ and CKM Unitarity}

\author{J.C. Hardy} 
\email{hardy@comp.tamu.edu}
\author{I.S. Towner}
\email{towner@comp.tamu.edu}
\affiliation{Cyclotron Institute, Texas A\&M University, College Station, Texas 77843}

\date{\today}

\begin{abstract}
A new critical survey is presented of all half-life, decay-energy and branching-ratio
measurements related to 20 superallowed $0^+$$\rightarrow 0^+$ $\beta$ decays.  Included
are 222 individual measurements of comparable precision obtained from 177 published references.  
Compared with our last review in 2008, we have added results from 24 new publications and
eliminated 9 references, the results from which having been superceded by much more precise
modern data.  We obtain world-average $ft$-values for each of the eighteen transitions that have
a complete set of data, then apply radiative and isospin-symmetry-breaking corrections to
extract ``corrected" $\F t$ values.  Fourteen of these $\F t$ values now have a precision of order
0.1\% or better.  In the process of obtaining these results we carefully evaluate the available
calculations of the isospin-symmetry-breaking corrections by testing the extent to which they
lead to $\F t$ values consistent with conservation of the vector current (CVC). Only one set of
calculations satisfactorily meets this condition.  The resultant average $\F t$ value, when combined
with the muon liftime, yields the up-down quark-mixing element of the Cabibbo-Kobayashi-Maskawa
(CKM) matrix, $V_{ud} = 0.97417 \pm 0.00021$.  The unitarity test on the top row of the matrix becomes
$|V_{ud}|^2 + |V_{us}|^2 + |V_{ub}|^2 = 0.99978 \pm 0.00055$ if the Particle Data Group recommended
value for $V_{us}$ is used.  However, recent lattice QCD calculations, not included yet in the PDG
evaluation, have introduced some inconsistency into kaon-decay measurements of $V_{us}$ and $V_{us}
/ V_{ud}$.  We examine the impact of these new results on the unitarity test and conclude that
there is no evidence of any statistically significant violation of unitarity.  Finally, from the
$\F t$-value data we also set limits on the possible existence of scalar interactions. 
\end{abstract}

\maketitle

\section{\label{s:intro} Introduction}

Precise measurements of the beta decay between nuclear analog states of spin, $J^{\pi} = 0^+$,
and isospin, $T = 1$, provide demanding and fundamental tests of the properties of the
electroweak interaction.  Collectively, these transitions sensitively probe the conservation
of the vector weak current, set tight limits on the presence of scalar currents and provide the
most precise value for $V_{ud}$, the up-down quark-mixing element of the Cabibbo-Kobayashi-Maskawa
(CKM) matrix.  This latter result has become a linchpin in the most demanding available test of the
unitarity of the CKM matrix, a property which is fundamental to the electroweak standard model.

We have published six previous surveys of $0^+$$\rightarrow 0^+$ superallowed transitions
\cite{TH73,HT75,Ko84,HT90,HT05,HT09}, the first having appeared over 40 years ago and the most recent, six
years ago.  In each, we published a complete survey of all relevant nuclear data that pertained to these
superallowed transitions and used the results to set limits on the weak-interaction parameters that
were important at the time.  Notably, since $V_{ud}$ became the quantity of greatest interest 25
years ago, its value as obtained from our surveys of superallowed decays has improved by a factor of five in
precision but has never strayed outside of the uncertainties quoted in preceding surveys.  This consistency
is testimony to the robustness of what is by now a very large body of nuclear data.

Since our last survey closed in September 2008, there has continued to be a great deal of activity in
this field, both in experiment and in theory.  And this activity in honing $V_{ud}$ has been matched by
efforts to make similar improvements in the value of $V_{us}$, the second important element in the top-row
unitarity sum.  (The third element, $V_{ub}$, is too small to play a significant role.)  Since the
value of $V_{us}$ has undergone some unexpected changes in the past decade and has not yet settled at a reliably stable
result, interest in the CKM unitarity test continues to stimulate work in the field. Since 2008, new
measurements relating to $0^+$$\rightarrow 0^+$ superallowed transitions have appeared in 24 publications,
and the new more-precise results they contain have made 9 of the references accumulated in 2008 entirely obsolete
and left 11 more with some results replaced, in all cases because new values had uncertainties a factor of ten
or more smaller.  Altogether this means that, of the references in this 2014 survey, about 15\% are new and, being
among the most precise, their influence is disproportionately greater than that.

In addition to new measurements, there have also been important theoretical contributions to the small
isospin-symmetry-breaking corrections that must be applied to the data in order to extract $V_{ud}$ and
the values of other weak-interaction parameters.  In the past six years, a number of different groups
have published their results for these terms, with calculations based upon a variety of different models.
The diversity of results has prompted development of a test that allows each set of correction terms to be judged
by its ability to produce $\F t$ values that are consistent with conservation of the vector current (CVC).  As part
of our survey, we apply this test to all sets that cover at least half the number of well-measured superallowed
transitions.  As a result, we have identified the only set that yields self-consistent $\F t$ values and it is this
set that we use in our ultimate analysis of the experimental data.

Overall, recent improvements have been numerous enough that we consider this to be an opportune time to
produce a new and updated survey of the nuclear data used to establish $V_{ud}$.  We incorporate data on a
total of 20 superallowed transitions and have continued the practice we began in 1984 \cite{Ko84} of updating
all original data to take account of the most modern calibration standards.  However many of the measurements
that required updating have been superseded by more precise modern measurements so there are fewer updated
old results in our new survey than there have been in the past.

Superallowed $0^+$$\rightarrow 0^+$ $\beta$ decay between $T=1$ analog states depends uniquely on
the vector part of the weak interaction and, according to the conserved vector current (CVC) hypothesis,
its experimental $ft$ value should be directly related to the vector coupling constant, a fundamental constant
which is the same for all such transitions.  In practice, the expression for $ft$ includes several
small ($\sim$1\%) correction terms.  It is convenient to combine some of these terms with the
$ft$ value and define a ``corrected" $\F t$ value.  Thus, we write \cite{HT09}
\be
\F t \equiv ft(1+\delta_R^{\prime}) (1 + \delta_{NS} - \delta_C)
= \frac{K}{2 \GV^2 (1 + \DRV )}
\label{Ftdef}
\ee
where $K/(\hbar c )^6 = 2 \pi^3 \hbar \ln 2 / (m_e c^2)^5 = 8120.2776(9) \times
10^{-10}$ GeV$^{-4}$s, $\GV $ is the vector coupling constant for semi-leptonic weak interactions,
$\delta_C$ is the isospin-symmetry-breaking correction and $\DRV$ is the transition-independent part
of the radiative correction.  The terms $\delta_R^{\prime}$ and $\delta_{NS}$ comprise the
transition-dependent part of the radiative correction, the former being a function only of the
electron's energy and the $Z$ of the daughter nucleus, while the latter, like $\delta_C$, depends in
its evaluation on the details of nuclear structure.  From this equation, it can be seen that
each measured transition establishes an individual value for $\GV$ and, if the CVC assertion is
correct that $\GV$ is not renormalized in the nuclear medium, all such values -- and all the
$\F t$ values themselves -- should be identical within uncertainties, regardless of the specific
nuclei involved.

Our procedure in this paper is to examine all experimental data related to 20 superallowed transitions,
comprising all those that have been well studied, together with other cases that are now coming under scrutiny
after becoming accessible to precision measurement in relatively recent years.  The methods used in data
evaluation are presented in Sec. \ref{s:data}, with the calculations and corrections required to extract
$\F t$ values from these data being described and applied in Sec. \ref{s:Ftvalu}.  Then in Sec. \ref{s:dcgeneral}
we take a careful look at the various sets of isospin-symmetry-breaking correction terms and explain our
choice of the set we use for the survey results.  Finally, in Sec. \ref{s:impact} we explore the impact of
these results on two weak-interaction issues: CKM unitarity and the possible existence of scalar interactions.
This is much the same pattern as we followed in our last two reviews \cite{HT05, HT09} so we will not describe
the formalism again in detail, referring the reader instead to those earlier works.

\section{\label{s:data} Experimental Data}

The $ft$-value that characterizes any $\beta$ transition depends on three measured quantities: the
total transition energy $Q_{EC}$, the half-life $t_{1/2}$ of the parent state, and the branching ratio
$R$ for the particular transition of interest.  The $Q_{EC}$-value is required to determine the statistical
rate function, $f$, while the half-life and branching ratio combine to yield the partial half-life, $t$.
In Tables~\ref{QEC}-\ref{reject} we present the measured values of these three quantities and supporting
information for a total of twenty superallowed transitions.  In all, there are 222 independent measurements
from 177 references.  Thus, on average each quantity has been measured with comparable precision three or
more times by different groups.  Such redundancy virtually eliminates the possibility of individual experimental
anomalies having a significant impact on the overall results.

\subsection{\label{eval} Evaluation principles}

In our treatment of the data, we considered all measurements formally published before September 2014.  We scrutinized
all the original experimental reports in detail.  Where necessary and possible, we used the information provided
there to correct the results for calibration data that have improved since the measurement was made.  If corrections
were evidently required but insufficient information was provided to make them, the results were rejected.  Of
the surviving results, only those with (updated) uncertainties that are within a factor of ten of the most precise
measurement for each quantity were retained for averaging in the tables.  Each datum appearing in the tables is
attributed to its original journal reference {\it via} an alphanumeric code comprising the initial two letters
of the first author's name and the two last digits of the publication date.  These codes are correlated
with the actual reference numbers, \cite{Ad83}-\cite{Zi87}, in Table~\ref{ref}.

The statistical procedures we have followed in analyzing the tabulated data are based on those used by the
Particle Data Group in their periodic reviews of particle properties (e.g. Ref. \cite{PDG14}) and adopted by
us in our previous surveys.  We gave a detailed description of those procedures in our 2004 survey \cite{HT05}
so will not repeat it here.

Our evaluation principles and associated statistical procedures constitute a very conservative approach to
the data.  Unless there is a clearly identifiable reason to reject a result, we include it in our data base
even if it deviates significantly from other measurements of the same quantity, the consequent non-statistical
spread in results being reflected in an increased uncertainty assigned to the average.  Wherever this
occurs, the factor by which the uncertainty has been increased, which is the square-root of the normalized $\chi^2$, is
listed in the ``scale" column of a table.  Occasionally a measurement with an acceptable uncertainty is nevertheless
excluded from our data base, in which case the reason for its exclusion is always listed in Table \ref{reject}.  For
example, there are a few publications that include a number of measurements -- a set of half-lives or $Q_{EC}$ values
-- most or all of which deviate substantially from other accepted measurements of the same quantities.  In those
cases, we consider that some systematic problem has been revealed, and exclude all the results from that publication.

\begingroup
\squeezetable
\setcounter{LTchunksize}{200}
\setlength{\LTcapwidth}{6.5in}
\begin{center}
\begin{longtable*}{lllllllll} \\
\caption {Measured results from which the decay transition energies, $Q_{EC}$, have been derived for superallowed
$\beta$-decays.  The lines giving the average superallowed $Q_{EC}$ values themselves are in bold print.
(See Table~\ref{ref} for the correlation between the alphanumeric reference code used in this table and the
actual reference numbers.)
\label{QEC}} \\[-2mm]

\hline\hline
&&&&&&&& \\

\multicolumn{2}{c}{Parent/Daughter}
 & Property\footnotemark[1] 
 & \multicolumn{3}{c}{Measured Energies used to determine $Q_{EC}$ (keV)}
 & \multicolumn{1}{c}{}
 & \multicolumn{2}{c}{Average value} \\[1mm]
\cline{4-6} 
\cline{8-9} 
& & & & & & & &  \\[-2mm]
   \multicolumn{2}{c}{nuclei} & 
 & \multicolumn{1}{c}{1}
 & \multicolumn{1}{c}{2} 
 & \multicolumn{1}{c}{3}
 & \multicolumn{1}{c}{} 
 & \multicolumn{1}{l}{~~~Energy (keV)} 
 & \multicolumn{1}{c}{scale} \\[1mm]

\hline\hline &&&&&&&& \\

\endfirsthead

\multicolumn{9}{l}
{\tablename\ \thetable{} (continued)} \\[2mm]
\hline\hline
\\

\multicolumn{2}{c}{Parent/Daughter}
 & Property\footnotemark[1] 
 & \multicolumn{3}{c}{Measured Energy (keV)}
 & \multicolumn{1}{c}{}
 & \multicolumn{2}{c}{Average value} \\[1mm]
\cline{4-6} 
\cline{8-9} 
& & & & & & & &  \\[-2mm]
   \multicolumn{2}{c}{nuclei} & 
 & \multicolumn{1}{c}{1}
 & \multicolumn{1}{c}{2} 
 & \multicolumn{1}{c}{3}
 & \multicolumn{1}{c}{}  
 & \multicolumn{1}{l}{~~~~Energy (keV)} 
 & \multicolumn{1}{c}{scale} \\[1mm]
\hline
& & & & & & & & \\[-1mm]
\endhead

\hline
\endfoot

\hline\hline
\vspace{-5mm}
\endlastfoot

& & & & & & & \\[-4mm]
$T_z = -1$: & & & & & & & \\
& & & & & & & \\[-1mm]
~~ $^{10}$C & $^{10}$B & $Q_{EC}(gs)$ & ~~3647.83 $\pm$ 0.34 ~$\,$[Ba84] & ~~3647.95 $\pm$ 0.12 [Ba98] & ~3648.12 $\pm$ 0.08 [Er11] & & &  \\
 & & & ~~3648.34 $\pm$ 0.51 ~$\,$[Kw13] & & & & ~$\:$3648.063 $\pm$ 0.064 & 1.0   \\
 & & $E_x(d0^+)$ & ~$\;$1740.15 $\pm$ 0.17 ~$\,$[Aj88] & 1740.068 $\pm$ 0.017 \footnotemark[2] & & & ~$\,$1740.069 $\pm$ 0.017 & 1.0  \\
 & & $\bm{Q_{EC}(sa)}$ & & & & & {\bf 1907.994 $\pm$ 0.067} & \\[2mm]
~~ $^{14}$O & $^{14}$N & $Q_{EC}(gs)$ & ~$\;$5143.30 $\pm$ 0.60 ~$\,$[Bu61] & ~~5145.05 $\pm$ 0.46 [Ba62] & $\,$5145.52 $\pm$ 0.48 [Ro70]
 & & & \\
 & & & ~$\;$5143.43 $\pm$ 0.37 ~$\,$[Wh77] & ~~5144.33 $\pm$ 0.17 [To03] & & & ~~~5144.32 $\pm$ 0.28 & 2.1  \\
 & & $E_x(d0^+)$ & ~$\!$2312.798 $\pm$ 0.011$\:$[Aj91] & & & & ~~2312.798 $\pm$ 0.011 & \\
 & & $\bm{Q_{EC}(sa)}$ & & & & & ~~{\bf 2831.23 $\pm$ 0.23 \footnotemark[3]} & {\bf 2.3} \\[2mm]
~~ $^{18}$Ne & $^{18}$F & $ME(p)$ & $\:$~~~5316.8 $\pm$ 1.5 ~~$\,$[Ma94] & ~~5317.63 $\pm$ 0.36 [Bl04b] & & & ~~~$\:$5317.58 $\pm$ 0.35 & 1.0 \\
 & & $ME(d)$ & $\:$~~~871.99 $\pm$ 0.73 ~[Bo64] & ~~~~~874.2 $\pm$ 2.2 ~$\,$[Ho64] & ~~~~875.2 $\pm$ 2.8 ~[Pr67] & & & \\
 & & & ~~~~~877.2 $\pm$ 3.0 ~~$\,$[Se73] & ~~~$\,$874.01 $\pm$ 0.60 [Ro75] & & & ~~~~~873.37 $\pm$ 0.59 & 1.3 \\
 & & $Q_{EC}(gs)$ & & & & & ~~~$\:$4444.21 $\pm$ 0.68 & \\
 & & $E_x(d0^+)$ & ~$\;$1041.55 $\pm$ 0.08 ~[Ti95] & & & & ~~~$\:$1041.55 $\pm$ 0.08 & \\
 & & $\bm{Q_{EC}(sa)}$ & & & & & ~~{\bf 3402.66 $\pm$ 0.69} & \\[2mm]
~~ $^{22}$Mg & $^{22}$Na & $ME(p)$ & ~~~$\;$-401.2 $\pm$ 3.0 ~~$\,$[Ha74c] & ~~~~-400.8 $\pm$ 1.2 \footnotemark[4]
 &  ~~$\;$-400.5 $\pm$ 1.0 ~$\,$[Pa05] & & ~~~~-400.67 $\pm$ 0.73 & 1.0 \\
  & & $ME(d)$ & ~$\:\,$-5184.3 $\pm$ 1.5 ~~$\,$[We68] & ~~$\,$-5182.5 $\pm$ 0.5 ~$\,$[Be68] & ~$\,$-5181.3 $\pm$ 1.7 ~[An70] & & & \\
 & & & ~$\:\,$-5183.2 $\pm$ 1.0 ~~$\,$[Gi72] & ~-5181.56 $\pm$ 0.16 [Mu04] & -5181.08 $\pm$ 0.30
 [Sa04] & & ~~~-5181.58 $\pm$ 0.23 & 1.7 \\
 & & $Q_{EC}(gs)$ & ~$\;$4781.64 $\pm$ 0.28 ~[Mu04] & ~~4781.40 $\pm$ 0.67 [Sa04] & & & ~~~$\;$4781.53 $\pm$ 0.24 & 1.0 \\
 & & $E_x(d0^+)$ & ~~~$\,$657.00 $\pm$ 0.14 ~[En98] & & & & ~~~~~$\,$657.00 $\pm$ 0.14 & \\
 & & $\bm{Q_{EC}(sa)}$ & & & & & ~~$\,${\bf 4124.53 $\pm$ 0.28} & \\[2mm] 
~~ $^{26}$Si & $^{26}$Al & $ME(p)$ & ~~~-7145.4 $\pm$ 3.0 ~~~[Ha74c] & ~~$\,$-7139.5 $\pm$ 1.0 ~~$\,$[Pa05] & ~~$\,$-7140.4 $\pm$ 2.9 [Kw10]
 & & ~~~~$\,$-7140.1 $\pm$ 1.2  & 1.3 \\
  & & $ME(d0^+)$ & -11981.96 $\pm$ 0.26 \footnotemark[5] & & & & $\,$~-11981.96 $\pm$ 0.26 & \\
 & & $\bm{Q_{EC}(sa)}$ & ~~$\,$4840.85 $\pm$ 0.10 ~[Er09a] & & & & ~~$\,${\bf 4840.86 $\pm$ 0.10} & {\bf 1.0} \\[2mm]
~~ $^{30}$S & $^{30}$P & $ME(p)$ & $\:$~~$\;$-14060 $\pm$ 15 ~~~[Mi67] & ~~~$\,$-14054 $\pm$ 25 ~~$\,$[Mc67] & ~~~-14068 $\pm$ 30 ~~[Ha68] & & & \\ 
 & & & ~-14063.4 $\pm$ 3.0 ~~$\,$[Ha74c] & & & & ~~~-14063.1 $\pm$ 2.9& 1.0 \\
 & & $ME(d)$ & ~~~$\,$-20203 $\pm$ 3 ~~~~~[Ha67] &  -20200.61 $\pm$ 0.40 [Re85]& & & ~$\,$-20200.65 $\pm$ 0.40 & 1.0 \\
 & & $Q_{EC}(gs)$ & ~~6141.61 $\pm$0.19 ~~[So11] & & & & ~~~~6141.59 $\pm$ 0.26 & 1.4 \\
 & & $E_x(d0^+)$ & $\:$~~~677.29 $\pm$ 0.07 ~[En98] & & & & ~~~~~$\,$677.29 $\pm$ 0.07 & \\
 & & $\bm{Q_{EC}(sa)}$ & & & & & ~~~{\bf 5464.30 $\pm$ 0.27} & \\[2mm]
~~ $^{34}$Ar & $^{34}$Cl & $ME(p)$ & ~-18380.2 $\pm$ 3.0 ~~$\,$[Ha74c] & ~-18378.4 $\pm$ 3.5 ~~$\,$[He01]
 & -18377.10 $\pm$ 0.41$\,$[He02] & & ~$\,$-18377.17 $\pm$ 0.40 & 1.0 \\
 & & $ME(d)$ & $\!$-24440.03 $\pm$ 0.06 \footnotemark[5] & & & & ~$\,$-24440.03 $\pm$ 0.06 & \\
 & & $\bm{Q_{EC}(sa)}$ & ~~6061.83 $\pm$ 0.08 ~[Er11] & & & & {\bf ~~$\,$6061.87 $\pm$ 0.19} & {\bf 2.5} \\[2mm] 
~~ $^{38}$Ca & $^{38}$K & $ME(p)$ & -22058.53 $\pm$ 0.28~$\,$[Ri07] & -22058.01 $\pm$ 0.65 $\,$[Ge07] & & &
 ~$\,$-22058.45 $\pm$ 0.26 & 1.0 \\
 & & $ME(d0^+)$ & -28670.58 $\pm$ 0.21 \footnotemark[5] & & & & ~$\,$-28670.58 $\pm$ 0.21 & \\
 & & $\bm{Q_{EC}(sa)}$ & ~~6612.12 $\pm$ 0.07 ~[Er11] & & & & ~~~{\bf 6612.12 $\pm$ 0.07} &  {\bf 1.0} \\[2mm]
~~ $^{42}$Ti & $^{42}$Sc & $\bm{Q_{EC}(sa)}$ & ~~7016.83 $\pm$ 0.25 ~[Ku09] & & & & ~~~{\bf 7016.83 $\pm$ 0.25} & \\[2mm]
$T_z = 0$: & & & & & & & & \\
& & & & & & & \\[-1mm]
~~ $^{26m}$Al & $^{26}$Mg & $Q_{EC}(gs)$ & ~$\;$4004.79 $\pm$ 0.55 ~[De69] & ~~4004.41 $\pm$ 0.10 \footnotemark[6]
 & ~~4004.37 $\pm$ 0.22 [Ge08] & & ~~~~$\,$4004.41 $\pm$ 0.09 & 1.0 \\
 & & $E_x(p0^+)$ & ~$\;$228.305 $\pm$ 0.013$\,$[En98] & & & & ~~~~$\,$228.305 $\pm$ 0.013 & \\
 & & $\bm{Q_{EC}(sa)}$ & ~~4232.19 $\pm$ 0.12 ~[Br94] & ~~4232.83 $\pm$ 0.13 ~[Er06b] & & & {\bf ~~$\,$$\,$4232.66 $\pm$ 0.12 \footnotemark[3]} & {\bf 2.1} \\[2mm]
~~ $^{34}$Cl & $^{34}$S & $\bm{Q_{EC}(sa)}$ &  ~~5491.65 $\pm$ 0.26 \footnotemark[7] & $\,$5491.662 $\pm$ 0.047 [Er09b]& & & ~{\bf 5491.662 $\pm$ 0.046} & {\bf 1.0} \\[2mm] 
~~ $^{38m}$K & $^{38}$Ar & $\bm{Q_{EC}(sa)}$ & ~~6044.38 $\pm$ 0.12 ~[Ha98] & $\,$6044.223 $\pm$ 0.041 [Er09b]] & & & $\:${\bf 6044.240 $\pm$ 0.048} & {\bf 1.2} \\[2mm]
~~ $^{42}$Sc & $^{42}$Ca & $\bm{Q_{EC}(sa)}$ & ~~6425.84 $\pm$ 0.17 \footnotemark[8] & ~~6426.13 $\pm$ 0.21 [Er06b] & &
 & ~~$\,${\bf 6426.28 $\pm$ 0.30 \footnotemark[3]} &     {\bf 3.0}  \\[2mm]
~~ $^{46}$V  & $^{46}$Ti & $\bm{Q_{EC}(sa)}$ & ~~7052.90 $\pm$ 0.40 [Sa05] & ~~7052.72 $\pm$ 0.31 [Er06b] & ~~7052.11 $\pm$ 0.27 $\,$[Fa09] & \\
 & & & ~~7052.44 $\pm$ 0.10 [Er11] & & & & ~~{\bf 7052.45 $\pm$ 0.10} & {\bf 1.1}  \\[2mm]
~~ $^{50}$Mn & $^{50}$Cr & $\bm{Q_{EC}(sa)}$ & ~~7634.48 $\pm$ 0.07 $\,$[Er08] & & & & $\,${\bf 7634.451 $\pm$ 0.066 \footnotemark[3]} &  {\bf 1.0} \\[2mm]
~~ $^{54}$Co & $^{54}$Fe & $\bm{Q_{EC}(sa)}$ & ~~8244.54 $\pm$ 0.10 $\,$[Er08] & & & & ~~{\bf 8244.37 $\pm$ 0.28 \footnotemark[3]} & {\bf 3.4} \\[2mm]
~~ $^{62}$Ga & $^{62}$Zn & $\bm{Q_{EC}(sa)}$ & ~~9181.07 $\pm$ 0.54 $\:$[Er06a] & & & & ~~{\bf 9181.07 $\pm$ 0.54}
 &  \\[2mm]
~~ $^{66}$As & $^{66}$Ge & $ME(p)$ & ~~~$\,$-52018 $\pm$ 30 ~~~[Sc07] & & & & ~~~~-52018 $\pm$ 30 & \\
 & & $ME(d)$ & ~-61607.0 $\pm$ 2.4 $\,$~~[Sc07] & & & & ~~-61607.0 $\pm$ 2.4 & \\
 & & $\bm{Q_{EC}(sa)}$ & ~~~~~~9550 $\pm$ 50 ~~~[Da80] & & & & {\bf ~~~~~9579 $\pm$ 26 } & {\bf 1.0} \\[2mm]
~~ $^{70}$Br & $^{70}$Se & $\bm{Q_{EC}(sa)}$ & ~~~~~~9970 $\pm$ 170$\,$~~[Da80] & & & & {\bf ~~~~~9970 $\pm$ 170}
 &  \\[2mm]
~~ $^{74}$Rb & $^{74}$Kr & $ME(p)$ & ~~~~-51905 $\pm$ 18 ~~$\,$[He02] & ~~-51915.2 $\pm$ 4.0 ~[Ke07] & -51916.5 $\pm$ 6.0 [Et11]&
 & ~~-51915.2 $\pm$ 3.3 & 1.0 \\
 & & $ME(d)$ & ~$\,$-62332.0 $\pm$ 2.1 ~~[Ro06] & & & & ~~-62332.0 $\pm$ 2.1 &  \\
 & & $\bm{Q_{EC}(sa)}$ & & & & & ~$\,${\bf 10416.8 $\pm$ 3.9} & \\[-2mm]
\footnotetext[1]{Abbreviations used in this column are as follows: ``$gs$", transition between ground states;
``$sa$", superallowed transition; ``$p$", parent; ``$d$", daughter; ``$ME$", mass excess; ``$E_x(0^+)$",
excitation energy of the $0^+$ (analog) state.  Thus, for example, ``$Q_{EC}(sa)$" signifies the $Q_{EC}$-value
for the superallowed transition, ``$ME(d)$", the mass excess of the daughter nucleus; and ``$ME(d0^+)$, the mass
excess of the daughter's $0^+$ state.}
\footnotetext[2]{Result based on references [Ba88] and [Ba89].}
\footnotetext[3]{Average result includes the results of $Q_{EC}$ pairs; see Table~\ref{Qdiff}.}
\footnotetext[4]{Result based on references [Bi03], [Se05] and [Je07].}
\footnotetext[5]{Result obtained from the $Q_{EC}$ value for the superallowed decay of $d0^+$, which appears elsewhere in this table, combined with the mass of its daughter taken from [Wa12].}
\footnotetext[6]{Result based on references [Is80], [Al82], [Hu82], [Be85], [Pr90], [Ki91] and [Wa92].}
\footnotetext[7]{Result based on references [Wa83], [Ra83] and [Li94].}
\footnotetext[8]{Result based on references [Zi87] and [Ki89].} 
\end{longtable*}
\end{center}
\endgroup

\vspace{-5cm}

\subsection{\label{tables} Data Tables}

The $Q_{EC}$-value data appear in Tables~\ref{QEC} and \ref{Qdiff}.  Of the 20 superallowed decays listed, nine
--- those of $^{10}$C, $^{14}$O, $^{26}$Al$^m$, $^{34}$Cl, $^{38}$K$^m$, $^{42}$Sc, $^{46}$V, $^{50}$Mn and $^{54}$Co ---
have stable daughter nuclei.  In past surveys, the corresponding $Q_{EC}$ values were predominantly obtained from
direct reaction measurements but, by now, all but the $^{14}$O $Q_{EC}$ value have been measured with a Penning trap.
Each of these latter measurements has determined the parent and daughter masses interleaved in a single experiment, thus
effectively measuring the $Q_{EC}$ value directly from the ratio of cyclotron frequencies. All direct measurements of a
$Q_{EC}$ value are identified in column
3 of Table~\ref{QEC} by ``$Q_{EC}(sa)$" and each individual result, whether reaction or Penning-trap based, is itemized
with its appropriate reference in the next three columns.  The weighted average of all measurements for a particular
decay appears in column 7, with the corresponding scale factor (see Sec.~\ref{eval}) in column 8.  Four of these cases, 
$^{10}$C, $^{34}$Cl, $^{38}$K$^m$ and $^{46}$V, have no further complications.  For the remaining five, however, in
addition to the individual $Q_{EC}$-value results, $Q_{EC}$-value differences have also been obtained via ($^3$He, $t$)
reactions on composite targets.  These difference measurements are presented in Table~\ref{Qdiff}.  They have been dealt
with in combination with the direct $Q_{EC}$-value measurements to obtain a best overall fit by a method described in
our 2004 survey \cite{HT05}.  The final average $Q_{EC}$ value for each transition appears in column 7 of Table~\ref{QEC}
and the average differences are in column 4 of Table~\ref{Qdiff}.  All are flagged with footnotes to indicate the interconnection.

\begin{table*}[t]
\caption{$Q_{EC}$-value differences for superallowed $\beta$-decay branches.  These data are also used as input to determine
some of the average $Q_{EC}$-values listed in Table~\ref{QEC}.   (See Table~\ref{ref} for the correlation between the
alphabetical reference code used in this table and the actual reference numbers.)
\label{Qdiff}}
\begin{ruledtabular}
\begin{tabular}{llll}
Parent   
 & \multicolumn{1}{l}{Parent}
 & \multicolumn{2}{c}{$Q_{EC2} - Q_{EC1}$ (keV)} \\[1mm]
\cline{3-4} \\[-2mm]
nucleus 1 
 & \multicolumn{1}{l}{nucleus 2}
 & \multicolumn{1}{c}{measurement} 
 & \multicolumn{1}{c}{average\footnotemark[1]} \\[1mm]
\hline \\[-1mm]
$^{14}$O & $^{26m}$Al & 1401.68 $\pm$ 0.13 [Ko87] & 1401.43 $\pm$ 0.26 \\
$^{26m}$Al & $^{42}$Sc & ~$\,$2193.5 $\pm$ 0.2 ~$\,$[Ko87] & 2193.62 $\pm$ 0.32 \\
$^{42}$Sc & $^{50}$Mn & ~$\,$1207.6 $\pm$ 2.3 ~$\,$[Ha74d] & 1208.17 $\pm$ 0.31 \\
$^{42}$Sc & $^{54}$Co & ~$\,$1817.2 $\pm$ 0.2 ~$\,$[Ko87] & 1818.10 $\pm$ 0.41 \\
$^{50}$Mn & $^{54}$Co & ~$\,$610.09 $\pm$ 0.17 $^{\rm[Ko87]}_{\rm[Ko97b]}$ & ~$\,$609.92 $\pm$ 0.29 \\[-4mm]
\footnotetext[1]{Average values include the results of direct $Q_{EC}$-value measurements: see Table~\ref{QEC}.}
\end{tabular}
\end{ruledtabular}
\end{table*}

There are two cases, $^{26}$Al$^m$ and $^{38}$K$^m$, in which the superallowed decay originates from an isomeric state.  
For the former, there are $Q_{EC}$-value measurements of comparable precision that correspond to the ground state as well
as to the isomer. Obviously, the two sets of measurements are simply related to one another by the excitation energy of
the isomeric state in the parent.  In Table~\ref{QEC} the set of measurements for the ground-state $Q_{EC}$-value and
for the excitation energy of the isomeric state appear in separate rows, each with its identifying property given in
column 3 and its weighted average appearing in column 7.  In the row below, the average value given in column 7 for the
superallowed transition is the weighted average not only of the direct superallowed $Q_{EC}$-value measurements in that
row, but also of the result derived from the two preceding rows.  Note that in all cases the $Q_{EC}$-value for the
superallowed transition appears in bold-face type.

For the remaining transitions, those which have unstable parents and daughters, the situation is somewhat more complicated. 
In some cases only the parent and daughter masses have been measured, either from transfer reactions or by Penning trap, but
not the $Q_{EC}$ value.  In other cases, the measured results for masses and $Q_{EC}$-values are of comparable precision
and so both must be included in the average.  Also, in most of the cases with $T_Z$ = -1 parents, the masses of the parent and
daughter nuclei are not sufficient to determine the $Q_{EC}$-value for the superallowed branch; that also requires the excitation
energy of the analog 0$^+$ state in the daughter.  If needed, all of these properties are identified in column 3 of
Table~\ref{QEC}, with the individual measurements of that property, their weighted average and a scale factor appearing in
columns to the right.  The average $Q_{EC}$-value listed for the corresponding superallowed transition is obtained from
these separate averages and appears in bold print. 

As in our previous surveys, we have not used the current Atomic Mass Evaluation tables \cite{Wa12} to derive the $Q_{EC}$-values
of interest.  Our method is to include all pertinent measurements for each property; typically, only a subset of the available
data is included as input to the mass tables.  Furthermore, we have examined each reference in detail and either accepted the
result, updated it to modern calibration standards or rejected it for cause.  The updating procedures are outlined, reference
by reference, in Table~\ref{update} and the rejected results are similarly documented in Table~\ref{reject}.  With a comparatively
small data set, we could afford to pay the kind of individual attention that is impossible when one is considering all nuclear
masses.

\begin{figure}[b]
\epsfig{file=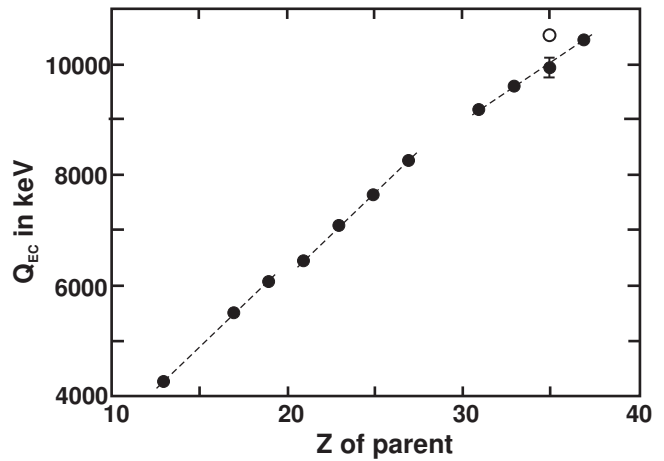,width=8.5cm}
\caption{The tabulated $Q_{EC}$ values for the $T_Z$ = 0 superallowed transitions in Table~\ref{QEC} (solid points) are plotted as a
function of the $Z$ of the parent nucleus.  For all cases other than that of $^{70}$Br the uncertainties are smaller than the plotted
points.  The open circle at $Z$ = 35 is the $Q_{EC}$ value measured by Penning trap \cite{Sa09} and assigned to the superallowed
transition from $^{70}$Br.  As explained in the text and noted in Table~\ref{reject}, this result has been omitted from Table~\ref{QEC}.  The dashed
lines are only to guide the eye; separate straight lines are drawn for each of the $sd$, $f_{7/2}$ and upper-$fp$ shells.}
\label{fig1}
\end{figure}

One of our omissions from Table~\ref{QEC} requires a more detailed explanation than could be included in Table~\ref{reject}.  
There are two reported measurements of the $^{70}$Br $Q_{EC}$ value.  The first is a rather old result in Da80 \cite{Da80}, which
came from a measurement of the positron end-point energy as recorded in a plastic scintillator; the second one was reported very
recently in Sa09 \cite{Sa09} and is based on Penning-trap measurements.  Since the latter's uncertainty is more than ten times smaller
than the former's, it would normally be the only one to appear in our table.  However, we have chosen to eliminate it and include only
the old imprecise measurement.  The
reason is made clear in Fig.~\ref{fig1}, where the Penning-trap result is seen to deviate from systematic behavior by some 500 keV.
Penning traps are clearly capable of measuring the mass of trapped ions to much higher accuracy than that, but it is not easy to identify
the nuclear state they are measuring.  It is likely in this case that the trap actually measured an isomeric state in $^{70}$Br rather
than its ground state.  Of course arguments based on systematics are not infallible either and, in any case, this $Q_{EC}$ value needs
to be determined more precisely.  Fresh experiments are certainly called far.

The half-life data appear in Table~\ref{t1/2} in similar format to Table~\ref{QEC}.  For obvious reasons,
half-life measurements do not lend themselves to being updated.  Consequently, a number of mostly pre-1970
measurements have been rejected because they were not analyzed with the ``maximum-likelihood" method.  The
importance of using this technique for precision measurements was not recognized until 1969 \cite{Fr69a}
and, without access to the primary data, there is no way a new analysis can be applied retroactively.  All
rejected half-life measurements are also documented in Table~\ref{reject}.

\begingroup
\squeezetable
\begin{table*}[t]
\caption{Half-lives, $t_{1/2}$, of superallowed $\beta$-emitters.  (See Table~\ref{ref} for the correlation between the
  alphabetical reference code used in this table and the actual reference numbers.)
\label{t1/2}}
\begin{ruledtabular}
\begin{tabular}{llllllll}
Parent   
 & \multicolumn{4}{c}{Measured half-lives, $t_{1/2}$ (ms)}
 & \multicolumn{1}{c}{}
 & \multicolumn{2}{c}{Average value} \\[1mm]
\cline{2-5} 
\cline{7-8} \\[-2mm]
nucleus 
 & \multicolumn{1}{c}{1}
 & \multicolumn{1}{c}{2} 
 & \multicolumn{1}{c}{3} 
 & \multicolumn{1}{c}{4} 
 & \multicolumn{1}{c}{}
 & \multicolumn{1}{c}{$t_{1/2}$ (ms)} 
 & \multicolumn{1}{c}{scale} \\[1mm]
\hline
 & & & & & & & \\[-1mm]
$T_z = -1$: & & & & & & & \\
~~ $^{10}$C  & ~~$\,$19280 $\pm$ 20 $\,$~~[Az74] & $\:$19295 $\pm$ 15 ~[Ba90] & ~19310 $\pm$ 4 ~~~[Ia08] & ~19282 $\pm$ 11 ~[Ba09] & & 19305.2 $\pm$ 7.1  & 2.0   \\
~~ $^{14}$O  & ~~$\,$70480 $\pm$ 150 ~[Al72] & $\:$70588 $\pm$ 28 ~[Cl73] &~70430 $\pm$ 180 [Az74] &~70684 $\pm$ 77 $\,$~[Be78] & & &  \\
 & ~~$\,$70613 $\pm$ 25 $\,$~~[Wi78] & $\:$70560 $\pm$ 49 ~[Ga01] & ~70641 $\pm$ 20 $\,$~[Ba04] & ~70696 $\pm$ 52 ~$\,$[Bu06] & & &  \\
 & ~~$\,$70623 $\pm$ 53 $\,$~~[Ta12] & $\:$70610 $\pm$ 30 ~[La13] & ~70632 $\pm$ 94 $\,$~[La13] & & &~~$\,$70619 $\pm$ 11 & 1.0 \\
~~ $^{18}$Ne & ~~~~1669 $\pm$ 4 ~~~~[Al75] & $\,$~~1687 $\pm$ 9 ~~[Ha75] & $\,$1665.6 $\pm$ 1.9 $\,$[Gr07] & $\,$1664.8 $\pm$ 1.1 $\,$[Gr13] & & $\,$~1665.4 $\pm$ 0.9 & 1.0            \\
~~ $^{22}$Mg & ~~~~3857 $\pm$ 9 ~~~~[Ha75] & 3875.5 $\pm$ 1.2 [Ha03]   & & & & $\,$~3875.2 $\pm$ 2.4 & 2.0 \\
~~ $^{26}$Si & ~~2245.3 $\pm$ 0.7 $\,$~[Ia10] & & & & & ~$\,$2245.3 $\pm$ 0.7 & \\
~~ $^{30}$S  & ~~1178.3 $\pm$ 4.8 $\,$~[Wi80] & 1175.9 $\pm$ 1.7 ~[So11] & & & & $\,$~1176.2 $\pm$ 1.6 & 1.0 \\
~~ $^{34}$Ar & ~~~844.5 $\pm$ 3.4 ~~[Ha74a] & ~843.8 $\pm$ 0.4 $\,$~[Ia06]  & & & & ~~~843.8 $\pm$ 0.4 & 1.0 \\
~~ $^{38}$Ca & ~~~$\,$443.8 $\pm$ 1.9 $\:$~[Bl10] & 443.77 $\pm$ 0.36 [Pa11] & & & & $\:$~443.77 $\pm$ 0.35  & 1.0          \\
~~ $^{42}$Ti & $\,$~~~~~\,202 $\pm$ 5 ~~~[Ga69] & 208.14 $\pm$ 0.45 [Ku09] & & & & ~~208.09 $\pm$ 0.55 & 1.2 \\[1mm]
$T_z = 0$: & & & & & & &  \\
~~ $^{26m}$Al &  ~~~~6346 $\pm$ 5 $\,$~~~~[Fr69a] & $\,$~~6346 $\pm$ 5 $\,$~~~[Az75] & 6339.5 $\pm$ 4.5 $\,$~[Al77] & 6346.2 $\pm$ 2.6 ~[Ko83] & &  &  \\
  & 6346.54 $\pm$ 0.76 ~[Fi11] & 6347.8 $\pm$ 2.5 ~[Sc11] & 6345.3 $\pm$ 0.9 ~[Ch13] & & & 6346.02 $\pm$ 0.54 & 1.0 \\
~~ $^{34}$Cl & ~~~~1526 $\pm$ 2 $\,$~~~~[Ry73] & 1525.2 $\pm$ 1.1 $\,$~[Wi76] & 1527.7 $\pm$ 2.2 $\,$~[Ko83]
 & 1526.8 $\pm$ 0.5 ~[Ia06] & & 1526.55 $\pm$ 0.44   &  1.0   \\
~~ $^{38m}$K & ~~~925.6 $\pm$ 0.7 $\,$~~[Sq75] & $\,$~922.3 $\pm$ 1.1 $\,$~[Wi76] & 921.71 $\pm$ 0.65 [Wi78]
 & 924.15 $\pm$ 0.31$\,$[Ko83] & &  &  \\
 & ~~~924.4 $\pm$ 0.6 $\,$~~[Ba00] & 924.46 $\pm$ 0.14 [Ba10] &  & & & $\,$~924.33 $\pm$ 0.27 & 2.3  \\
~~ $^{42}$Sc &  ~$\,$680.98 $\pm$ 0.62 ~[Wi76] & 680.67 $\pm$ 0.28 [Ko97a]  &    &  &  & $\,$~680.72 $\pm$ 0.26  &  1.0      \\
~~ $^{46}$V &  ~$\,$422.47 $\pm$ 0.39 ~[Al77] & 422.28 $\pm$ 0.23 [Ba77a] & 422.57 $\pm$ 0.13 [Ko97a] & 422.66 $\pm$ 0.06 [Pa12] &  & 422.622 $\pm$ 0.053 & 1.2    \\
~~ $^{50}$Mn &  ~~~284.0 $\pm$ 0.4 $\,$~~[Ha74b] & $\,$~282.8 $\pm$ 0.3 $\,$~[Fr75] & 282.72 $\pm$ 0.26 [Wi76]
 & 283.29 $\pm$ 0.08 [Ko97a] & & &   \\
 & ~$\,$283.10 $\pm$ 0.14 ~[Ba06] & & & & & $\,$~283.21 $\pm$ 0.11 & 1.7  \\
~~ $^{54}$Co & $\,$~~$\,$193.4 $\pm$ 0.4 $\,$~~[Ha74b] & $\,$~193.0 $\pm$ 0.3 $\,$~[Ho74] & 193.28 $\pm$ 0.18 [Al77]
 & 193.28 $\pm$ 0.07 [Ko97a] & & 193.271 $\pm$ 0.063 & 1.0        \\
~~ $^{62}$Ga & ~$\,$115.84 $\pm$ 0.25 ~[Hy03] & 116.19 $\pm$ 0.04 [Bl04a] & 116.09 $\pm$ 0.17 [Ca05]
 & 116.01 $\pm$ 0.19 [Hy05] & & &         \\
             & 116.100 $\pm$ 0.025  [Gr08] & & & & & 116.121 $\pm$ 0.040 & 1.9    \\
~~ $^{66}$As &  ~~~95.78 $\pm$ 0.39 ~[Al78] & $\,$~95.77 $\pm$ 0.28 [Bu88] & ~~~~~97 $\pm$ 2 ~~~~[Ji02]  &  &
 & ~~~95.79 $\pm$ 0.23 & 1.0         \\
~~ $^{70}$Br &  ~~~~$\,$80.2 $\pm$ 0.8 $\,$~~[Al78] & $\,$~78.54 $\pm$ 0.59 [Bu88] &    &  &  & ~~~79.12 $\pm$ 0.79 & 1.7       \\
~~ $^{74}$Rb &  ~~~64.90 $\pm$ 0.09 ~[Oi01] & 64.761 $\pm$ 0.031$\,$[Ba01] &    &   &  & $\,$~64.776 $\pm$ 0.043 &  1.5   \\

\end{tabular}
\end{ruledtabular}
\end{table*}
\endgroup

Finally, the branching-ratio measurements are presented in Table~\ref{R}.  The decays of the $T_z = 0$
parents are the most straightforward since, in all these cases, the superallowed branch accounts for
$>$99.5\% of the total decay strength.  Thus, even imprecise measurements of the weak non-superallowed
branches can be subtracted from 100\% to yield the superallowed branching ratio with good relative precision.
For the higher-Z parents of this type, particularly $^{62}$Ga and heavier, it has been shown theoretically
\cite{Ha02} and experimentally (\cite{Fi08} for $^{62}$Ga, and \cite{Pi03,Du13} for $^{74}$Rb) that numerous
very-weak Gamow-Teller transitions occur, which, in total, can carry significant decay strength.  Where such
unobserved transitions are expected to exist but have not already been accounted for in the quoted references,
we have used a combination of experiment and theory to arrive at an upper limit for the unobserved strength,
with uncertainties being adjusted accordingly.

The branching ratios for decays from $T_z = -1$ parents are much more challenging to determine, since the
superallowed branch is usually one of several strong branches -- with the notable exception of $^{14}$O --
and, in two of the measured cases, it actually has a branching ratio of less than 10\%.  For the decays of
$^{22}$Mg and $^{38}$Ca, the superallowed branching ratio has been experimentally determined and the result
published, so no special treatment was required for them.  However, the decays of $^{18}$Ne, $^{26}$Si,
$^{30}$S, $^{34}$Ar and $^{42}$Ti had to be treated differently.  In each case, the absolute branching
ratio for a single $\beta$-transition has been measured. The branching ratios for other $\beta$-transitions
then had to be determined from the relative intensities of $\beta$-delayed $\gamma$ rays in the daughter.  
The relevant $\gamma$-ray intensity measurements appear in Table~\ref{BDG}, with their averages then being
used to determine the superallowed branching-ratio averages shown in bold type in Table~\ref{R}.  These
cases are also flagged with a footnote in that table.

\begingroup
\squeezetable
\begin{table*}
\caption{Measured results from which the branching ratios, R, have been derived for superallowed $\beta$-transitions.
The lines giving the average superallowed branching ratios themselves are in bold print. ( See Table~\ref{ref} for
the correlation between the alphabetical reference code used in this table and the actual reference numbers.)
\label{R}}
\begin{ruledtabular}
\begin{tabular}{llllllll}
\multicolumn{2}{c}{Parent/Daughter}
 & Daughter state
 & \multicolumn{2}{c}{Measured Branching Ratio, R (\%)}
 & \multicolumn{1}{c}{}
 & \multicolumn{2}{c}{Average value} \\[1mm]
\cline{4-6} 
\cline{7-8} \\[-1mm]
   \multicolumn{2}{c}{nuclei}
 & \multicolumn{1}{c}{$E_x$ (MeV)}
 & \multicolumn{1}{c}{1}
 & \multicolumn{1}{c}{2} 
 & \multicolumn{1}{c}{} 
 & \multicolumn{1}{c}{R (\%)} 
 & \multicolumn{1}{c}{scale} \\[1mm]

\hline
& & & & & & & \\[-1mm]
 $T_z = -1$: & & & & & & & \\
~~ $^{10}$C & $^{10}$B & 2.16 & 0$^{+0.0008}_{-0}$ [Go72] & & & 0$^{+0.0008}_{-0}$ &   \\
 & & {\bf 1.74} & 1.468 $\pm$ 0.014 [Ro72]& 1.473 $\pm$ 0.007 [Na91] & & & \\
 & & & 1.465 $\pm$ 0.009 [Kr91] & 1.4625 $\pm$ 0.0025 [Sa95] & & & \\
 & & & 1.4665 $\pm$ 0.0038 [Fu99] & & & {\bf 1.4646 $\pm$ 0.0019} & {\bf 1.0} \\
~~ $^{14}$O & $^{14}$N & gs & 0.68 $\pm$ 0.10 [Sh55,To05] & 0.74 $\pm$ 0.05 [Fr63,To05]] & & & \\
 & & & 0.54 $\pm$ 0.02 [Si66,To05] & & & 0.571 $\pm$ 0.068 & 3.7 \\
 & & 3.95 & 0.062 $\pm$ 0.007 [Ka69] & 0.058 $\pm$ 0.004 [Wi80] & & & \\
 & & & 0.053 $\pm$ 0.002 [He81] & & & 0.0545 $\pm$ 0.0019 & 1.1 \\
 & & {\bf 2.31} & & & & {\bf 99.374 $\pm$ 0.068} & \\
~~ $^{18}$Ne & $^{18}$F & {\bf 1.04} & 9 $\pm$ 3 [Fr63] & 7.69 $\pm$ 0.21\footnotemark[1] [Ha75] & & {\bf 7.70 $\pm$ 0.21} & {\bf 1.0} \\
~~ $^{22}$Mg & $^{22}$Na & {\bf 0.66} & 54.0 $\pm$ 1.1 [Ha75] & 53.15 $\pm$ 0.12 [Ha03] & & {\bf 53.16 $\pm$ 0.12} & {\bf 1.0} \\
~~ $^{26}$Si & $^{26}$Al & 1.06 & 21.8 $\pm$ 0.8 [Ha75] & 21.21 $\pm$ 0.64 [Ma08] & & 21.44 $\pm$ 0.50 & 1.0 \\
 & & {\bf 0.23} & & & & {\bf 75.49 $\pm$ 0.57\footnotemark[1]} & \\
~~ $^{30}$S & $^{30}$P & gs & 20 $\pm$ 1 [Fr63] & & & 20 $\pm$ 1 &  \\
 & & {\bf 0.68} & & & & {\bf 77.4 $\pm$ 1.0\footnotemark[1]} & \\
~~ $^{34}$Ar & $^{34}$Cl & 0.67 & 2.49 $\pm$ 0.10 [Ha74a] & & & 2.49 $\pm$ 0.10 & \\
 & & {\bf gs} & & & & {\bf 94.45 $\pm$ 0.25\footnotemark[1]} & \\
~~ $^{38}$Ca & $^{38}$K & {\bf 0.13} & 77.28 $\pm$ 0.16 [Pa14] & & & {\bf 77.28 $\pm$ 0.16} & \\
~~ $^{42}$Ti & $^{42}$Sc & 0.61 & 56 $\pm$ 14 [Al69] & 51.1 $\pm$ 1.1 [Ku09] & & 51.1 $\pm$ 1.1 & 1.0 \\
 & & {\bf gs} & & & & {\bf 47.7 $\pm$ 1.3\footnotemark[1]} & \\[1mm]
 $T_z = 0$: & & & & & & &  \\
~~ $^{26m}$Al & $^{26}$Mg & {\bf gs} & $>$99.997 [Ki91] & $>$99.9985 [Fi12] & & {\bf 100.0000}$\bm{^{+0}_{-0.0015}}$ & \\
~~ $^{34}$Cl & $^{34}$S & {\bf gs} & $>$99.988 [Dr75] & & & {\bf 100.000}$\bm{^{+0}_{-0.012}}$ & \\
~~ $^{38m}$K & $^{38}$Ar & 3.38 & $<$0.0019 [Ha94] & $<$0.0008 [Le08] & & 0.0000$^{+0.0008}_{-0}$ & \\
 & & gs($^{38}$K)\footnotemark[2] & 0.0330 $\pm$ 0.0043 [Le08] & & & 0.0330 $\pm$ 0.0043  & \\
 & & {\bf gs} & & & & {\bf 99.9670} $\bm{^{+0.0043}_{-0.0044}}$ & \\
~~ $^{42}$Sc & $^{42}$Ca & 1.84 & 0.0063 $\pm$ 0.0026 [In77] & 0.0022 $\pm$ 0.0017 [De78] & & & \\
 & & & 0.0103 $\pm$ 0.0031 [Sa80] & 0.0070 $\pm$ 0.0012 [Da85] & & 0.0059 $\pm$ 0.0014 & 1.6 \\
  & & {\bf gs} & & & & {\bf 99.9941 $\pm$ 0.0014} & \\
~~$^{46}$V & $^{46}$Ti & 2.61 & 0.0039 $\pm$ 0.0004 [Ha94] & & & 0.0039 $\pm$ 0.0004 & \\
 & & 4.32 & 0.0113 $\pm$ 0.0012 [Ha94] & & & 0.0113 $\pm$ 0.0012 & \\
 & & $\Sigma$GT\footnotemark[3] & $<$0.01 & & & 0.00$^{+0.01}_{-0}$ & \\
 & & {\bf gs} & & & & {\bf 99.9848}$\bm{^{+0.0013}_{-0.0042}}$ & \\
~~$^{50}$Mn & $^{50}$Cr & 3.63 & 0.057 $\pm$ 0.003 [Ha94] & & & 0.057 $\pm$ 0.003 & \\
 & & 3.85 & $<$0.0003 [Ha94] & & & $0.0000^{+0.0003}_{-0}$ & \\
 & & 5.00 & 0.0007 $\pm$ 0.0001 [Ha94] & & & 0.0007 $\pm$ 0.0001 & \\
 & & {\bf gs} & & & & {\bf 99.9423 $\pm$ 0.0030} & \\
~~$^{54}$Co & $^{54}$Fe & 2.56 & 0.0045 $\pm$ 0.0006 [Ha94] & & & 0.0045 $\pm$ 0.0006 & \\
 & & $\Sigma$GT\footnotemark[3] & $<$0.03 & & & 0.00$^{+0.03}_{-0}$ & \\
 & & {\bf gs} & & & & {\bf 99.9955}$\bm{^{+0.0006}_{-0.0300}}$ & \\
~~$^{62}$Ga & $^{62}$Zn & $\Sigma$GT\footnotemark[3] & 0.142 $\pm$ 0.008 [Fi08] & 0.107 $\pm$ 0.024 [Be08] &
 & 0.139 $\pm$ 0.011 & 1.4 \\
 & & {\bf gs} & & & & {\bf 99.862 $\pm$ 0.011} & \\
~~$^{74}$Rb & $^{74}$Kr & $\Sigma$GT\footnotemark[3] & 0.5 $\pm$ 0.1 [Pi03] & 0.455 $\pm$ 0.031 [Du13] & & 0.459 $\pm$ 0.030 & \\
 & & {\bf gs} & & & & {\bf 99.541 $\pm$ 0.030} & \\[-3mm]

\footnotetext[1]{Result also incorporates data from Table~\ref{BDG}}.
\footnotetext[2]{The decay of $^{38m}$K includes a weak $\gamma$-ray branch to the $^{38}$K ground state, which competes with the $\beta$ decay.}
\footnotetext[3]{designates total Gamow-Teller transitions to levels not explicitly listed; in cases where upper limits are shown, they were derived
with the help of calculations in [Ha02] or with refined versions of those calculations.}
\end{tabular}
\end{ruledtabular}
\end{table*}
\endcenter
\endgroup

\begingroup
\squeezetable
\begin{table*}
\caption{Relative intensities of $\beta$-delayed $\gamma$-rays in the superallowed $\beta$-decay daughters.  These data
are used to determine some of the branching ratios presented in Table~\ref{R}.  (See Table~\ref{ref}
for the correlation between the alphabetical reference code used in this table and the actual reference numbers.)
\label{BDG}}
\begin{ruledtabular}
\begin{tabular}{llllllll}
\multicolumn{2}{c}{Parent/Daughter}
 & \multicolumn{1}{c}{daughter}
 & \multicolumn{2}{c}{Measured $\gamma$-ray Ratio}
 & \multicolumn{1}{c}{}
 & \multicolumn{2}{c}{Average value} \\[1mm]
\cline{4-6} 
\cline{7-8} \\[-2mm]
   \multicolumn{2}{c}{nuclei}
 & \multicolumn{1}{c}{ratios\footnotemark[1]}
 & \multicolumn{1}{c}{1}
 & \multicolumn{1}{c}{2} 
 & \multicolumn{1}{c}{} 
 & \multicolumn{1}{c}{Ratio} 
 & \multicolumn{1}{c}{scale} \\[1mm]

\hline
& & & & & & & \\[-1mm]
$^{18}$Ne & $^{18}$F & $\gamma_{660}/\gamma_{1042}$ & ~$\,$0.0169 $\pm$ 0.0004 ~[He82] & ~$\,$0.0172 $\pm$ 0.0005 ~$\,$[Ad83] & & & \\
 & & & 0.01733 $\pm$ 0.00012 [Gr13] & & & 0.01729 $\pm$ 0.00011 & 1.0 \\
$^{26}$Si & $^{26}$Al & $\gamma_{1622}/\gamma_{829}$ & ~~~0.149 $\pm$ 0.016 ~~~[Mo71] & ~~~0.134 $\pm$ 0.005 ~~$\,$[Ha75]& & & \\
 & & & ~$\,$0.1245 $\pm$ 0.0023 ~$\,$[Wi80]& ~$\,$0.1301 $\pm$ 0.0062 ~[Ma08] & & ~$\,$0.1269 $\pm$ 0.0026 & 1.3 \\
 & & $\gamma_{1655}/\gamma_{829}$ & 0.00145 $\pm$ 0.00032 [Wi80] & & & 0.00145 $\pm$ 0.00032 & \\
 & & $\gamma_{1843}/\gamma_{829}$ & ~~~0.013 $\pm$ 0.003 ~~~[Mo71] & ~~~0.016 $\pm$ 0.003 ~~$\,$[Ha75] & & & \\
 & & & 0.01179 $\pm$ 0.00027 [Wi80] & & & 0.01183 $\pm$ 0.00027 & 1.0 \\
 & & $\gamma_{2512}/\gamma_{829}$ &0.00282 $\pm$ 0.00010 [Wi80] & & & 0.00282 $\pm$ 0.00010 & \\
 & & $\gamma_{\rm total}/\gamma_{829}$ & & & & ~$\,$0.1430 $\pm$ 0.0026 & \\
$^{30}$S & $^{30}$P & $\gamma_{709}/\gamma_{677}$ & ~~~0.006 $\pm$ 0.003 ~~~[Mo71] & ~$\,$0.0037 $\pm$ 0.0009 ~[Wi80] & & ~$\,$0.0039 $\pm$ 0.0009 & 1.0 \\
 & & $\gamma_{2341}/\gamma_{677}$ & ~~~0.033 $\pm$ 0.002 ~~~[Mo71] & ~$\,$0.0290 $\pm$ 0.0006 ~[Wi80] & & ~$\,$0.0293 $\pm$ 0.0011 & 1.9 \\
 & & $\gamma_{3019}/\gamma_{677}$ & 0.00013 $\pm$ 0.00006 [Wi80] & & & 0.00013 $\pm$ 0.00006 & \\
 & & $\gamma_{\rm total}/\gamma_{677}$ & & & & ~$\,$0.0334 $\pm$ 0.0014 & \\
$^{34}$Ar & $^{34}$S & $\gamma_{461}/\gamma_{666}$ & ~~~~$\,$0.28 $\pm$ 0.16 ~~~~$\:$[Mo71] & ~~~0.365 $\pm$ 0.036 ~~$\,$[Ha74a] & & ~~~0.361 $\pm$ 0.035 & 1.0 \\
 & & $\gamma_{2580}/\gamma_{666}$ & ~~~~$\,$0.38 $\pm$ 0.09 ~~~~$\:$[Mo71] & ~~~0.345 $\pm$ 0.010 ~~$\,$[Ha74a] & & ~~~0.345 $\pm$ 0.010 & 1.0 \\
 & & $\gamma_{3129}/\gamma_{666}$ & ~~~~$\,$0.67 $\pm$ 0.08 ~~~~$\:$[Mo71] & ~~~0.521 $\pm$ 0.012 ~~$\,$[Ha74a] & & ~~~0.524 $\pm$ 0.022 & 1.8 \\
 & & $\gamma_{\rm total}/\gamma_{666}$ & & & & ~~~1.231 $\pm$ 0.043 & \\
$^{42}$Ti & $^{42}$Sc & $\gamma_{2223}/\gamma_{611}$ & ~~~0.012 $\pm$ 0.004 ~~~[Ga69] & & & ~~~0.012 $\pm$ 0.004 & \\
 & & $\gamma_{\rm total}/\gamma_{611}$ & ~~~0.023 $\pm$ 0.012 [Ga69,En90] & & & ~~~0.023 $\pm$ 0.012 & \\[-3mm]

\footnotetext[1]{$\gamma$-ray intensities are denoted by $\gamma_{E}$, where $E$ is the $\gamma$-ray energy in keV.}
\end{tabular}
\end{ruledtabular}
\end{table*}
\endgroup

\begingroup
\squeezetable
\begin{table*}[!]
\caption{References for which the original decay-energy results have been updated to incorporate the most recent calibration standards.  (See Table~\ref{ref} for the correlation between the alphabetical reference code used in this table and the actual reference numbers.)
\label{update}}
\vskip 1mm
\begin{ruledtabular}
\begin{tabular}{lll}
  \multicolumn{1}{l}{References (parent nucleus)\footnotemark[1]}
& \multicolumn{1}{l}{}
& \multicolumn{1}{l}{Update procedure} \\[1mm]
\hline
\\[-1mm]
\textbullet~Bo64$\,$($^{18}$Ne), Ba84$\,$($^{10}$C), Br94$\,$($^{26m}$Al) &~~~~ & \textbullet~We have converted all original ($p,n$) threshold measurements to $Q$-values \\ 
 Ba98$\,$($^{10}$C), Ha98$\,$($^{38m}$K), To03$\,$($^{14}$O) & &  using the most recent mass excesses [Wa12]. \\[1mm]
\textbullet~Wh77$\,$($^{14}$O) & & \textbullet~This ($p,n$) threshold measurement has been adjusted to reflect more \\
 & & recent calibration $\alpha$-energies [Ry91] before being converted to a $Q$-value. \\[1mm]
\textbullet~Pr67$\,$($^{18}$Ne) & & \textbullet~Before conversion to a $Q$-value, this ($p,n$) threshold was adjusted to reflect a  \\
 & & new value for the $^7$Li($p,n$) threshold [Wh85], which was used as calibration. \\[1mm]
\textbullet~Bu61$\,$($^{14}$O), Ba62$\,$($^{14}$O) & & \textbullet~These $^{12}$C($^3$He,$n$) threshold measurements have been adjusted for updated \\
 & & calibration reactions based on current mass excesses [Wa12]. \\[1mm]
\textbullet~Ha74d$\,$($^{34}$Cl) & & \textbullet~This ($^3$He,$t$) reaction $Q$-value was calibrated by the $^{27}$Al($^3$He,$t$) reaction \\
 & & to excited states in $^{27}$Si; it has been revised according to modern mass \\
 & & excesses [Wa12] and excited-state energies [En98]. \\[1mm]
\textbullet~Ba88 and Ba89$\,$($^{10}$C) & & \textbullet~These measurements of excitation energies in $^{10}$B have
been updated to \\
 & & modern $\gamma$-ray standards [He00]. \\[1mm] 
\textbullet~Ki89$\,$($^{42}$Sc) & & \textbullet~This $^{41}$Ca($p,\gamma$) reaction $Q$-value was measured relative to that for $^{40}$Ca($p,\gamma$); \\
 & & we have slightly revised the result based on modern mass excesses [Wa12]. \\[1mm]
\textbullet~Ha74c$\,$($^{22}$Mg, $^{26}$Si, $^{30}$S, $^{34}$Ar) & & \textbullet~These ($p,t$) reaction $Q$-values have been adjusted to reflect the current $Q$- \\
 & & value for the $^{16}$O($p,t$) reaction [Wa12], against which they were calibrated. \\[-2mm]    

\footnotetext[1]{These references all appear in Table~\ref{QEC} under the appropriate parent nucleus.}
\end{tabular}
\end{ruledtabular}
\end{table*}
\endgroup

\begingroup
\squeezetable
\begin{table*}[h]
\caption{References from which some or all results have been rejected even though their quoted uncertainties
qualified them for inclusion.  (See Table~\ref{ref} for the correlation between the alphabetical reference
code used in this table and the actual reference numbers.)
\label{reject}}
\vskip 1mm
\begin{ruledtabular}
\begin{tabular}{llll}
  \multicolumn{2}{l}{References (parent nucleus)}
& \multicolumn{1}{l}{}
& \multicolumn{1}{l}{Reason for rejection} \\[1mm]
\hline \\[-2mm]
1. & Decay-energies: &~~~~ & \\[1mm]
 & \textbullet~Pa72$\,$($^{30}$S) & & \textbullet~No calibration is given for the measured ($p,t$) reaction $Q$-values; update \\
 & & & is clearly required but none is possible. \\[1mm]
 & \textbullet~No74$\,$($^{22}$Mg) & & \textbullet~Calibration reaction $Q$-values have changed but calibration process is too \\
 & & & complex to update. \\[1mm]
 & \textbullet~Ro74$\,$($^{10}$C) & & \textbullet~P.H. Barker (co-author) later considered that inadequate attention had \\
 & & & been paid to target surface purity [Ba84]. \\[1mm]
 & \textbullet~Ba77b$\,$($^{10}$C) & & \textbullet~P.H. Barker (co-author) later stated [Ba84] that the ($p,t$) reaction $Q$-value \\
 & & & could not be updated to incorporate modern calibration standards. \\[1mm]
 & \textbullet~Vo77$\,$($^{14}$O, $^{26}$Al$^m$, $^{34}$Cl, $^{42}$Sc, $^{46}$V, $^{50}$Mn, $^{54}$Co)
 & & \textbullet~Most of the results in this reference disagree significantly with more recent \\
 & & & and accurate measurements.  A detailed justification for rejection is presented in \\
 & & & our 2009 survey \cite{HT09}. \\[1mm]
 & \textbullet~Wh81 and Ba98$\,$($^{14}$O) & & \textbullet~The result in [Wh81] was updated in [Ba98] but then eventually withdrawn \\
 & & & by P.H. Barker (co-author) in [To03]. \\
 & \textbullet~Sa09 ($^{70}$Br) & & \textbullet~The result is inconsistent with $Q_{EC}$-value systematics.  See text (Sec.\ref{tables}).  \\[2mm]
2. & Half-lives: & & \\[1mm]
 & \textbullet~He61$\,$($^{14}$O), Ba62$\,$($^{14}$O), Fr63$\,$($^{14}$O), & & \textbullet~Quoted uncertainties are too small, and results likely biased, in light of  \\
 & Fr65$\,$($^{42}$Sc, $^{50}$Mn), Si72$\,$($^{14}$O) & &  statistical difficulties more recently understood (see [Fr69a]).  In particular, \\
 &  & & ``maximum-likelihood" analysis was not used. \\[1mm]
 & \textbullet~Ha72a$\,$($^{34}$Cl, $^{42}$Sc) & & \textbullet~All four quoted half-lives are systematically higher than more recent and \\
 & & & accurate measurements. \\[1mm]
 & \textbullet~Ro74$\,$($^{10}$C) & & \textbullet~P.H. Barker (co-author) later considered that pile-up had been \\
 & & & inadequately accounted for [Ba90]. \\[1mm]
 & \textbullet~Ch84$\,$($^{38m}$K) & & \textbullet~``Maximum-likelihood" analysis was not used.  \\
 & \textbullet~Ma08$\,$($^{26}$Si) & & \textbullet~No account was taken of the beta-detection-efficiency difference \\
 & & & between the parent and daughter activities. See [Ia10] for a more detailed \\
 & & & explanation.  \\[2mm]
3. & Branching-ratios: & & \\[1mm]
 & \textbullet~Fr63$\,$($^{26}$Si)& & \textbullet~Numerous impurities present; result is obviously wrong. \\[1mm]

\end{tabular}
\end{ruledtabular}
\end{table*}
\endgroup

\begingroup
\squeezetable
\begin{table*}
\caption{Reference key, relating alphabetical reference codes used in Tables~\ref{QEC}-\ref{reject} to the actual reference numbers.
\label{ref}}
\vskip 1mm
\begin{ruledtabular}
\begin{tabular}{llllllllllll}
  Table & Reference & Table & Reference & Table & Reference & Table & Reference & Table & Reference & Table & Reference \\ 
  code & number & code & number & code & number & code & number & code & number & code & number \\
\hline \\[-2mm]
  Ad83  & \cite{Ad83}  &
  Aj88  & \cite{Aj88}  &
  Aj91  & \cite{Aj91}  &
  Al69  & \cite{Al69}  &
  Al72  & \cite{Al72}  &
  Al75  & \cite{Al75}  \\
  Al77  & \cite{Al77}  &
  Al78  & \cite{Al78}  &
  Al82  & \cite{Al82}  &
  An70  & \cite{An70}  &
  Az74  & \cite{Az74}  &
  Az75  & \cite{Az75}  \\
  Ba62  & \cite{Ba62}  &
  Ba77a & \cite{Ba77a} &
  Ba77b & \cite{Ba77b} &
  Ba84  & \cite{Ba84}  &
  Ba88  & \cite{Ba88}  &
  Ba89  & \cite{Ba89}  \\
  Ba90  & \cite{Ba90}  &
  Ba98  & \cite{Ba98}  &
  Ba00  & \cite{Ba00}  &
  Ba01  & \cite{Ba01}  &
  Ba04  & \cite{Ba04}  &
  Ba06  & \cite{Ba06}  \\
  Ba09  & \cite{Ba09}  &
  Ba10  & \cite{Ba10}  &
  Be68  & \cite{Be68}  &
  Be78  & \cite{Be78}  &
  Be85  & \cite{Be85}  &
  Be08  & \cite{Be08}  \\
  Bi03  & \cite{Bi03}  &
  Bl04a & \cite{Bl04a} &
  Bl04b & \cite{Bl04b} &
  Bl10  & \cite{Bl10}  &
  Bo64  & \cite{Bo64}  &
  Br94  & \cite{Br94}  \\
  Bu61  & \cite{Bu61}  &
  Bu88  & \cite{Bu88}  &
  Bu06  & \cite{Bu06}  &
  Ca05  & \cite{Ca05}  &
  Ch84  & \cite{Ch84}  &
  Ch13  & \cite{Ch13}  \\
  Cl73  & \cite{Cl73}  &
  Da80  & \cite{Da80}  &
  Da85  & \cite{Da85}  &
  De69  & \cite{De69}  &
  De78  & \cite{De78}  &
  Dr75  & \cite{Dr75}  \\
  Du13  & \cite{Du13}  &
  En90  & \cite{En90}  &
  En98  & \cite{En98}  &
  Er06a & \cite{Er06a} &
  Er06b & \cite{Er06b} &
  Er08  & \cite{Er08}  \\
  Er09a & \cite{Er09a} &
  Er09b & \cite{Er09b} &
  Er11  & \cite{Er11}  &
  Et11  & \cite{Et11}  &
  Fa09  & \cite{Fa09}  &
  Fi08  & \cite{Fi08}  \\
  Fi11  & \cite{Fi11}  &
  Fi12  & \cite{Fi12}  &
  Fr63  & \cite{Fr63}  &
  Fr65  & \cite{Fr65}  &
  Fr69a & \cite{Fr69a} &
  Fr75  & \cite{Fr75}  \\
  Fu99  & \cite{Fu99}  &
  Ga69  & \cite{Ga69}  &
  Ga01  & \cite{Ga01}  &
  Ge07  & \cite{Ge07}  &
  Ge08  & \cite{Ge08}  &
  Gi72  & \cite{Gi72}  \\
  Go72  & \cite{Go72}  &
  Gr07  & \cite{Gr07}  &
  Gr08  & \cite{Gr08}  &
  Gr13  & \cite{Gr13}  &
  Ha67  & \cite{Ha67}  &
  Ha68  & \cite{Ha68}  \\
  Ha72a & \cite{Ha72a} &
  Ha74a & \cite{Ha74a} &
  Ha74b & \cite{Ha74b} &
  Ha74c & \cite{Ha74c} &
  Ha74d & \cite{Ha74d} &
  Ha75  & \cite{Ha75}  \\
  Ha94  & \cite{Ha94}  &
  Ha98  & \cite{Ha98}  &
  Ha02  & \cite{Ha02}  &
  Ha03  & \cite{Ha03}  &
  He61  & \cite{He61}  &
  He81  & \cite{He81}  \\ 
  He82  & \cite{He82}  &
  He00  & \cite{He00}  &
  He01  & \cite{He01}  &
  He02  & \cite{He02}  &
  Ho64  & \cite{Ho64}  &
  Ho74  & \cite{Ho74}  \\
  Hu82  & \cite{Hu82}  &
  Hy03  & \cite{Hy03}  &
  Hy05  & \cite{Hy05}  &
  Ia06  & \cite{Ia06}  &
  Ia08  & \cite{Ia08}  &
  Ia10  & \cite{Ia10}  \\
  In77  & \cite{In77}  &
  Is80  & \cite{Is80}  &
  Je07  & \cite{Je07}  &
  Ji02  & \cite{Ji02}  &
  Ka69  & \cite{Ka69}  &
  Ke07  & \cite{Ke07}  \\
  Ki89  & \cite{Ki89}  &
  Ki91  & \cite{Ki91}  &
  Ko83  & \cite{Ko83}  &
  Ko87  & \cite{Ko87}  &
  Ko97a & \cite{Ko97a} &
  Ko97b & \cite{Ko97b} \\
  Kr91  & \cite{Kr91}  &
  Ku09  & \cite{Ku09}  &
  Kw10  & \cite{Kw10}  &
  Kw13  & \cite{Kw13}  &
  La13  & \cite{La13}  &
  Le08  & \cite{Le08}  \\
  Li94  & \cite{Li94}  &
  Ma94  & \cite{Ma94}  &
  Ma08  & \cite{Ma08}  &
  Mc67  & \cite{Mc67}  &
  Mi67  & \cite{Mi67}  &
  Mo71  & \cite{Mo71}  \\
  Mu04  & \cite{Mu04}  &
  Na91  & \cite{Na91}  &
  No74  & \cite{No74}  &
  Oi01  & \cite{Oi01}  &
  Pa72  & \cite{Pa72}  &
  Pa05  & \cite{Pa05}  \\
  Pa11  & \cite{Pa11}  &
  Pa12  & \cite{Pa12}  &
  Pa14  & \cite{Pa14}  &
  Pi03  & \cite{Pi03}  &
  Pr67  & \cite{Pr67}  &
  Pr90  & \cite{Pr90}  \\
  Ra83  & \cite{Ra83}  &
  Re85  & \cite{Re85}  &
  Ri07  & \cite{Ri07}  &
  Ro70  & \cite{Ro70}  &
  Ro72  & \cite{Ro72}  &
  Ro74  & \cite{Ro74}  \\
  Ro75  & \cite{Ro75}  &
  Ro06  & \cite{Ro06}  &
  Ry73  & \cite{Ry73}  &
  Ry91  & \cite{Ry91}  &
  Sa80  & \cite{Sa80}  &
  Sa95  & \cite{Sa95}  \\
  Sa04  & \cite{Sa04}  &
  Sa05  & \cite{Sa05}  &
  Sa09  & \cite{Sa09}  &
  Sc07  & \cite{Sc07}  &
  Sc11  & \cite{Sc11}  &
  Se73  & \cite{Se73}  \\
  Se05  & \cite{Se05}  &
  Sh55  & \cite{Sh55}  &
  Si66  & \cite{Si66}  & 
  Si72  & \cite{Si72}  &
  So11  & \cite{So11}  & 
  Sq75  & \cite{Sq75}  \\
  Ta12  & \cite{Ta12}  &
  Ti95  & \cite{Ti95}  & 
  To03  & \cite{To03}  &
  To05  & \cite{To05}  &
  Vo77  & \cite{Vo77}  &
  Wa83  & \cite{Wa83}  \\
  Wa92  & \cite{Wa92}  &
  Wa12  & \cite{Wa12}  &
  We68  & \cite{We68}  & 
  Wh77  & \cite{Wh77}  & 
  Wh81  & \cite{Wh81}  & 
  Wh85  & \cite{Wh85}  \\
  Wi76  & \cite{Wi76}  &
  Wi78  & \cite{Wi78}  &
  Wi80  & \cite{Wi80}  &
  Zi87  & \cite{Zi87}  &
    
\end{tabular}
\end{ruledtabular}
\end{table*}
\endgroup

\section{\label{s:Ftvalu} The $\F t$ values}

With the input data now settled, we can proceed to derive the $ft$ values for the 20 superallowed transitions
included in the tables.  We calculate the statistical rate function $f$ using the same code as in our previous
survey.  The basic methodology for the calculation is described in the Appendix to our 2004 survey \cite{HT05},
with refinements applied to incorporate excitation of the daughter atom, as explained in Appendix A of our 2008
survey \cite{HT09}.  Our final $f$ values for the $T$ = 1 transitions of interest here are recorded in the
second column of Table~\ref{Ft}.  They were evaluated with the $Q_{EC}$ values and their uncertainties taken from
column 7 of Table~\ref{QEC}.

The third column of Table~\ref{Ft} lists (as percentages) the electron-capture fraction, $P_{EC}$, calculated
for each of the 20 superallowed transitions.  The method of calculation was described in our 2004 survey \cite{HT05},
to which the reader is referred for more details.  The partial half-life, $t$, for each transition is then
obtained from its total half-life, $t_{1/2}$, branching ratio, $R$, and electron-capture fraction according
to the following formula:
\be
t = \frac{t_{1/2}}{R} \left ( 1 + P_{EC} \right ).
\label{partialt}
\ee
The resultant values for the partial half-lives and the corresponding $ft$ values appear in columns 4 and 5
of the table.

To obtain the $\F t$ from each $ft$ value, we use Eq.~(\ref{Ftdef}) to apply the small transition-dependent
correction terms, $\delta_R^{\prime}$, $\delta_{NS}$ and $\delta_C$.  The values we use for $\delta_R^{\prime}$
appear in column 6 of Table~\ref{Ft} while those of $\delta_C$-$\delta_{NS}$, the combination of the other two terms
that appears in Eq.~(\ref{Ftdef}), are given in column 7. Finally, column 8 of the table contains the derived
$\F t$ values and, at the bottom of the column, their average, $\overline{\F t}$.  The three theoretical corrections applied here, 
together with the transition-independent radiative correction $\DRV$, which is ultimately needed to extract $V_{ud}$,
will be described in more detail in the following section.

\begingroup
\squeezetable
\begin{table*}[!]
\begin{center}
\caption{Derived results for superallowed Fermi beta decays.
\label{Ft}}
\vskip 1mm
\begin{ruledtabular}
\begin{tabular}{rrrrrrrr}
& & & & & & & \\[-3mm]
Parent & & $P_{EC}$ &  Partial half-life & &
&  &  \\
nucleus & \multicolumn{1}{c}{$f$} & (\%) &  \multicolumn{1}{c}{$t(ms)$} &  
\multicolumn{1}{c}{$ft(s)$} &
\multicolumn{1}{c}{$\delta_R^{\prime}$ (\%)} & $\delta_C - \delta_{NS}$ (\%) &
 \multicolumn{1}{c}{$\F t (s)$}  \\[1mm] 
\hline
& & & & & & & \\[-2mm]
\multicolumn{2}{l}{$T_z = -1$:} & & & & & & \\
$^{10}$C & $2.30169 \pm 0.00070$ & 0.299 & $1322100 \pm 1800$~ & $3043.0 \pm 4.3$~~~&
1.679 & $0.520 \pm 0.039$ & $3078.0 \pm 4.5$\footnotemark[1]~~  \\
$^{14}$O & $42.771 \pm 0.023$~$\,$~ & 0.088 & $71126 \pm 50$~~~~ & $3042.2 \pm 2.7$~~ &
1.543 & $0.575 \pm 0.056$ & $3071.4 \pm 3.2$\footnotemark[1]~~  \\
$^{18}$Ne & $134.64 \pm 0.17$~~~~ & 0.081 & $21640 \pm 590\,$~~ & $2914 \pm 79$~~~ &
1.506 & $0.850 \pm 0.052$ & $2932.8 \pm 80$~~~~ \\
$^{22}$Mg & $418.37 \pm 0.17$~~~~ & 0.069 & $7295 \pm 17$~~~~ & $3051.9 \pm 7.2$~~ &
1.466 & $0.605 \pm 0.030$ & $3077.9 \pm 7.3$\footnotemark[1]~~  \\
$^{26}$Si & $1028.03 \pm 0.12$~~~~  & 0.064 & $2976 \pm 23$~~~~ & $3059 \pm 23$~~~ &
1.438 & $0.650 \pm 0.034$ & $3083 \pm 23$~~~~ \\
$^{30}$S  & $1976.71 \pm 0.56$~~~~  & 0.066 & $1520 \pm 21$~~~~ & $3005 \pm 41$~~~ &
1.423 & $1.040 \pm 0.032$ & $3016 \pm 41$~~~~ \\
$^{34}$Ar & $3410.97 \pm 0.61$~~~~  & 0.069 & $894.0 \pm 2.4$~~~ & $3049.6 \pm 8.1$~~ &
1.412 & $0.875 \pm 0.058$ & $3065.6 \pm 8.4$\footnotemark[1]~~ \\
$^{38}$Ca & $5328.88 \pm 0.30$~~~~ & 0.075 & $574.7 \pm 1.3$~~~ & $3062.3 \pm 6.8$~~  & 
1.414 & $0.940 \pm 0.072$ & $3076.4 \pm 7.2$\footnotemark[1]~~  \\
$^{42}$Ti & $7130.5 \pm 1.4$~~~~~$\,$ & 0.087 & $437 \pm 12\,$~~~ & $3114 \pm 84$~~~ &
1.428 & $1.175 \pm 0.080$ & $3121 \pm 84$~~~~ \\[2mm]
\multicolumn{2}{l}{$T_z = 0 $:} & & & & & & \\
$^{26m}$Al & $478.232 \pm 0.081\,$ & 0.083 & $6351.26 ^{+0.54}_{-0.55}$~~$\:$  & $3037.38 \pm 0.58$~ &
1.478 & $0.305 \pm 0.027$ & $3072.9 \pm 1.0$\footnotemark[1]~~  \\
$^{34}$Cl & $1996.003 \pm 0.096$~~ & 0.080 & $1527.77 ^{+0.44}_{-0.47}$~~$\:$ & $3049.43 ^{+0.88}_{-0.95}$~~ & 
1.443 & $0.735 \pm 0.048$  & $3070.7 ^{+1.7}_{-1.8}$\footnotemark[1]~~~  \\
$^{38m}$K & $3297.39 \pm 0.15$~~ & 0.085 & $925.42 \pm 0.28\,$~ & $3051.45 \pm 0.92$$\:$&
1.440 & $0.770 \pm 0.056$ & $3071.6 \pm 2.0$\footnotemark[1]~~  \\
$^{42}$Sc & $4472.23 \pm 1.15$~~ & 0.099 & $681.44 \pm 0.26\,$~ & $3047.5 \pm 1.4$~~ &
1.453 & $0.630 \pm 0.059$ & $3072.4 \pm 2.3$\footnotemark[1]~~  \\
$^{46}$V & $7209.25 \pm 0.54 $~~ & 0.101 & $423.113 ^{+0.053}_{-0.068}$$\:$~ & $3050.32 ^{+0.44}_{-0.46}$~$\:$ &
1.445 & $0.655 \pm 0.063$ & $3074.1 \pm 2.0$\footnotemark[1]~~  \\
$^{50}$Mn & $10745.97 \pm 0.50$~~ & 0.107 & $283.68 \pm 0.11\,$~ & $3048.4 \pm 1.2$~~ &
1.444 & $0.685 \pm 0.055$ & $3071.2 \pm 2.1$\footnotemark[1]~~  \\
$^{54}$Co & $15766.7 \pm 2.9\,$~~~ & 0.111 & $193.493 ^{+0.063}_{-0.086}$~$\:$ &$3050.7 ^{+1.1}_{-1.5}~~~$ &
1.443 & $0.805 \pm 0.068$ & $3069.8 ^{+2.4}_{-2.6}$\footnotemark[1]~~~  \\
$^{62}$Ga & $26400.3 \pm 8.3\,$~~~ & 0.135 & $116.440 \pm 0.042$ & $3074.0 \pm 1.5$~~ & 
1.459 & $1.52 \pm 0.21$~ & $3071.5 \pm 6.7$\footnotemark[1]~~  \\
$^{66}$As & $32120 \pm 460$~~~ & 0.153 &       &     & 
1.468 & $1.61 \pm 0.40$~ &       \\
$^{70}$Br & $38600 \pm 3600\,$~ & 0.173 &       &    &
1.49 & $1.78 \pm 0.25$~ &       \\
$^{74}$Rb & $47281 \pm 93$~~~ & 0.191 & $65.199 \pm 0.047$ & $3082.7 \pm 6.5$~~ &
1.50 & $1.69 \pm 0.27$~ & $3076 \pm 11\footnotemark[1]$~~~  \\[5mm]
& & & & & \multicolumn{2}{r}{Average (best 14), $\overline{\F t}$} & $3072.27 \pm 0.62\,$ \\
& & & & & \multicolumn{2}{r}{$\chi^2/\nu$} & \multicolumn{1}{c}{0.52} \\

\footnotetext[1]{Values used to obtain $\overline{\F t}$}
\end{tabular}
\end{ruledtabular}
\end{center}
\end{table*}
\endgroup

\subsection{\label{ss:theo} Theoretical Corrections}

Of the four theoretical correction terms $\DRV$, $\delta_R^{\prime}$, $\delta_{NS}$ and $\delta_C$ that appear in
Eq.~(\ref{Ftdef}) the first three are radiative corrections, and the fourth is the isospin-symmetry-breaking
correction.  In the following two subsections, one for each category of correction, we briefly describe all four
correction terms and give the sources and justification of the values we use for them.

\subsubsection{\label{sss:radcorr} Radiative corrections}

In a $\beta$-decay half-life experiment, the rate measured includes not
only the bare decay but also radiative decay processes, such as
bremsstrahlung.  Since it is the half-life of the bare $\beta$-decay
process that is required for the $ft$ value, the measured result
has to be amended with a radiative-correction calculation.  The
principal graphs to be evaluated are the one-photon bremsstrahlung,
the $\gamma W$-box and $ZW$-box diagrams.  For calculational convenience
it is standard to separate the contributions from these graphs into
contributions at high photon energies (short distances) and low photon
energies (long distances).

The short-distance correction includes
the $ZW$-box and the high-energy part of the $\gamma W$-box diagrams
and is evaluated by ignoring the hadronic structure and using
free-quark Lagrangians.  This contribution therefore is universal,
being independent of which particular nucleus is involved in the
$\beta$ decay.  We denote this contribution $\DRV$ and, it being universal,
we place it on the right-hand side of Eq.~(\ref{Ftdef}).
The current best value, which we adopt from Marciano and Sirlin \cite{MS06}, is
\be
\DRV = (2.361 \pm 0.038) \% .
\label{DRV}
\ee

The long-distance correction includes the brems-strahlung and the
low-energy part of the $\gamma W$-box diagram; it requires a
model calculation of the hadronic structure.  Contributions to this
correction have been
calculated \cite{Si67,SZ86,JR87,Si87,CWS04} to order $\alpha$, $\alpha^2$
and $Z \alpha^2$, and estimated from the leading-log term in order
$Z^2 \alpha^3$, where $\alpha$ is the fine-structure constant.  In the
latter two orders, the positron in the $\gamma W$-box
and bremsstrahlung diagrams interacts with the Coulomb field of the
nucleus.  We have listed the contributions from each order in Table
V of Ref.~\cite{TH08} so here we only show their sums, $\delta_R^{\prime}$,
for all the transitions of interest.  These appear in column 2 of Table~\ref{t:theocorr},
the table which collects all the theoretical correction terms, and also, for convenience,
in column 6 of Table~\ref{Ft}, which collects all the input to the left-hand
side of Eq.~(\ref{Ftdef}).

In contrast with our previous surveys, we list no uncertainties on the individual
$\delta_R^{\prime}$ values.  In the past, we have taken the uncertainty on each
transition's $\delta_R^{\prime}$ value to be equal to the entire $Z^2 \alpha^3$
contribution.   We have then treated the uncertainty as being statistical, adding
it in quadrature to the experimental uncertainty to obtain the total uncertainty
on the $\F t$ value for that transition.  The latter was in turn handled statistically
in the derivation of an average $\overline{\F t}$ for all the transitions.

We have now revised this prescription in two ways.  First, we reduce the magnitude
of the uncertainty on $\delta_R^{\prime}$ to one-third of the $Z^2 \alpha^3$ term, and
secondly we treat it as a systematic, rather than a statistical effect.  Our previous choice
for the magnitude originated 25 years ago with Sirlin \cite{SZ86}, who at the time did
not include the $Z^2 \alpha^3$ term in the radiative correction itself but used it
only as an estimate of its uncertainty.  One year later, though, he chose instead to
include it in the correction itself \cite{Si87}, while apparently neglecting any
contribution to the uncertainty.  We now believe that our choice to combine both these
approaches by including the $Z^2 \alpha^3$ term in $\delta_R^{\prime}$ and also assigning
it to be the latter's uncertainty was being overly cautious.  Furthermore, because the
uncertainty is associated with the $Z^2 \alpha^3$ term, it is expected to be a smooth
function of $Z^2$ and thus to behave systematically since any shift in the value of
$\delta_R^{\prime}$ must affect all $\F t$ values in the same direction.

We then proceed as follows:  We evaluate the individual transition $\F t$ values without
including any uncertainties associated with $\delta_R^{\prime}$, and obtain an average
$\overline{\F t}$.  Then we shift all the individual $\delta_R^{\prime}$ terms up and
down by one-third of the $Z^2 \alpha^3$ contribution, recalculate the $\F t$ values and
determine $\overline{\F t}$ for both.  The shifts in the value of the latter -- $\pm 0.36
{\rm s}$ for the data in Table~\ref{Ft} -- becomes the systematic uncertainty assigned to
$\overline{\F t}$ to account for the uncertainty in $\delta_R^{\prime}$.  Note that our
choice to take one-third of the $Z^2 \alpha^3$ term is rather arbitrary, but has the
benefit that it is still conservative and at the same time results in the uncertainty in
$\delta_R^{\prime}$ having an impact on the overall result that is comparable to its impact
in our previous survey \cite{HT09}.

We turn now to the third radiative term $\delta_{NS}$, which arises from an evaluation of the
low-energy part of the $\gamma W$-box graph for an axial-vector weak interaction.  If it is
assumed that the $\gamma N$ and $WN$-vertices are both with the same nucleon, $N$, then
the evaluated box graph becomes proportional to the Fermi $\beta$-decay operator
yielding a universal correction already included in $\DRV$.

If instead the $\gamma$- and $W$-interactions in the $\gamma W$-box graph for an axial-vector
current are with different nucleons in the nucleus, then the evaluation involves two-nucleon operators,
which necessitates a nuclear-structure calculation.  This component of the radiative correction
we denote by $\delta_{NS}$ and list its values in column 3 of Table~\ref{t:theocorr}.  The
values and their uncertainties have been taken from Table VI in Ref.~\cite{TH08}.  For this
correction term, a number of model calculations were carried out for each nucleus \cite{TH08}
and the uncertainties listed were chosen to encompass the spread in the results from these
calculations.  Therefore the uncertainty is nucleus-specific and, as such, can be treated as
statistical and not systematic. We thus combine it in quadrature with the experimental errors
in determining the $\F t$-value uncertainties.

\subsubsection{\label{sss:dcspecific}Isospin-symmetry-breaking correction}

In this section we describe only the set of isospin-symmetry-breaking corrections, $\delta_C$, 
that we have used in deriving the corrected $\F t$ values given in Table~\ref{Ft}.  A discussion
of other alternative calculations of $\delta_C$ -- and our reasons for rejecting them -- is
postponed to Sect.~\ref{s:dcgeneral}.  The set we have selected follows from a semi-phenomenological
approach based on the shell-model combined with Woods-Saxon radial-functions.  This model, 
which we designate as SM-WS, has been described in detail by us in \cite{TH08}, where also the
results for $\delta_C$ are tabulated.  We describe the model only briefly here, while making two minor
updates to our previous results.

The calculation is done in two parts, which is made possible by our dividing $\delta_C$ into two terms:
\be 
\delta_C = \delta_{C1} + \delta_{C2} .
\label{c1c2} 
\ee
The idea is that $\delta_{C1}$ follows from a tractable shell-model calculation that does not include
significant nodal mixing, while $\delta_{C2}$ corrects for the nodal mixing that would be present
if the shell-model space were much larger.

For $\delta_{C1}$, a modest shell-model space (usually one major oscillator shell) is employed, in
which Coulomb and other charge-dependent terms are added to the charge-independent effective
Hamiltonian customarily used for the shell model.  These charge-dependent additional terms are
separately adjusted for each superallowed $\beta$ transition in order to reproduce the $b$- and
$c$-coefficients of the isobaric multiplet mass equation (IMME) for the triplet of $T=1$, $0^+$
states that includes the parent and daughter states of the transition.

Since the Coulomb force is long range, its influence in configuration space extends much further
than the single major oscillator shell included in the calculation of $\delta_{C1}$.  To incorporate
the effects of multi-shell mixing, we note first that its principal impact is to change the structure
of the radial wave function by introducing mixing with radial functions that have more nodes.  
Since this mixing primarily affects protons, it results in proton radial functions that differ
from the neutron ones so, when the overlap is computed, its departure from unity determines the
value of $\delta_{C2}$.  The radial functions themselves are derived from a Woods-Saxon potential. 
Again there is a case-by-case adjustment in the Woods-Saxon potentials to ensure that the different
measured proton and neutron separation energies in the $\beta$-decay parents and daughters are
correctly reproduced.

\begin{table*}
\caption{Corrections $\delta_R^{\prime}$, $\delta_{NS}$ and $\delta_C$ that
are applied to experimental $ft$ values to obtain $\F t$ values.
\label{t:theocorr}}
\begin{ruledtabular}
\begin{tabular}{rddddd}
\multicolumn{1}{r}{Parent} &
\multicolumn{1}{r}{$\delta_R^{\prime}$} &
\multicolumn{1}{r}{$\delta_{NS}$} &
\multicolumn{1}{r}{$\delta_{C1}$} &
\multicolumn{1}{r}{$\delta_{C2}$} &
\multicolumn{1}{r}{$\delta_{C}$} \\
\multicolumn{1}{r}{nucleus} &
\multicolumn{1}{r}{$(\%)$} &
\multicolumn{1}{r}{$(\%)$} &
\multicolumn{1}{r}{$(\%)$} &
\multicolumn{1}{r}{$(\%)$} &
\multicolumn{1}{r}{$(\%)$} \\[1mm]
\hline
& & & & & \\[-1mm]
$T_z = -1:$ & & & & & \\
\nuc{10}{C} &  1.679 & -0.345(35) & 0.010(10) & 0.165(15) & 0.175(18) \\ 
\nuc{14}{O} &  1.543 & -0.245(50) & 0.055(20) & 0.275(15) & 0.330(25) \\
\nuc{18}{Ne} & 1.506 & -0.290(35) & 0.155(30) & 0.405(25) & 0.560(39) \\
\nuc{22}{Mg} & 1.466 & -0.225(20) & 0.010(10) & 0.370(20) & 0.380(22) \\
\nuc{26}{Si} & 1.439 & -0.215(20) & 0.030(10) & 0.405(25) & 0.435(27) \\
\nuc{30}{S} &  1.423 & -0.185(15) & 0.155(20) & 0.700(20) & 0.855(28) \\
\nuc{34}{Ar} & 1.412 & -0.180(15) & 0.030(10) & 0.665(55) & 0.695(56) \\
\nuc{38}{Ca} & 1.414 & -0.175(15) & 0.020(10) & 0.745(70) & 0.765(71) \\
\nuc{42}{Ti} & 1.427 & -0.235(20) & 0.105(20) & 0.835(75) & 0.940(78) \\[3mm]
$T_z = 0:$ & & & & & \\
\nuc{26m}{Al}& 1.478 & 0.005(20)  & 0.030(10) & 0.280(15) & 0.310(18) \\
\nuc{34}{Cl} & 1.443 & -0.085(15) & 0.100(10) & 0.550(45) & 0.650(46) \\
\nuc{38m}{K} & 1.440 & -0.100(15) & 0.105(20) & 0.565(50) & 0.670(54) \\
\nuc{42}{Sc} & 1.453 & 0.035(20)  & 0.020(10) & 0.645(55) & 0.665(56) \\
\nuc{46}{V} &  1.445 & -0.035(10) & 0.075(30) & 0.545(55) & 0.620(63) \\
\nuc{50}{Mn} & 1.444 & -0.040(10) & 0.035(20) & 0.610(50) & 0.645(54) \\
\nuc{54}{Co} & 1.443 & -0.035(10) & 0.050(30) & 0.720(60) & 0.770(67) \\
\nuc{62}{Ga} & 1.459 & -0.045(20) & 0.275(55) & 1.20(20)  & 1.48(21) \\
\nuc{66}{As} & 1.468 & -0.060(20) & 0.195(45) & 1.35(40)  & 1.55(40) \\
\nuc{70}{Br} & 1.486 & -0.085(25) & 0.445(40) & 1.25(25)  & 1.70(25) \\
\nuc{74}{Rb} & 1.499 & -0.075(30) & 0.115(60) & 1.50(26)  & 1.62(27) \\
\end{tabular}
\end{ruledtabular}
\end{table*}

The SM-WS calculations of Towner and Hardy \cite{TH08} must clearly be classified as semi-phenological.  A
number of transition-specific nuclear properties have been fitted in their determination of $\delta_C$.  In
contrast, most of the alternative models discussed in Sect.~\ref{s:dcgeneral} are first-principles theory
calculations.  They have no local phenomenological constraints and therefore are not capable of offering
the precision required for our $\F t$ analysis.  Nevertheless, they play a very important role in confirming
that the semi-phenomenological results are not inconsistent with the broad features predicted by a
first-principles calculation.

The values we use for $\delta_{C1}$ are given in column 4 of Table~\ref{t:theocorr}.  They differ slightly
from the published values in \cite{TH08} because a new compilation of IMME coefficients by MacCormick and
Audi \cite{MA14}, based on the 2012 atomic mass evaluation \cite{AME12}, has made small changes to these
coefficients.  Also, there is still insufficient experimental data for nuclei in the upper $pf$ shell,
so the compilation of coefficients for $T$=1 multiplets ends at $A$=58.  For the four superallowed decays
from $^{62}$Ga to $^{74}$Rb, we have used an extrapolation formula \cite{MA14} to estimate the $c$-coefficients and
then used the relation $Q_{ec} = -b-c$ to obtain the $b$-coefficient for each case.  With these new
IMME coefficients, we re-evaluated all the $\delta_{C1}$ values, yielding the results shown in Table~\ref{t:theocorr}.

The values we use for $\delta_{C2}$ are shown in column 5 of Table~\ref{t:theocorr} and are the same as
those previously published \cite{TH08} for all but four cases where an update has been effected.  To explain the origin
of the update we need to explain some details of the $\delta_{C2}$ calculation:  As already mentioned, the
radial functions used to calculate the radial overlap are taken to be eigenfunctions of a phenomenological
Woods-Saxon potential.  The radius parameter of this potential is determined by our requiring that the
charge density constructed from the proton eigenfunctions of the potential yields a root-mean-charge radius
in agreement with the experimental value measured by electron scattering \cite{deV87}.  However, in most
cases the experimental charge radius is known only for the stable isotope of the element of interest, whereas
our need is for the radius of the unstable beta-decaying isotope.  Thus, we add an estimated isotope shift 
to the nearby measured rms radius and apply a generous uncertainty.  This uncertainty is only one of three 
contributions to the final uncertainty quoted for each $\delta_{C2}$ value.  The other two account for:
{\it a}) the scatter in the results from three different methodologies, and {\it b}) the scatter in the
results from different shell-model interactions used to compute the required spectroscopic amplitudes \cite{TH02}. 

The issue of the appropriate experimental charge radius has not been revisited since our 2002 work \cite{TH02}.
Since then, a new compilation of charge radii has been published \cite{An04}, including not only results from electron
scattering, but also values obtained from muonic-atom $X$-rays, $K_{\alpha}$ isotope shifts and optical shifts.
In this compilation, radii are given for three of the beta-decaying isotopes of relevance to our superallowed
beta-decay studies: $^{18}$Ne, $^{34}$Ar and $^{38m}$K.  In addition, more recently, collinear-laser spectroscopy
on the neutron-deficient Rb isotopes enabled the charge radius of $^{74}$Rb to be determined from its hyperfine
splitting \cite{Ma11}.  For these four cases, therefore, we have recomputed the $\delta_{C2}$ correction. 

For the lightest three cases, $^{18}$Ne, $^{34}$Ar and $^{38m}$K, the change in the rms charge radius was sufficient
to produce a noticeable shift in the $\delta_{C2}$ value, though not outside our previously quoted uncertainty.
Unfortunately, though, the reduction in the error on the rms charge radius did not significantly lower the overall
uncertainty assigned to $\delta_{C2}$ because in all three cases the uncertainty is dominated by the spread in the
results obtained from the three different methodologies.  For the heaviest case, $^{74}$Rb, the revision in the
rms radius was small, so it made no change in the value of $\delta_{C2}$ but did reduce its uncertainty.  However, even though
the uncertainty in the radius was reduced by a factor of ten, it only led to a 20\% reduction in the uncertainty for
$\delta_{C2}$.  For all four cases the revised results appear in Table~\ref{t:theocorr}.  

The sum of $\delta_{C1}$ and $\delta_{C2}$ is shown in the last column of Table~\ref{t:theocorr}.  As with
$\delta_{NS}$, uncertainties have been assigned to $\delta_C$ which are nucleus-specific.  They represent the spread of
results obtained with different shell-model interactions and different methodologies, as well as uncertainties in rms
radii and IMME coefficients: all for the specific nuclei involved in each transition.  We therefore treat them
subsequently as statistical uncertainties.

\subsection{Consistency of $\F t$ values}
\label{ss:cons}

The experimental and theoretical results appearing in columns 2-7 of Table~\ref{Ft} have been treated as input to
Eqs.~(\ref{Ftdef}) and (\ref{partialt}), from which we derive the $\F t$ values listed in column 8.  All the input
uncertainties that appear in the table are combined in quadrature to obtain the $\F t$-value uncertainties.  

\begin{figure}[t]
\epsfig{file=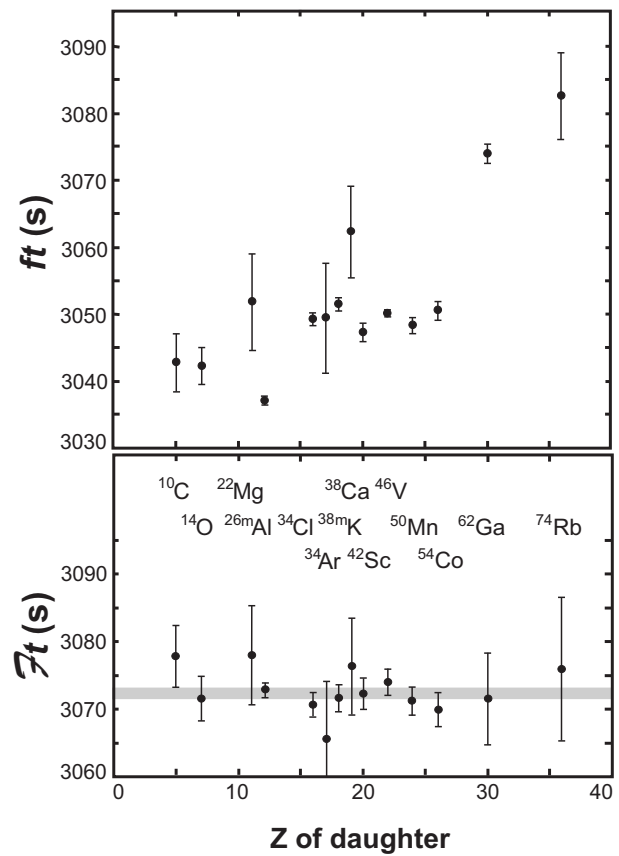,width=8cm}
\caption{In the top panel are plotted the uncorrected experimental $ft$ values as a function of the
charge on the daughter nucleus.  In the bottom panel, the corresponding $\F t$ values are given; they
differ from the $ft$ values by the inclusion of the correction terms $\delta_R^{\prime}$, $\delta_{NS}$
and $\delta_C$.  The horizontal grey band in the bottom panel gives one standard deviation around the
average $\overline{\F t}$ value.}
\label{fig-2}
\end{figure}

There are now 14 superallowed transitions whose $\F t$ values have uncertainties less than $\pm$0.4\%, with the best
case, $^{26}$Al$^m$, being known an order of magnitude better than that.  The uncorrected $ft$ values and the fully
corrected $\F t$ values for these transitions are plotted as a function of the daughter $Z$ values in the top and
bottom panels, respectively, of Fig.~\ref{fig-2}.  Readily evident is the remarkable efficacy of the applied
corrections in converting the scatter of the $ft$ values into a self-consistent set of $\F t$ values. Such consistency
is an expectation of CVC and an essential prerequisite if the data are to be used in any further probes of the Standard
Model.  The results in column 8 of Table~\ref{Ft} and the bottom panel of Fig.~\ref{fig-2} clearly satisfy the test,
the weighted average of the 14 most precise results being given at the bottom of column 8 along with the corresponding
chi-square per degree of freedom of $\chi^2/\nu = 0.52$.  

Although the $\F t$-value results depend on theoretical correction terms in addition to primary experimental data, we
treat all of the uncertainties in Table~\ref{Ft} as being statistical in nature for reasons explained in Sec.~\ref{ss:theo}.  
This leaves the uncertainty associated with $\delta^{\prime}_R$, which is derived as a systematic effect in
Sec.~\ref{sss:radcorr}, still to be applied to the average $\overline{\F t}$.  Thus, the final result for $\overline{\F t}$
becomes 
\bea
\overline{\F t} & = & 3072.27 \pm 0.62_{stat} \pm 0.36_{\delta_R^{\prime}}~s
\nonumber \\
& = & 3072.27 \pm 0.72 ~s,
\label{Ftavg}
\eea  
where the two uncertainties have been combined in quadrature on the second line.  Since $\F t$ is proportional to the
square of the vector coupling constant, $\GV$, then Eq.~(\ref{Ftavg}) can be said to confirm the constancy of $\GV$
-- and to verify this key component of the CVC hypothesis -- at the level of $1.2\times10^{-4}$. 

Compared with the results of our last survey \cite{HT09}, the value of $\overline{\F t}$ in Eq.~(\ref{Ftavg})
is somewhat higher, though well within the previous error bars, and carries a smaller uncertainty.  The value
of $\chi^2/\nu$ associated with the current $\overline{\F t}$ result is higher than the corresponding value
in 2008 but this undoubtedly reflects the fact that one additional transition has been added and the data
for some of the other transitions are more precise today than they were six years ago.  In any case, the
confidence level for the new result remains very high: 91\%.

\subsection{\label{ss:errors} Uncertainty budgets}

\begin{figure}[t]
\epsfig{file=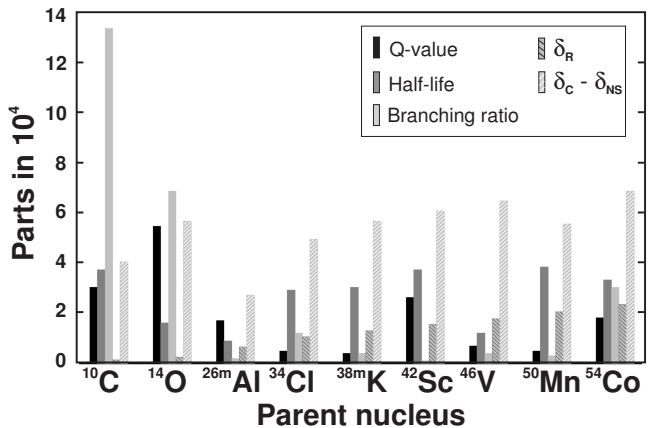,width=8.5cm}
\caption{Summary histogram of the fractional uncertainties attributable to each 
experimental and theoretical input factor that contributes to the final
$\protect\F t$ values for the ``traditional nine" superallowed transitions.  The
bars for $\delta^\prime_R$ are only a rough guide to the effect on each transition of this term's systematic
uncertainty.  See text. }
\label{fig-3}
\end{figure}

\begin{figure}[b]
\epsfig{file=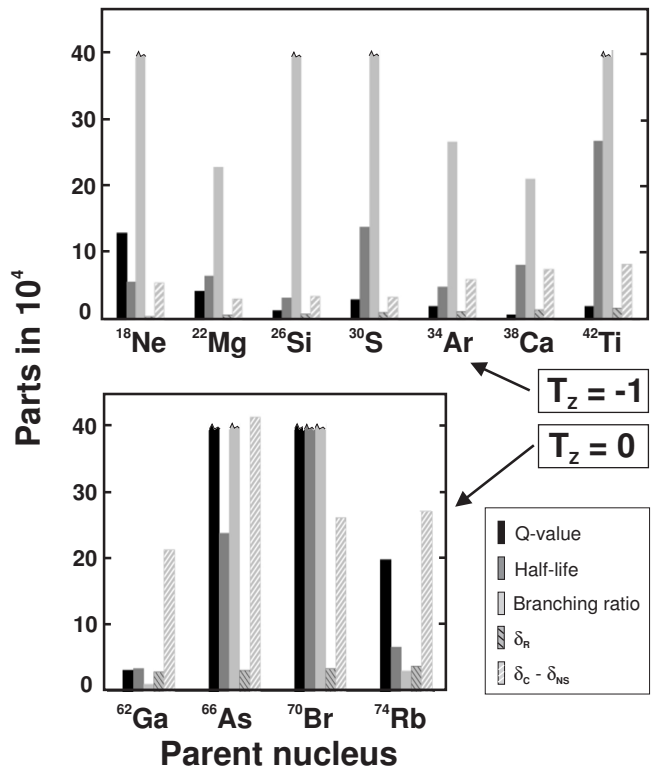,width=8.5cm}
\caption{Summary histogram of the fractional uncertainties attributable to each 
experimental and theoretical input factor that contributes to the final
$\protect\F t$ values for the eleven other superallowed transitions.  Where
the error is cut off with a jagged line at 40 parts in $10^4$, no useful experimental
measurement has been made.  The
bars for $\delta^\prime_R$ are only a rough guide to the effect on each transition of this term's systematic
uncertainty.  See text. }
\label{fig-4}
\end{figure}

We show the contributing factors to the individual $\F t$-value fractional uncertainties in two figures.  The first,
Fig.~\ref{fig-3}, encompasses the nine cases with stable daughter nuclei.  Their experimental parameters have been
measured with increasing precision for many years, so we refer to these as the "traditional nine".  The remaining
eleven cases, of which five now approach the traditional nine in precision, appear in Fig.~\ref{fig-4}.  In both
figures, the first three bars in each group of five show the contributions from experiment, while the last two
correspond to theory.  Although we are now treating the contribution from $\delta^{\prime}_R$ as a systematic
uncertainty that is applied to the final average $\overline{\F t}$, nevertheless we show a bar as a rough guide to
the significance of the $\delta^{\prime}_R$ uncertainty for each transition.  In each case, we take the height
of that bar to correspond to one-third the size of the $Z^2\alpha^3$ term in the expression for $\delta^{\prime}_R$
(see Sec.~\ref{sss:radcorr}).

From Fig.~\ref{fig-3}, it can be seen that for seven of the nine transitions plotted there -- all but those from
$^{10}$C and $^{14}$O -- the contributions from their three experimental uncertainties are substantially smaller than the 
corresponding contributions from the theoretical uncertainty due to the combined nuclear-structure-dependent corrections, 
($\delta_C - \delta_{NS}$).  The same can be said for the transitions from $^{62}$Ga and $^{74}$Rb, which appear among
the $T_Z=0$ cases illustrated in Fig.~\ref{fig-4}, although for these two cases the theoretical uncertainties are 3-10
times larger than they are for the lighter nuclei because of nuclear-model ambiguities.

There is good reason for these nine cases to have particularly small experimental uncertainties.  They are all transitions from
$T_Z$=0 parent nuclei, which populate even-even daughters in which there are no, or very few, $1^+$ states at low
enough energy to be available for competing Gamow-Teller decays.  Thus, the branching ratios for the superallowed
transitions are all $>$$99\,\%$ and have very small associated uncertainties, the largest being for the decays of $^{54}$Co and
$^{74}$Rb, which both have a $3\times10^{-4}$ fractional uncertainty.  In both cases, this is because they are
predicted to have Gamow-Teller branches that are too weak to have been observed but numerous enough that their total
strength is not negligible.  To account for such competition, one must first make a sensitive search for weak branches
and then resort to an estimate of the strength of the branches that could have been missed at the level of experimental
sensitivity achieved.  Such estimates are currently based on shell-model calculations, as first suggested in \cite{Ha02},
and obviously they introduce some additional uncertainty.

The presence of numerous weak Gamow-Teller branches becomes an increasingly significant issue for the heavier-mass nuclei,
which have increasingly large $Q_{EC}$ values.  For cases with $A \geq 62$, they present a major experimental challenge if
they are to be fully characterized.  To date this has been accomplished for the decays of $^{62}$Ga \cite{Fi08,Be08} and
$^{74}$Rb \cite{Du13} but at considerable effort.  It remains to be seen if the same level of precision will ultimately be
achievable for $^{66}$As and $^{70}$Br, the two other cases in the bottom panel of Fig.~\ref{fig-4}, or for the even heavier
$T_Z$=0 parents that extend beyond $^{74}$Rb up to $^{98}$In.

\begin{table*}[t]
\caption{Recent $\delta_C$ calculations (in per cent units)
based on models labelled     
SM-WS (shell-model, Woods-Saxon), SM-HF (shell-model, Hartree-Fock),
RPA (random phase approximation), IVMR (isovector monopole resonance)
and DFT (density functional theory).  Also given is the $\chi^2 / \nu$,
chi-square per degree of freedom, from the confidence test 
discussed in the text.
\label{t:CVCtest}}
\begin{ruledtabular}
\begin{tabular}{rddddddd}
& & & \multicolumn{3}{c}{RPA}
& & \multicolumn{1}{r}{ }\\
\cline{4-6} \\[-2mm]
& \multicolumn{1}{r}{SM-WS} &
\multicolumn{1}{r}{SM-HF} &
\multicolumn{1}{r}{PKO1} &
\multicolumn{1}{r}{DD-ME2} &
\multicolumn{1}{r}{PC-F1} &
\multicolumn{1}{r}{IVMR\footnotemark[1]} &
\multicolumn{1}{r}{DFT} \\[1mm]
\hline
& & & & & & & \\[-1mm]
$T_z = -1:$ & & & & & & & \\
\nuc{10}{C}  & 0.175 & 0.225 & 0.082 & 0.150 & 0.109 & 0.147 & 0.650 \\ 
\nuc{14}{O}  & 0.330 & 0.310 & 0.114 & 0.197 & 0.150 &       & 0.303 \\    
\nuc{22}{Mg} & 0.380 & 0.260 &       &       &       &       & 0.301 \\   
\nuc{34}{Ar} & 0.695 & 0.540 & 0.268 & 0.376 & 0.379 &       &       \\   
\nuc{38}{Ca} & 0.765 & 0.620 & 0.313 & 0.441 & 0.347 &       &       \\   
$T_z = 0:$ & & & & & & & \\
\nuc{26m}{Al}& 0.310 & 0.440 & 0.139 & 0.198 & 0.159 &       & 0.370 \\   
\nuc{34}{Cl} & 0.650 & 0.695 & 0.234 & 0.307 & 0.316 &       &       \\   
\nuc{38m}{K} & 0.670 & 0.745 & 0.278 & 0.371 & 0.294 & 0.434 &       \\   
\nuc{42}{Sc} & 0.665 & 0.640 & 0.333 & 0.448 & 0.345 &       & 0.770 \\   
\nuc{46}{V}  & 0.620 & 0.600 &       &       &       &       & 0.580 \\    
\nuc{50}{Mn} & 0.645 & 0.610 &       &       &       &       & 0.550 \\    
\nuc{54}{Co} & 0.770 & 0.685 & 0.319 & 0.393 & 0.339 &       & 0.638 \\   
\nuc{62}{Ga} & 1.475 & 1.205 &       &       &       &       & 0.882 \\  
\nuc{74}{Rb} & 1.615 & 1.405 & 1.088 & 1.258 & 0.668 &       & 1.770 \\[3mm]
$\chi^2 / \nu$ & 1.4 & 6.4 & 4.9 & 3.7 & 6.1 &  &  4.3\footnotemark[2] \\
\end{tabular}
\end{ruledtabular}
\footnotetext[1]{Rodin \protect\cite{Ro13} also computes $\delta_C = 0.992 \%$
for both \nuc{66}{As} and \nuc{70}{Br}.}
\footnotetext[2]{The result for $^{62}$Ga has not been included in the least-squares fit from which this value for $\chi^2 / \nu$ has been obtained.}
\end{table*}

The decays of $^{10}$C, $^{14}$O, and all the transitions depicted in the top panel of Fig.~\ref{fig-4} originate from
$T_Z$ = -1 parent nuclei and populate odd-odd daughters in which there are low-lying $1^+$ states strongly fed by Gamow-Teller
decay.  These branches are of comparable intensity to the superallowed one so they -- or the superallowed branch itself -- 
must be measured directly with high relative precision, a very difficult proposition.  The outcome is branching-ratio
uncertainties that exceed all the other contributions to the $\F t$-value uncertainties, experimental or theoretical, 
for these cases.  (Measurements of weak competing branches in the $T_Z$ = 0
cases discussed in the previous paragraph require high sensitivity but not high relative precision since the total
Gamow-Teller branching is more than a factor of 100 weaker than the superallowed branch for all of them.)  Advances in
experimental techniques for measuring branching ratios have improved the situation in recent years \cite{Ha03,Pa14} and
will improve it even more within the next few years.  Nevertheless, it is unlikely that these cases will ever equal the
overall level of precision already achieved for the $T_Z$ = 0 parent decays.  Their value lies instead in testing the
calculated corrections for isospin symmetry breaking \cite{Pa14} as will be described in Sec.~\ref{ss:mirror}.

\section{\label{s:dcgeneral} Isospin-symmetry Breaking}

Our own isospin-symmetry-breaking calculations, which take a semi-phenomenological approach based on the shell-model 
together with Woods-Saxon radial-functions (denoted SM-WS), have been discussed in Sect.~\ref{sss:dcspecific}.  The
results obtained there for $\delta_C$ are listed in the last column of Table~\ref{t:theocorr} and are repeated for
comparison purposes in the second column of Table~\ref{t:CVCtest}.  Those are not the only calculations of $\delta_C$.
There are a number of others that have appeared in the literature, of which we outline some more recent entries here.

\subsection{\label{ss:otherdc} Other $\delta_C$ calculations}

{\bf SM-HF}.  Ormand and Brown \cite{OB89} were the first to suggest that the  calculation of the radial
overlap -- {\it i.e.} the $\delta_{C2}$ component of $\delta_C$  -- might be better served if a mean-field
Hartree-Fock potential were used rather than the phenomenological Woods-Saxon potential.  The most recent
calculation of this type is by Hardy and Towner \cite{HT09} and their results are listed in column 3 of
Table~\ref{t:CVCtest}.  They considered the initial and final states for the $\beta$ decay to be a core of
$(A-1)$ nucleons, to which the last proton in the parent or the last neutron in the daughter is bound.  They
performed a spherical Skyrme-Hartree-Fock calculation for the $(A-1)$ nucleus with three different Skyrme
interactions to obtain the mean fields for the binding potentials.  One difficulty with this approach is that
the Hartree-Fock procedure when carried out for a nucleus with $N \ne Z$ introduces spurious isospin mixing.  
No attempt was made to remove the spurious terms.

In a recent exploratory work, Le Bloas, Bonneau, Quentin and Bartel \cite{Le11} hope to get around this
difficulty by performing the Hartree-Fock calculation in the even-even $N=Z$ nucleus with $A-2$ nucleons.  
The initial and final state for $\beta$ decay are then constructed by adding a proton and neutron to the
Hartree-Fock core for the parent, and two neutrons for the daughter, while ensuring that the two additional
particles are in time-reversal-invariant conjugate pairs.  In preliminary calculations, they estimate a
lower bound on the $\delta_C$ value to be of order 0.15 to $0.20 \%$.

{\bf RPA}. In this approach, a spherical Hartree-Fock calculation is performed on the even-even member of
the pair of nuclei involved in a superallowed $\beta$-decay transition: the parent nucleus in the decay of
a $T_z$ = -1 nucleus, or the daughter nucleus in the decay of a $T_Z$ = 0 nucleus.  The odd-odd nucleus is
then treated as a particle-hole excitation built on the even-even Hartree-Fock state.  The calculation is
carried out in the charge-exchange random-phase approximation (RPA), the lowest-energy state in the RPA
spectrum being identified as the isobaric analogue state, the state involved in the superallowed $\beta$
transition.  Isospin symmetry breaking is introduced by the presence of a Coulomb interaction augmented by
explicit charge-symmetry breaking interactions included in the two-body force used in the Hartree-Fock
calculation.

First calculations of this type were performed by Sagawa \etal \cite{SGS86} with zero-range Skyrme interactions.
These were improved upon by Liang \etal \cite{LGM09} who replaced zero-range interactions with finite-range
meson-exchange potentials in a relativistic rather than non-relativistic treatment.  In a variation of this
method, density-dependent meson-nucleon vertices were introduced in a Hartree (only) computation with nonlocal
interactions.  Liang \etal \cite{LGM09} have given results for nine different interactions, out of which we have
selected two, labelled PKO1 and DD-ME2, to display in columns 4 and 5 of Table~\ref{t:CVCtest}.  Yet another
variant on this technique from Li, Yao and Chen \cite{LYC11} replaced the finite-range meson-exchange residual
interaction with a relativistic point-coupling energy functional in an otherwise identical calculation.  One of
their results, labelled PC-F1, is given in column 6 of Table~\ref{t:CVCtest}. 

{\bf IVMR}.  A connection between isospin-symmetry breaking and the location of the giant isovector monopole
resonance (IVMR) has been demonstrated by Auerbach \cite{Au09}.  This has recently been exploited by Rodin \cite{Ro13}
to show that $\delta_C$ is related to the reciprocal of the square of an energy parameter that characterizes the
IVMR strength distributions.  The proportionality coefficient in this relation is determined by basic properties of
the ground state of the even-even nucleus, and should be reliably calculated in any realistic nuclear model.  Rodin
gives a few results using an RPA model.  These are recorded in column 7 of Table~\ref{t:CVCtest}.

{\bf DFT}.  The issue of spurious mixing in Hartree-Fock models has been fully addressed in the work of Satula \etal
\cite{Sa12} who use density functional theory (DFT).  They start with a self-consistent Slater-determinant state
obtained from a solution of the axially-deformed Skyrme-Hartree-Fock equations.  That state violates both rotational
and isospin symmetries, so their strategy is to restore rotational invariance and remove spurious isospin mixing by a
re-diagonalization of the Hamiltonian in a basis that conserves those quantities.  The numerical effort required to
project out good angular-momentum and isospin is such that modern Skyrme force parameterizations, which
include a density dependent three-body term, had to be rejected.  This effectively left them only one choice: 
the old Skyrme V (SV) force of Beiner \etal \cite{Be75}.

The results from Satula \etal \cite{Sa12} produce an unacceptably large correction, $\delta_C \sim 10 \%$, for $A = 38$.
This they attribute to the poor spectroscopic properties of the SV force which, in this case, shift the energy of
the $2s_{1/2}$ subshell to be close to the Fermi surface. A similar effect is probably evident in the $A = 34$ case
as well, where a large correction, $\delta_C \sim 1 \%$, is obtained.  Thus, in recording the DFT results in column 8
of Table~\ref{t:CVCtest} we have left out the $A = 34$ and 38 cases.  

\subsection{\label{ss:CVCtest} CVC test for $\delta_C$ corrections}

The sets of isospin-symmetry-breaking corrections $\delta_C$ recorded in Table~\ref{t:CVCtest} encompass a wide range
of nuclear models and assumptions.  Evidently, some test is required in order to assess the quality of each set and
determine its relative merit.  The test we use \cite{TH10} is based on the premise that the CVC hypothesis is valid
and thus the corrected $\F t$ values for all measured transitions must be statistically consistent with one another.
The success of any calculated set of $\delta_C$ values is thus judged by its ability to produce $\F t$ values that
are mutually consistent.  If we write the average of these $\F t$ values as $\overline{\F t}$, then it follows from
Eq.~(\ref{Ftdef}) that for each individual transition in the set we can write
\be
\delta_C = 1 + \delta_{NS} - \frac{\overline{\F t}}{ft (1 + \delta_R^{\prime})} .
\label{dcexpt}
\ee
For any set of corrections to be acceptable, the calculated value of $\delta_C$ for each transition must satisfy this
equation.  The test, for a set of $\delta_C$ values spanning $n$ superallowed $\beta$ transitions, is therefore to
treat $\overline{\F t}$ as a single adjustable parameter and use it to bring the $n$ results from the right-hand side
of Eq.~(\ref{dcexpt}), which are based primarily on experiment, into the best possible agreement with the $n$ calculated
values of $\delta_C$.  The normalized $\chi^2/\nu$ of the fit, where $\nu = n - 1$ is the number of degrees of freedom, 
then provides a figure of merit.

A few of the sets of $\delta_C$ calculations had estimated uncertainties included in their original publications but
most did not.  So, to be able to test all sets on an equal footing, we have assigned no uncertainties to any of the
theoretical $\delta_C$ values on the left-hand side of Eq.~(\ref{dcexpt}).  We do assign uncertainties to the
``experimental'' quantities on the right-hand side however.  The $ft$ values themselves are taken from Table~\ref{Ft}
but in some cases their error bars have been slightly reduced: where an experimental input from Tables~\ref{QEC},
\ref{t1/2} or \ref{R} involved component measurements that were incompatible enough to require a scale factor, we
have turned this scale factor off.  Our intention is to make the uncertainty on the $ft$ value as purely statistical
as possible.  The uncertainty on $\delta_{NS}$ is taken from Table~\ref{t:theocorr}, and that for $\delta_R^{\prime}$
is set at one third of the order $Z^2 \alpha^3$ contribution as discussed in Sect.~\ref{sss:radcorr}. The $\chi^2 / \nu$
values obtained for each set of $\delta_C$ values are given in the last line of Table~\ref{t:CVCtest}.  Note that in
testing the DFT calculation, we have dropped the value for \nuc{62}{Ga} as it is anomalously low for a heavy-mass
nucleus and, if left, would give a far too pessimistic assessment of the overall success of this calculation.  
 
\begin{figure}[t]
\epsfig{file=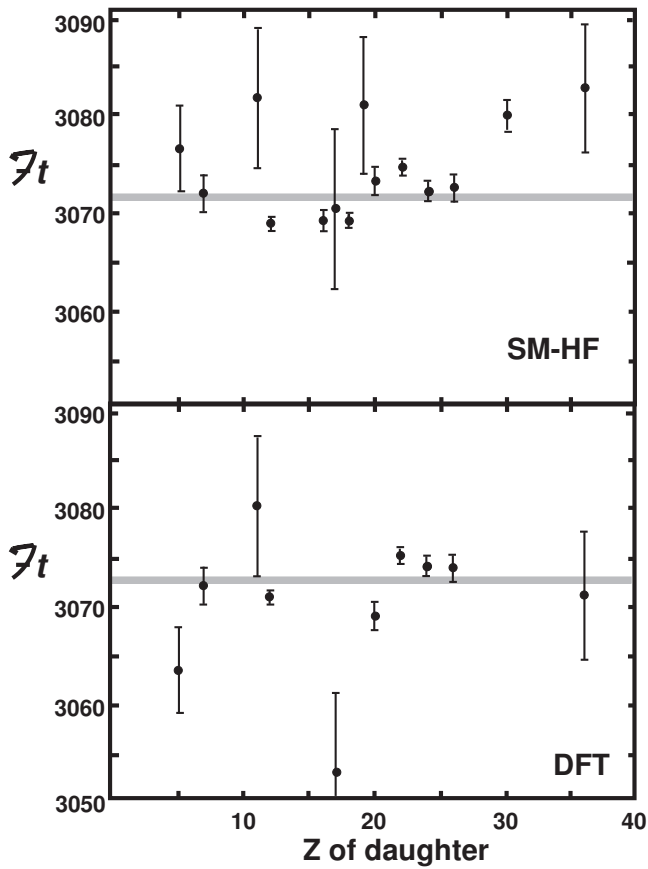,width=8.5cm}
\caption{$\F t$ values as obtained with two alternative models used to calculate $\delta_C$: SM-HF (top) and DFT (bottom).
The average $\F t$ value in each case is given by the gray band.  The corresponding $\chi^2/\nu$ for SM-HF is 6.4; and for
DFT is 4.3 (see Table~\ref{t:CVCtest}).  Obviously neither model satisfies the CVC expectation that all $\F t$ values should
agree within statistics.}   
\label{fig-5}
\end{figure}

The most obvious outcome of this analysis is that the SM-WS model has a normalized $\chi^2$ smaller by almost a factor
of three than any of the other cases cited, and is the only one with an acceptable confidence level: 17\% compared to
$<$0.01\% for all other cases.  To illustrate how the alternative models perform we show in Fig.~\ref{fig-5} a plot of
the $\F t$ values as determined with two of the alternative models for $\delta_C$ used: SM-HF and DFT.  A comparison with
the bottom panel of Fig.~\ref{fig-2}, which used the SM-WS model, emphasizes how far the scattered results in Fig.~\ref{fig-5}
are from agreement with CVC expectations.  It is for this reason that the SM-WS $\delta_C$ values are the ones we have
used to derive the $\F t$ values in Table~\ref{Ft}.

\begin{table}
\caption{Values of the calculated difference in the isospin-symmetry breaking corrections $(\delta_C^b - \delta_C^a)$ in
percent units for the mirror $\beta$ decays of $^{38}$Ca and $^{38m}$K.  Included are all the models listed in
Table~\ref{t:CVCtest} that include a $\delta_C$ value for both nuclides.  Note that the uncertainties for SM-WS and SM-HF
were derived from the model calculations themselves (see text); for the other three models no uncertainties exist, so
we assigned them a similar uncertainty.  Also listed are the derived $ft^a / ft^b$ ratios and the experimental result.  
The $ft$-ratio uncertainties incorporate the contributions from $\delta^{\prime}_R$, $\delta_{NS}$ and $\delta_C$, 
although the $\delta_C$ uncertainties predominate.
\label{t:mirror}}
\begin{ruledtabular}
\begin{tabular}{rdd}
&  \multicolumn{1}{r}{$\delta_C^b - \delta_C^a$} &
\multicolumn{1}{r}{$ft^a / ft^b$} \\[1mm]
\hline
& & \\[-1mm]
SM-WS & -0.095(34) & 1.0020(4) \\
SM-HF & 0.125(38) & 0.9998(4) \\
PKO1 & -0.035(30) & 1.0014(4) \\
DD-ME2 & -0.070(30) & 1.0017(4) \\
PC-F1 & -0.053(30) & 1.0015(4) \\[2mm]
Expt \protect\cite{Pa14} & & 1.0036(22) \\
\end{tabular}
\end{ruledtabular}
\end{table}

\subsection{\label{ss:mirror} Mirror test for $\delta_C$ corrections}

With the recent addition of the $\beta$ decay of \nuc{38}{Ca} \cite{Pa14} to the superallowed data set, an 
opportunity has been created for the first time to make a high-precision comparison of the $ft$ values from
a pair of mirror superallowed decays, \nuc{38}{Ca} $\rightarrow$ \nuc{38m}{K} and \nuc{38m}{K} $\rightarrow$
\nuc{38}{Ar}.  The ratio of mirror $ft$ values is very sensitive to the model used to calculate the
isospin-symmetry-breaking correction, $\delta_C$, and hence serves as another test of the merits of the
available calculations.

Starting again with the CVC premise that the corrected $\F t$ values defined in Eq.~(\ref{Ftdef}) must be
nucleus independent, we can write the ratio of experimental $ft$ values for a pair of mirror superallowed
transitions as follows:
\be
\frac{ft^a}{ft^b} = 1 + (\delta_R^{\prime b} - \delta_R^{\prime a})
+ (\delta_{NS}^b - \delta_{NS}^a) - (\delta_C^b - \delta_C^a) ,
\label{ftratio}
\ee
where superscript ``$a$" denotes the decay of the $T_z = -1$ parent (\nuc{38}{Ca} $\rightarrow$ \nuc{38m}{K}
in the current example) and ``$b$" denotes the mirror decay of the $T_z = 0$ parent (\nuc{38m}{K} $\rightarrow$
\nuc{38}{Ar}).  The advantage offered by Eq.~(\ref{ftratio}) is that the theoretical uncertainty on the
differences $(\delta_R^{\prime b} - \delta_R^{\prime a})$, $(\delta_{NS}^b - \delta_{NS}^a)$ and $(\delta_C^b -
\delta_C^a)$ is significantly less than the uncertainties on the corrections individually for the SM-WS and SM-HF
calculations.  This is a consequence of the way that these uncertainties were determined.  For example, each value for
$\delta_C$ was taken to be the average of the results obtained from different parameter sets and the quoted
``statistical'' uncertainty reflected the scatter in those results.  If the same approach is used to derive the mirror
differences of the correction terms $(\delta_C^b - \delta_C^a)$, the scatter among the results from different
parameter sets is less than the scatter observed in either $\delta_C^b$ or $\delta_C^a$. 

In Table~\ref{t:mirror} we list $(\delta_C^b - \delta_C^a)$ for those cases from Table~\ref{t:CVCtest} where a value
for $\delta_C$ is given for both \nuc{38}{Ca} and \nuc{38m}{K}.  In the cases of SM-WS and SM-HF the uncertainty was
derived, as just explained, by the same methods as those described in the source publications, \cite{TH08} and
\cite{HT09}.  For the RPA calculations, the source publications list no uncertainties, so we have arbitrarily assigned
an uncertainty that is similar to the ones obtained for the shell-model-based calculations.  We also list the ratio of
$ft$ values obtained from Eq.~(\ref{ftratio}) using $(\delta_R^{\prime b} - \delta_R^{\prime a}) = 0.026(1) \%$ and
$(\delta_{NS}^b - \delta_{NS}^a) = 0.075(20) \%$.  For comparison, the experimental result \cite{Pa14} for the ratio
appears at the bottom of the column.

The SM-WS and RPA calculations all agree with this result, while the SM-HF calculation differs by two standard
deviations.  This is immediately attributable to the positive sign that the SM-HF calculation obtains for $(\delta_C^b
- \delta_C^a)$, which is unlike the results from all the other model calculations.  Intuitively one expects a negative
sign for this difference since nucleus ``$a$" (\nuc{38}{Ca} in this case) has one more proton than nucleus ``$b$'' 
(\nuc{38m}{K}), and the Coulomb force is expected to be the dominant influence on the isospin-symmetry breaking.  This
strongly suggests that there is a problem either with the Hartree-Fock protocol adopted in \cite{HT09} or with the
degree of spurious isospin mixing that this protocol inevitably includes.

For now, this test can only be applied to the $A=38$ mirror pair since no other pairs have been completely
characterized with the necessary precision.  However, every well-known superallowed transitions from a $T_Z = 0$ parent
nuclus has a mirror transition from the $T_Z = -1$ member of the same isospin triplet; the latter is simply not fully
characterized yet.  For two of these $T_Z = -1$ parents, \nuc{26}{Si} and \nuc{34}{Ar}, the mirror transition only lacks
a precise enough branching-ratio measurement in order to become useful to test the $\delta_C$ calculations.  In one more
case, that of \nuc{42}{Ti}, the half-life and the branching ratio both remain to be determined with sufficient precision.  
It is likely, though, that all three of these cases will be completed before long \cite{Pa14},  Of course, there are many more
possible mirror pairs -- with $A$ = 46, 50, 54, 62 ... -- but the $T_Z = -1$ parent nuclei are farther from
stability and certainly present a considerable challenge to precise measurement.

\section{\label{s:impact} Impact on weak-interaction physics}

\subsection{\label{ss:Vud} Value of $V_{ud}$}

We have now tested all available $\delta_C$ calculations and demonstrated that only the results from the SM-WS
semi-phenomenological model satisfy the CVC condition that the corrected $\F t$ values be statistically consistent with
one another.  It is this set of correction terms that we consequently used to derive the results in Table~\ref{Ft}, which
led to the value for the average $\overline{\F t}$ and its uncertainty that appears in Eq.~(\ref{Ftavg}).  In our past
two surveys \cite{HT05,HT09}, we imposed a further systematic uncertainty to account for differences between the two models
available at the time to calculate $\delta_C$.  Specifically, we calculated the $\F t$ values twice, once with SM-WS and
once with SM-HF $\delta_C$ values and then averaged the two resulting $\overline{\F t}$ values together.  The added systematic
uncertainty was taken to be equal to half the difference between them.  But all that happened before we had tests to
evaluate the merits of the models!  

Not only do we now have such tests but, in the six years since our last survey, new measurements have improved the $ft$-value
data so that the test provides real discrimination.  Given the failure of SM-HF to satisfy either the CVC test in
Sec.~\ref{ss:CVCtest} or the mirror test in Sec.~\ref{ss:mirror}, there is no longer any justification to consider it as a
viable alternative to the SM-WS model.  Furthermore, there is no other model calculation to replace it.  Not only do all the
other models fail the CVC test but, in fact, not one of them is currently capable of calculating $\delta_C$ values for the
full set of measured transitions (see Table~\ref{t:CVCtest}).  We therefore now consider the $\overline{\F t}$ value in
Eq.~(\ref{Ftavg}) to be our final result.

We can now use this result for $\overline{\F t}$ to determine the vector coupling constant, $\GV$, from Eq.~(\ref{Ftdef}).  
The value of $\GV$ itself is of little interest but, together with the weak interaction constant for the purely leptonic
muon decay, $\GF$, it yields the much more interesting up-down element of the Cabibbo-Kobayashi-Maskawa (CKM) quark-mixing
matrix.  The basic relationship is $V_{ud}=\GV/\GF$, which in terms of $\overline{\F t}$ becomes
\bea
|V_{ud}|^2 & = & \frac{K}{2 \GF^2 (1 + \DRV ) \overline{\F t}}
\nonumber \\[2mm]
& = & \frac{2915.64 \pm 1.08}{\overline{\F t}}~,
\label{Vud2}
\eea
where we have used the Particle Data Group \cite{PDG14} value for the weak interaction coupling constant from muon decay, $\GF 
/(\hbar c )^3 = 1.1663787(6)\times 10^{-5}$ GeV$^{-2}$; and the value for $\DRV$, the nucleus-independent radiative correction, is
taken from Eq.~(\ref{DRV}).  Substituting the result for $\overline{\F t}$ from Eq.~(\ref{Ftavg}) we obtain
\be
|V_{ud}|^2 = 0.94900 \pm 0.00042.
\label{Vudsq}
\ee
Note that the total uncertainty here is almost entirely due to the uncertainties contributed by the theoretical corrections.  By
far the largest contribution, 0.00035, arises from the uncertainty in $\DRV$; 0.00018 comes from the nuclear-structure-dependent
corrections ($\delta_C - \delta_{NS}$), and 0.00011 is attributable to $\delta_R^{\prime}$.  Only 0.00009 can be considered to be
experimental in origin.  The latter contribution has decreased by nearly a factor of two since our last survey but, because of
the dominance of the theoretical uncertainties, which have not changed significantly, the overall improvement in $|V_{ud}|^2$ is
much less pronounced.

\begin{figure}
\epsfig{file=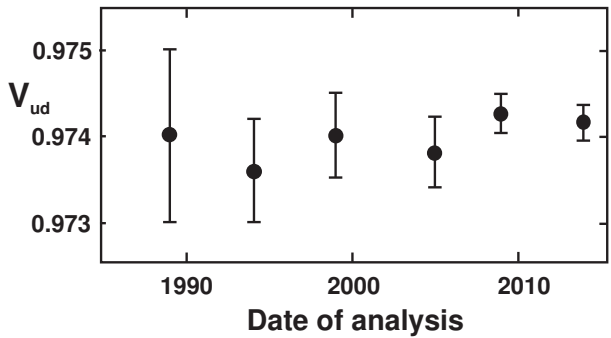,width=8cm}
\caption{Values of $V_{ud}$ as determined from superallowed $0^+$$\rightarrow 0^+$ $\beta$ decays
plotted as a function of analysis date, spanning the past two-and-a-half decades.  In order, from the earliest
date to the most recent, the values are taken from Refs.~\cite{HT90}, \cite{TH95}, \cite{TH99},
\cite{HT05}, \cite{HT09} and this work.}
\label{fig-6}
\end{figure}

The corresponding value of $V_{ud}$ is
\be
|V_{ud}| = 0.97417(21)~,
\label{Vudvalu}
\ee
a result which is consistent with, but a bit more precise than, values we have obtained in previous analyses of superallowed $\beta$
decay.  To emphasize the consistency and steady improvement that has characterized the value of $V_{ud}$ as derived from nuclear
$\beta$ decay, in Fig.~\ref{fig-6} we plot our new result together with $V_{ud}$ values published at various times over the past
two-and-a-half decades \cite{HT90,TH95,TH99,HT05,HT09}.

\subsection{\label{ss:Vus} Value of $V_{us}$}

The recommended value for $V_{us}$, the second largest top-row element of the CKM matrix, is derived from the decays, both leptonic and
semi-leptonic, of the kaon.  Other determinations based on hyperon decays and hadronic tau decays, which do not currently have
sufficient precision to challenge the results from kaon decays, will not be considered here.

\begin{table*}
\caption{Lattice QCD values for $f_+(0)$ and $f_K / f_{\pi}$ appear in columns 2 and 3, respectively, distinguished by the number of quark flavors
present in the simulations.  The values corresponding to $N_f = 2$ and $N_f = 2+1$ are averages taken from FLAG \protect\cite{FLAG14}.  The results
for $N_f = 2+1+1$ are from more recent publications \cite{Ba14, Ba14a}.   The deduced values of $|V_{us}|$ (for $K_{\ell 3}$ decays) and those of
$|V_{us}| / |V_{ud}|$ (for $K_{\ell 2}$ decays) appear in columns 4 and 5, respectively.  The unitarity sums in column 6 incorporate $V_{ud}$ from
Eq.~(\protect\ref{Vudvalu}) and $V_{ub}$ from Eq.~(\protect\ref{Vubvalu}).  Their residuals, $\Delta_{CKM}$, are in column 7 and, if unitarity is
not met within the quoted uncertainty, the number of standard deviations, $\sigma$, of the discrepancy appears in column 8.  Rows 7 to 9 give
$|V_{us}|$ obtained by our fitting three data -- $|V_{ud}|$ from $\beta$ decay, $|V_{us}|$ from $K_{\ell 3}$ decays, and $|V_{us}|/|V_{ud}|$ from
$K_{\ell 2}$ decays -- with two free parameters, $|V_{ud}|$ and $|V_{us}|$, for each of the specified values of $N_f$.  The Particle Data Group $|V_{us}|$
value \protect\cite{PDG14} is given in the last row.
\label{t:Vus}}
\begin{ruledtabular}
\newcolumntype{d}[1]{D{.}{.}{#1}}
\begin{tabular}{rd{4}d{4}d{4}d{4}d{5}d{5}d{1}}
& \multicolumn{1}{d{0}}{f_+(0)} &
\multicolumn{1}{d{0}}{f_K / f_{\pi}} &
\multicolumn{1}{d{0}}{|V_{us}|} & 
\multicolumn{1}{d{0}}{|V_{us}| / |V_{ud}|} &
\multicolumn{1}{d{0}}{|V_u|^2} &
\multicolumn{1}{d{0}}{\Delta_{CKM}} &
\multicolumn{1}{d{0}}{\sigma} \\[1mm]
\hline
& & & & & & & \\[-2mm]
$N_f=2+1+1$ & 0.9704(32)\footnotemark[1] 
& & 0.2232(9) & & 0.9988(6) &  -0.0012(6)  & 2.1 \\
$N_f = 2+1$ & 0.9661(32) & & 0.2241(9) & & 0.9992(6) & -0.0008(6) & 1.4 \\
$N_f = 2$ & 0.9560(84) & & 0.2265(20) & & 1.0003(10) & 0.0003(10) & \\[3mm]
$N_f=2+1+1$ & & 1.1960(25)\footnotemark[2] 
& & 0.2308(6) & 0.9996(5) & -0.0004(5) & \\
$N_f=2+1$ & & 1.192(5) & & 0.2315(10) & 0.9999(6) & -0.0001(6) & \\
$N_f = 2$ & & 1.205(18) & & 0.2290(34) & 0.9988(15) &  -0.0012(15) \\[3mm]
$N_f=2+1+1$ & & & 0.2243(8) & & 0.9993(8) & -0.0007(8) &    \\
$N_f = 2+1$ & & & 0.2247(7) & & 0.9995(6) & -0.0005(6) &     \\
$N_f = 2$   & & & 0.2256(17) & & 0.9999(9) & -0.0001(9) & \\[3mm]
PDG 14 & & & 0.2253(8) & & 0.9998(6) & -0.0002(6) & \\
\end{tabular}
\end{ruledtabular}
\footnotetext[1]{This recent result from \protect\cite{Ba14} replaces the FLAG average \protect\cite{FLAG14}, which is less precise. }
\footnotetext[2]{This recent result from \protect\cite{Ba14a} with symmetrized uncertainty replaces the FLAG average \protect\cite{FLAG14}, which is less precise.}
\end{table*}

For the semi-leptonic $K \rightarrow \pi \ell \nu_{\ell}$ ($K_{\ell 3}$) decays, there are four separate decay channels that may be
studied: charged kaons or neutral kaons (long or short) decaying to either electrons or muons.  Results from these experiments have been
evaluated by the FlaviaNet group \cite{Fl10}, with updates discussed at the CKM14 conference by Moulson \cite{Mo14}.  Extracted from
the experimental data is the product
\be
f_+(0) |V_{us}| = 0.2165(4) ,
\label{fVus}
\ee
where $f_+(0)$ is the semi-leptonic-decay form factor at zero-momentum transfer.  Its value is close to unity.  In the exact $SU(3)$
symmetry limit, the CVC hypothesis would require its value to be exactly one but, in $K_{\ell 3}$ decays, $SU(3)$ symmetry is broken at
second order and a theoretical calculation is required to estimate the extent of the symmetry breaking.  Today, lattice QCD calculations
are used for this purpose, replacing former semi-analytic methods based on chiral perturbation theory.  

For purely leptonic kaon decays, $K^{\pm} \rightarrow \mu^{\pm} \nu$ ($K_{\ell 2}$), it is their ratio to leptonic pion decays, 
$\pi^{\pm} \rightarrow \mu^{\pm} \nu$, that is measured since hadronic uncertainties can be minimized in the ratio.  The resulting
experimental output is the ratio of CKM matrix elements $|V_{us}|/|V_{ud}|$ multiplied by the ratio of decay constants
$f_K / f_{\pi}$.  The current recommended value from Moulson \cite{Mo14} is 
\be
\frac{f_K}{f_{\pi}} \frac{|V_{us}|}{|V_{ud}|} = 0.2760(4) .
\label{fKfpiV}
\ee
Again a lattice QCD calculation is required to evaluate the ratio of decay constants.

In the past few years, there has been a rapid expansion in large-scale numerical simulations in lattice QCD aimed at determining the
low-energy constants of flavor physics.  A Flavor Lattice Averaging Group (FLAG) formed in 2007 has been enlarged and the first report
from the expanded group has just been released \cite{FLAG14}.  Their recommended values for the low-energy constants depend on $N_f$,
the number of dynamical quark flavors included in the lattice simulations.  The earliest results with $N_f=2$ included just up and down
quarks; but, more recently, strange quarks were added, so those calculations are designated by $N_f = 2 + 1$.  Most recently, calculations
with $N_f = 2 + 1 + 1$ have been reported, in which charm quarks are incorporated as well.  The FLAG group gives results separately for
$N_f =2$, $N_f = 2+ 1$ and $N_f = 2+1+1$, arguing that they have no {\it a priori} way to estimate quantitatively the differences among
results produced in simulations with different numbers of dynamical quarks.

In Table~\ref{t:Vus} we give recommended values for $f_+(0)$ and $f_K / f_{\pi}$ that lead to values of $|V_{us}|$ in rows 1 to 3, and
$|V_{us}| / |V_{ud}|$ in rows 4 to 6.  The entries for $N_f=2$ and $N_f = 2 + 1$ are the FLAG averages from \cite{FLAG14}.  Those for
$N_f = 2+1+1$ are more recent results \cite{Ba14, Ba14a}, which appeared after the deadline for inclusion in the FLAG averages.  Since
they are more precise than the FLAG averages, we have followed Moulson \cite{Mo14} in using them instead.
  
With $|V_{us}|$ from rows 1-3, $|V_{us}| / |V_{ud}|$ from rows 4-6, and $|V_{ud}|$ from superallowed $\beta$ decay as given in
Eq.~(\ref{Vudvalu}), we have three pieces of data from which to determine two parameters, $|V_{ud}|$ and $|V_{us}|$.  To do so, we performed
non-linear least squares fits to obtain the values of $|V_{us}|$ given in rows 7 to 9 in Table~\ref{t:Vus}.  Note that the $\chi^2$ of these
three fits were 2.6, 1.2 and $<$$1$, respectively, so the uncertainties shown for the first two $|V_{us}|$ values have been scaled appropriately.
The corresponding value of $|V_{ud}|$ in each case undergoes very little change from its input value, although its uncertainty increases:  
For the $N_f = 2 + 1 + 1$ case, its central value shifts by two units in the last digit quoted, to 0.97415(34); for $N_f = 2 + 1$, it shifts by
one unit, to 0.97416(23); and for $N_f=2$, there is no change at all.

In the last line of the Table~\ref{t:Vus} we give the most up-to-date Particle Data Group (PDG) value \cite{BM13} for $|V_{us}|$, which was arrived at
without the very recent $N_f = 2 + 1 +1$ lattice calculations \cite{Ba14,Ba14a}.  To obtain their average result, the PDG used $|V_{us}| =
0.2253(14)$ from their analysis of $K_{\ell 3}$ decays, and $|V_{us}| / |V_{ud}| = 0.2313(10)$ from the leptonic decays of kaons and pions.  Note
that both values are statistically consistent with all the results for the same quantities given in rows 1-3 and 4-6.  The same can be
said for the comparison of the average PDG result for $|V_{us}|$ on the bottom line with our fitted results on the three lines above it.

\subsection{\label{ss:CKMu} CKM Unitarity}

The standard model does not prescribe the individual elements of the CKM matrix -- they must be determined experimently -- but absolutely
fundamental to the model is the requirement that the matrix be unitary.  To date, the most demanding test of CKM unitarity comes from the
sum of the squares of the top-row elements,
\be
|V_u|^2 \equiv |V_{ud}|^2 + |V_{us}|^2 + |V_{ub}|^2
= 1 + \Delta_{CKM} ,
\label{unitary}
\ee
which should equal exactly one: {\it i.e.} the residual, $\Delta_{CKM}$, should be exactly zero.  Since $|V_{ud}|^2$ accounts for $95 \%$
of this sum, its precision (and accuracy) is of paramount importance.  Our analysis in this survey demonstrates that the current value for
$|V_{ud}|$ given in Eq.~(\ref{Vudvalu}) is solidly based on a robust body of diverse data and, although its precision has improved
continuously, it has not shifted outside of previously quoted error bars in a quarter century.  It now has a precision of $0.02 \%$, 
which makes it the most precisely determined element in the CKM matrix by far. 
 
Values for $|V_{us}|$ discussed in Sect.~\ref{ss:Vus} and shown in Table~\ref{t:Vus} typically have a quoted precision of $0.4 \%$, of which
approximately one-third is experimental and two-thirds theoretical, the latter arising from the precision attained in recent lattice QCD
simulations.  The third element of the top row, $V_{ub}$, is very small and hardly impacts on the unitarity test at all.  Its value from the
Particle Data Group \cite{PDG14} compilation is
\be
|V_{ub}| = (4.15 \pm 0.49) \times 10^{-3} .
\label{Vubvalu}
\ee

Our approach to the unitarity test in the past has always been to combine our result for $|V_{ud}|$ with the PDG evaluated results for
$|V_{us}|$ and $|V_{ub}|$.  If we do that -- taking $|V_{ud}|$ from Eq.~(\ref{Vudvalu}), $|V_{us}|$ from the bottom line of Table~\ref{t:Vus}
and $|V_{ub}|$ from Eq.~(\ref{Vubvalu}) -- we obtain the result
\be
|V_u|^2 = 0.99978\pm0.00055~.
\label{VuPDG}
\ee
But this cannot be the final word since the PDG evaluation does not include results from the most recent lattice calculations,
which are used to extract $|V_{us}|$ and $|V_{us}| / |V_{ud}|$ from kaon-decay measurements.

So how well is the unitarity relation satisfied if the new results are included?  In what follows we will take $|V_{ud}|$ solely from
Eq.~(\ref{Vudvalu}) and $|V_{ub}|$ from Eq.~(\ref{Vubvalu}).  We shall then examine how unitarity depends on the possible choices for
$|V_{us}|$.  First, from Table~\ref{t:Vus} we can see that if $|V_{us}|$ is determined solely from $K_{\ell 3}$ experiments, then successive
improvements in lattice calculations have led to increasing values of $f_+(0)$ and steadily smaller values of $|V_{us}|$ (see rows 1-3).  
This has led to increasing tension with unitarity.  The most recent lattice calculation with $N_f = 2 + 1+ 1$ quark flavors by Bazavov
\etal \cite{Ba14} leaves unitarity unsatisfied by 2.1 standard deviations.  This is certainly a provocative outcome.  Even with only
$N_f = 2 + 1$ flavors in the calculation there is some tension, with $\Delta_{CKM}$ being 1.4 standard deviations away from zero.

By contrast, $K_{\ell 2}$ experiments yielding the ratio $|V_{us}| / |V_{ud}|$ and a unitarity sum calculated via
\be 
|V_u|^2 = |V_{ud}|^2 \left ( 1 + \left | \frac{V_{us}}{V_{ud}} \right |^2 \right )
+ |V_{ub}|^2 = 1 + \Delta_{CKM}~,
\label{unitary2}
\ee
agree perfectly with unitarity regardless of the number of quark flavors in the decay-constant calculation (see rows 4-6).

Evidently there is some incompatibility between the $|V_{us}|$ value determined from $K_{\ell 3}$ decays and that determined from $K_{\ell 2}$
decays.  Even so, one possible way forward is to combine the $K_{\ell3}$ and $K_{\ell 2}$ results to arrive at a compromise value for $|V_{us}|$.  
This we have done, with the results appearing in rows 7-9 in Table~\ref{t:Vus}.  The unitarity test is satisfied for all three cases but this
has come at a price: For the more modern lattice calculations, the uncertainties have been increased.

Now let's examine a number of specific scenarios:

{\em 1) Kaon experiments correct, unitarity satisfied.}  Accepting that Eqs.~(\ref{fVus}) and (\ref{fKfpiV}) are correct and $|V_u|^2=1$, we can
solve for $f_+(0)$ and $f_K / f_{\pi}$ to obtain $f_+(0) = 0.9590(43)$ and $f_K/f_{\pi} = 1.191(5)$.  These values agree with all lattice
calculations for $f_K / f_{\pi}$, but only agree with the older $N_f=2$ results for $f_+(0)$.  

{\em 2) $f_K/f_{\pi}$ correct, $K_{\ell 2}$ experiment correct, unitarity satisfied.}  Accepting the $N_f=2+1+1$ value for $f_K / f_{\pi}$ and
the $K_{\ell 2}$ result in Eq.~(\ref{fKfpiV}), we can solve for $|V_{ud}|$, obtaining $|V_{ud}| = 0.97438(12)$, which agrees with the 
superallowed $\beta$-decay result in Eq.~(\ref{Vudvalu}).  Clearly the $K_{\ell 2}$ data and corresponding lattice calculations are fully
compatible with nuclear $\beta$ decay and unitarity.

{\em 3) $f_+(0)$ correct, $K_{\ell 3}$ correct, unitarity satisfied.}  Given the $N_f = 2+1+1$ value for $f_+(0)$ and the $K_{\ell 3}$
experimental result in Eq.(\ref{fVus}), we can solve for $|V_{ud}|$, obtaining $|V_{ud}| = 0.97477(20)$, which differs from the $\beta$-decay
value by 2.1 standard deviations.  Is there any way the $|V_{ud}|$ value in Eq.~(\ref{Vudvalu}) could possibly be shifted to this value?  It
can be seen in Eq.(\ref{Vud2}) that $|V_{ud}|^2$ is inversely proportional to both $\overline{\F t}$ and $(1 + \DRV )$.  For $\overline{\F t}$
to account for such a shift, it would have to decrease by six standard deviations.  That is unlikely enough but, since all 14 measured
transitions agree with one another and with CVC, all 14 would have to undergo the same shift, a virtual impossibility.  The only other
possibility is a shift in the nucleus-independent radiative correction, $\DRV$, which would have to be reduced from $2.36(4) \%$ to $2.24 \%$.
This is a change equal to three times the stated uncertainty which, while not impossible, is rather unlikely.

{\em 4) $f_+(0)$, $f_K / f_{\pi}$ correct, $K_{\ell 3}$, $K_{\ell 2}$ correct, unitarity not satisfied.}  With $|V_{us}|$ determined from
$K_{\ell 3}$ decays and $|V_{us}| / |V_{ud}|$ from $K_{\ell 2}$ decays, each with the $N_f=2+1+1$ lattice coupling constants, a value of
$|V_{ud}|$ can be obtained from their ratio.  The result, $|V_{ud}| = 0.9670(44)$, has a somewhat larger error bar than other determinations
from kaon physics because no constraint to satisfy unitarity has been imposed.  Nevertheless the result is two of its standard deviations
away from the nuclear $\beta$-decay value for $|V_{ud}|$ and the unitarity sum is likewise not satisfied, with $|V_u|^2 = 0.985(9)$ and a deficit,
$\Delta_{CKM} = -0.015(9)$, of 1.8 standard deviations.  For the $\beta$-decay value of $|V_{ud}|$ to be shifted into agreement with this
kaon-derived value would require the nucleus-independent radiative correction $\DRV$ to be increased from $2.36(4) \%$ to $3.88 \%$, forty
times its stated uncertainty.  Surely this can be ruled out!

One must conclude that there is no definitive answer for $|V_{us}|$ as of now since the two approaches to its measurement from kaon decay
are not completely consistent with one another.  On balance, though, the result for $|V_{us}| / |V_{ud}|$ obtained from $K_{\ell 2}$ and
pion decays seems the most reliable since it shows the greatest consistency as the lattice calculations have improved, which reinforces the
idea that systematic errors are reduced when a ratio is used.  If we then accept the $N_f = 2+1+1$ result on line 4 of Table~\ref{t:Vus} and
combine it with our result for $|V_{ud}|$ from Eq.~(\ref{Vudvalu}), we get $|V_{us}| = 0.2248(6)$ and a unitary sum of $|V_u|^2 = 0.99956(49)$.

\subsection{\label{ss:scalars} Scalar currents}

\subsubsection{Fundamental scalar current}
\label{sss:fsc}

The standard model prescribes the weak interaction to be an equal mix of vector ($V$) and axial-vector ($A$) interactions that maximizes
parity violation.  Searches for physics beyond the standard model therefore seek evidence that parity is not maximally violated (due to the
presence of right-hand currents) or that the interaction is not pure $V - A$ (due to the presence of scalar or tensor currents).  The data
in this survey allow us to contribute to the search for a scalar interaction since, if present, it would have a measurable effect on
superallowed $0^+$$\rightarrow 0^+$ $\beta$ transitions.

A scalar interaction would generate an additional term \cite{HT05} to the shape-correction function, which forms part of the integrand of the
statistical rate function, $f$, an integral over the $\beta$-decay phase space.  The additional term takes the form $(1 +  b_F \gamma_1 / W)$, 
where $W$ is the total electron energy in electron rest-mass units, and $\gamma_1 = \sqrt{(1 - (\alpha Z)^2)}$.  The strength of the scalar
interaction is contained in the unknown constant, $b_F$, which is called the Fierz interference term \cite{JTW57}.  Thus, the impact of a 
scalar interaction on the $\F t$ values would be to introduce a dependence on $\langle 1 / W \rangle$, the average inverse decay energy of
each $\beta^+$ transition.  No longer would the $\F t$ values be constant over the whole range of nuclei but they would instead exhibit a
smooth dependence on $\langle 1 / W \rangle$.  Since $\langle 1 / W \rangle$ is largest for the lightest nuclei, and decreases monotonically with
increasing $Z$ and $A$, the largest deviation of $\F t$ from constancy would occur for the cases of \nuc{10}{C} and \nuc{14}{O}.

\begin{figure}[t]
\epsfig{file=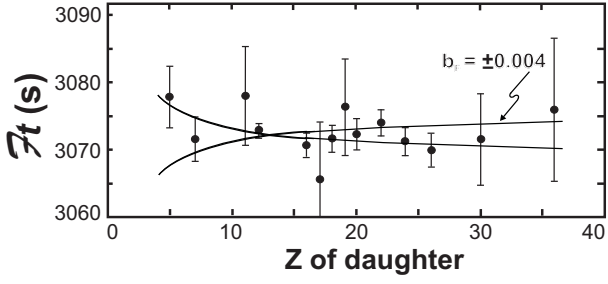,width=8cm}
\caption{Corrected $\F t$ values from Table~\ref{Ft} plotted as a function of the charge on the daughter nucleus, $Z$.  The curved lines represent
the approximate loci the $\F t$ values would follow if a scalar current existed with $b_F = \pm 0.004$.}
\label{fig-7}
\end{figure}

We have re-evaluated the statistical rate function, $f$, for each transition using a shape-correction function that includes the presence of the
scalar interaction via a Fierz interference term, $b_F$, which we treat as an adjustable parameter.  We then obtained a value of $b_F$ that
minimized the $\chi^2$ in a least-squares fit to the expression $\F t =$ constant.  The result we obtained is
\be 
b_F = - 0.0028 \pm 0.0026,
\label{bFvalu}
\ee
a marginally larger result than the value from our last survey \cite{HT09} but with the same uncertainty.  Note that the uncertainty quoted here
is one standard deviation ($68 \%$ CL), as obtained from the fit.  In Fig.~\ref{fig-7} we illustrate the sensitivity of this analysis by plotting
the measured $\F t$ values together with the loci of $\F t$ values that would be expected if $b_F = \pm 0.004$.  There is no statistically
compelling evidence for $b_F$ to be non-zero.

The result in (\ref{bFvalu}) can also be expressed in terms of the coupling constants that Jackson, Treiman and Wyld \cite{JTW57} introduced to
write a general form for the weak-interaction Hamiltonian.  Since we are dealing only with Fermi superallowed transitions, we can restrict
ourselves to scalar and vector couplings, for which the Hamiltonian becomes
\bea
H_{S+V} &  =  & (\overline{\psi}_p \psi_n)
        (C_S \overline{\phi}_e \phi_{\overline{\nu}_e}
        + C_S^{\prime} \overline{\phi}_e \gamma_5 \phi_{\overline{\nu}_e})
\nonumber \\
& &
+ \left ( \overline{\psi}_p
\gamma_{\mu}  \psi_n \right )
\left [C_V \overline{\phi}_e \gamma_{\mu} 
( 1 + \gamma_5 ) \phi_{\overline{\nu}_e} \right ]  ,
\label{JTWSV}
\eea
in the notation and metric of \cite{JTW57}.  We have taken the vector current to be maximally parity violating, as indicated by experiment.
The complexity of the relationship between $b_F$ and the couplings $C_S$, $C_S^{\prime}$ and $C_V$ depends on what assumptions are made about the
properties of the scalar current.  If we take the most restrictive conditions, that the scalar and vector currents are time-reversal invariant
($i.e.$ $C_S$ and $C_V$ are real) and that the scalar current, like the vector current, is maximally parity violating ($i.e.$ $C_S = 
C_S^{\prime}$), then we can write\footnote[1]{More correctly we write $C_S / C_V = \pm b_F / 2$, with the upper sign for $\beta^-$ transitions and
the lower sign for $\beta^+$ transitions.  Since all the superallowed Fermi transitions are positron emitters, we will display only the lower sign
in our equations.  The sign change comes about because $\overline{\psi}_p C_S \psi_n$ changes sign under charge conjugation relative to
$\overline{\psi}_p C_V \gamma_4\psi_n$. }
\be
\frac{C_S}{C_V} = - \frac{b_F}{2} = + 0.0014 \pm 0.0013 .
\label{CSCV}
\ee 
This limit from superallowed $\beta$ decay is, by far, the tightest limit available on the presence of a scalar current under the assumptions stated.

If we remove the condition that the scalar current be maximally parity violating, then the expression contains two unknowns,
\be
b_F = \frac{-2C_V (C_S + C_S^{\prime})}{2|C_V|^2+|C_S|^2+|C_S^{\prime}|^2}
\simeq - \left ( \frac{C_S}{C_V} + \frac{C_S^{\prime}}{C_V} \right ),
\label{CSCSpCV}
\ee
and cannot be solved individually for $C_S / C_V$ and $C_S^{\prime} / C_V$.  However, the $\beta$-$\nu$ angular-correlation coefficient, $a$, for a
superallowed $0^+ \rightarrow 0^+$ $\beta$ transition provides another independent measure of $C_S$ and $C_V$.  In that case
\bea
a & = & \frac{2|C_V|^2 - |C_S|^2 - |C_S^{\prime}|^2}
{2|C_V|^2 + |C_S|^2 + |C_S^{\prime}|^2}
\nonumber \\
& \simeq & 1 - \left ( \frac{|C_S|^2}{|C_V|^2} +
\frac{|C_S^{\prime}|^2}{|C_V|^2} \right )~,
\label{aenuCSCV}
\eea
which, together with Eq.~(\ref{CSCSpCV}), can be used to set limits on both $C_S / C_V$ and $C_S^{\prime} / C_V$.

In our previous survey \cite{HT09}
we combined our result for $b_F$ with the result from a $\beta$-$\nu$ correlation measurement in the superallowed emitter \nuc{38m}{K} \cite{Go05}. Our
new value for $b_F$ in Eq.~\ref{bFvalu} is so little changed from our previous one that we quote the same $68 \%$ confidence limits for $C_S / C_V$ and
$C_S^{\prime} / C_V$: $viz.$
\be
\frac{|C_S|}{|C_V|} \leq 0.065 ~~~~~
\frac{|C_S^{\prime}|}{|C_V|} \leq 0.065~.
\label{CSCSp}
\ee
The reader is referred to Fig.~8 in \cite{HT09} for a visual representation of these results and their derivation.

A review of the limits obtained on exotic weak-interaction couplings from precision $\beta$-decay experiments has recently been produced by
Naviliat-Cuncic and Gonz\'{a}lez-Alonso \cite{NG13}.  

\subsubsection{\label{sss:induced} Induced scalar currents}

If we consider only the vector part of the weak interaction for composite spin-1/2 nucleons, then the most general form the interaction can take is
written \cite{BB82}
\be
H_V = \overline{\psi}_p \left ( g_V \gamma_{\mu} - f_M \sigma_{\mu \nu}
q_{\nu} + i f_S q_{\mu} \right ) \psi_n ~
\overline{\phi}_e \gamma_{\mu} ( 1 + \gamma_5) \phi_{\overline{\nu}_e}
\label{HVs}
\ee
with $q_{\mu}$ being the four-momentum transfer between hadrons and leptons.  The values of the coupling constants $g_V$ (vector), $f_M$ (weak
magnetism) and $f_S$ (induced scalar) are pre-determined if the CVC hypothesis -- that the weak vector current is just an isospin rotation of
the electromagnetic vector current -- is correct.  In particular, because CVC implies that the vector current is divergenceless, the induced scalar
term $f_S$ should be identically zero.  With the data from superallowed $\beta$ decay it is possible to test this prediction of CVC by setting an
experimental limit on the value of $f_S$.

We showed in \cite{HT05} that the Hamiltonian in Eq.~(\ref{HVs}) can be reorganized to match exactly the form given by Jackson, Treiman and Wyld,
Eq.~(\ref{JTWSV}), with $C_S$ simply replaced by $m_e f_S$ and $C_V$ by $g_V$.  Here $m_e$ is the electron rest mass, which is $m_e = 1$ in
electron rest-mass units.  Thus the value for $C_S/C_V$ in Eq.~(\ref{CSCV}) applies equally well to the ratio $m_e f_S / g_V$.  Expressed as a
limit at the 90\% confidence level, we obtain
\be
\left | \frac{m_e f_S}{g_V} \right | < 0.0035 ~.
\label{fs90}
\ee
This result is a vindication for the CVC hypothesis that predicts $g_V=1$ and $f_S=0$.  Our $90 \%$ confidence limit confirms this prediction at
the level of 35 parts in $10^4$.

\section{\label{s:concl} Conclusions}

It has been six years since our previous survey of superallowed $0^+ \rightarrow 0^+$ $\beta$ decay.  In that time, substantial progress has been
made both in improving the precision of previously measured $ft$ values and in adding a new transition, a particularly interesting one that
completes for the first time a mirror pair of $0^+ \rightarrow 0^+$ transitions, \nuc{38}{Ca} $\rightarrow$ \nuc{38m}{K} and \nuc{38m}{K}
$\rightarrow$\nuc{38}{Ar}.  The principal outcome of the new data is to have improved our ability to discriminate among the various calculations
of the isospin-symmetry-breaking correction, $\delta_C$.  This is an important step forward, since these correction terms contribute more to the
overall uncertainty on $|V_{ud}|^2$ than do the experimental measurements themselves.  Even so, neither is the biggest contributor to the $|V_{ud}|^2$
uncertainty.  

\begin{figure}[t]
\epsfig{file=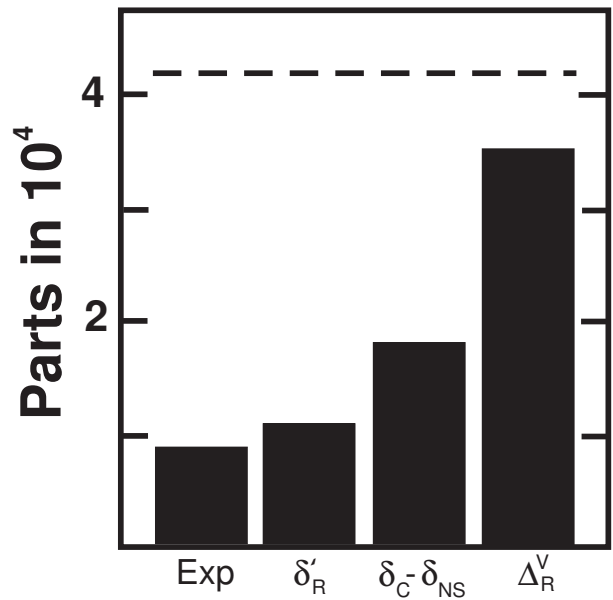,width=8cm}
\caption{Uncertainty budget for $|V_{ud}|^2$ as obtained from superallowed $0^+ \rightarrow 0^+$ $\beta$ decay.  The contributions are separated
into four categories: experiment, the transition-dependent part of the radiative correction ($\delta_{R}^{\prime}$), the nuclear-structure-dependent
terms ($\delta_C-\delta_{NS}$) and the transition-independent part of the radiative correction $\DRV$.}
\label{fig-8}
\end{figure}

It is instructive to look at the complete uncertainty budget for $|V_{ud}|^2$ in Fig.~\ref{fig-8}, where the four major contributions are displayed in
units of parts in $10^4$.  By now, experiment has completely outstripped theory in its remarkable precision.  But just as important as its precision
is the fact that it includes over 220 independent measurements covering 14 separate transitions, each with a $Q_{EC}$ value, half-life and branching
ratio that has been determined, in most cases, multiple times.  And all 14 transitions yield $\F t$ values that are statistically consistent with one
another.  This is indeed a robust body of data, completely insensitive to the possible presence of a few aberrant measurements. 

As Fig.~\ref{fig-8} makes clear, by far the largest contributor to the $|V_{ud}|^2$ uncertainty is the calculated radiative correction $\DRV$.  If any real
improvement in the unitarity test from $0^+ \rightarrow 0^+$ $\beta$ decay is to be achieved in future, it must come first from improved calculations of
$\DRV$.  Furthermore, since $\DRV$ is common to all other approaches to the measurement of $|V_{ud}|$ -- from neutron decay, $T=1/2$ nuclear mirror decays
and pion decays -- it provides an ultimate precision limit to them all, albeit well below the experimental uncertainties which currently dominate those
measurements.  In 2008, we identified improvements to $\DRV$ as the highest priority theoretical goal, and it remains so today.  The impact of any improvement
would be immediate: If the $\DRV$ uncertainty were cut in half, the $|V_{ud}|^2$ uncertainty would be reduced by 30\%.

The nuclear-structure-dependent corrections, ($\delta_C-\delta_{NS}$), are the second most important contributors to the overall uncertainty assigned to
$|V_{ud}|^2$.  Their contribution has been slightly reduced since 2008 as a result of improved experimental precision which, as already noted, has 
made possible a discriminating test for the efficacy of any set of calculated isospin-symmetry-breaking corrections, $\delta_C$.  As a result, we have been
able to select the only set in good agreement with the expectation of CVC that all measured transitions should have the same $\F t$ values within statistical
uncertainties.  This is an example of experiment contributing to the reduction of a theoretical uncertainty.  Further benefits from the same approach can
also be anticipated in future with the completion of more mirror pairs of $0^+ \rightarrow 0^+$ transitions -- at $A$ = 26, 34 and 42, for example -- and
with even higher precision in the already well-known $ft$ values.

Of course, the motivation for improving $|V_{ud}|$ is to tighten the uncertainty on CKM unitarity as a probe for physics beyond the standard model.  This
would obviously benefit from a resolution of the current conflict in the determinations of $|V_{us}|$.  Nevertheless, regardless of which current value for
$|V_{us}|$ one accepts, its contribution to the uncertainty on the unitarity sum is from 15-35\% less than that of our current value of $|V_{ud}|$.  (The
$relative$ precision of $|V_{ud}|$ is, however, more than an order-of-magnitude tighter than that of $|V_{us}|$.)  Thus, any improvement in $|V_{ud}|$ will have a
direct beneficial impact on the uncertainty of the unitarity sum.

There is another important outcome of the superallowed $\F t$ values that often gets less attention than it deserves: the experimental limit that it yields
on the possible occurrence of a scalar interaction.  The limit set here on the ratio of scalar-to-vector currents is the tightest available anywhere and it
can clearly be improved.  As a glance at Fig.~\ref{fig-7} will show, the two lightest superallowed transitions -- those from $^{10}$C and $^{14}$O -- are
crucial in setting the limit on a scalar interaction.  Both have relatively large uncertainties.  Both also present experimental challenges, particularly
in the measurement of their branching ratios.  There is no doubt, though, that an appreciable improvement in their $\F t$ values would pay off handsomely.

\begin{acknowledgments}

This material is based upon work supported by the U.S. Department of Energy, Office of Science, Office of Nuclear Physics, under
Award Number DE-FG03-93ER40773, and by the Robert A. Welch Foundation under Grant No.\,A-1397.

\end{acknowledgments}


\begin{thebibliography}{99999}

\bibitem{TH73}
I.S. Towner and J.C. Hardy, \np {\bf A205}, 33 (1973).

\bibitem{HT75}
J.C. Hardy and I.S. Towner, Nucl. Phys. {\bf A254}, 221 (1975).

\bibitem{Ko84}
V.T. Koslowsky, E. Hagberg, J.C. Hardy, H. Schmeing, R.E. Azuma
and I.S. Towner,
in {\it Proc. 7th Int. Conf. on atomic masses
and fundamental constants, Darmstadt-Seeheim}, ed. O. Klepper (T.H.
Darmstadt, 1984) p. 572

\bibitem{HT90}
J.C. Hardy, I.S. Towner, V.T. Koslowsky, E. Hagberg and H. Schmeing, Nucl. Phys. {\bf A509}, 429 (1990).

\bibitem{HT05}
J.C. Hardy and I.S. Towner, \prc {\bf 71}, 055501 (2005) and \prl {\bf 94}, 092502 (2005).

\bibitem{HT09}
J.C. Hardy and I.S. Towner, \pr C {\bf 79}, 055502 (2009).

\bibitem{Ad83}
E.G. Adelberger, M.M. Hindi, C.D. Hoyle, H.E. Swanson, R.D. Von Lintig and W.C. Haxton, Phys. Rev. C {\bf 27}, 2833 (1983); this reference replaces
the result reported in E.G. Adelberger, C.D. Hoyle, H.E. Swanson and R.D. Von Lintig, Phys. Rev. Lett. {\bf 46}, 695 (1981).

\bibitem{Aj88}
F. Ajzenberg-Selove, Nucl. Phys. {\bf A490}, 1 (1988).

\bibitem{Aj91}
F. Ajzenberg-Selove, Nucl. Phys. {\bf A523}, 1 (1991).

\bibitem{Al69}
A.M. Aldridge, K.W. Kemper and H.S. Plendl, Phys. Lett. {\bf 30B}, 165 (1969)

\bibitem{Al72}
D.E. Alburger, Phys. Rev. C {\bf 5}, 274 (1972).

\bibitem{Al75}
D.E. Alburger and F.P. Calaprice, Phys. Rev. C {\bf 12}, 1690 (1975).

\bibitem{Al77}
D.E. Alburger and D.H. Wilkinson, Phys. Rev. C {\bf 15}, 2174 (1977); this reference replaces the $^{46}$V half-life from \cite{Wi76}.

\bibitem{Al78}
D.E. Alburger, Phys. Rev. C {\bf 18}, 1875 (1978).

\bibitem{Al82}
P.F.A. Alkemade, C. Alderliesten, P. De Wit and C. Van der Leun, Nucl. Instr and Meth. {\bf 197}, 383 (1982).

\bibitem{An70}
A. Antilla, M. Bister and E. Arminen, Z. Phys. {\bf 234}, 455 (1970).

\bibitem{Az74}
G. Azuelos, J.E. Crawford and J.E. Kitching, Phys. Rev. C {\bf 9}, 1213 (1974).

\bibitem{Az75}
G. Azuelos and J.E. Kitching, Phys. Rev. C {\bf 12}, 563 (1975).

\bibitem{Ba62}
R.K. Barden, C.A. Barnes, W.A. Fowler and P.G. Seeger, Phys. Rev. {\bf 127}, 583 (1962)

\bibitem{Ba77a}
P.H. Barker, C.J. Scofield, R.J. Petty, J.M. Freeman, S.D. Hoath, W.E. Burcham and G.T.A. Squier, Nucl. Phys. {\bf A275}, 37 (1977); the same
result also appears in G.T.A. Squier, W.E. Burcham, S.D. Hoath, J.M. Freeman, P.H. Barker and R.J. Petty, Phys. Lett. {\bf 65B}, 122 (1976).

\bibitem{Ba77b}
P.H. Barker and J.A. Nolen, Proc. Int. Conf. on Nucl. Structure, Tokyo, Japan, 1977.

\bibitem{Ba84}
P.H. Barker and R.E. White, Phys. Rev. C {\bf 29}, 1530 (1984).

\bibitem{Ba88}
P.H. Barker and S.M. Ferguson, Phys. Rev. C {\bf 38}, 1936 (1988).

\bibitem{Ba89}
S.C. Baker, M.J. Brown and P.H. Barker, Phys. Rev. C {\bf 40}, 940 (1989).

\bibitem{Ba90}
P.H. Barker and G.D. Leonard, Phys. Rev. C {\bf 41}, 246 (1990).

\bibitem{Ba98}
P.H. Barker and P.A. Amundsen, Phys. Rev. C{\bf 58}, 2571 (1998); this reference updates the $^{10}$C $Q_{EC}$-value from \cite{Ba89};
its value for the $^{14}$O $Q_{EC}$-value was later withdrawn in \cite{To03}.

\bibitem{Ba00}
P.H. Barker and M.S. Wu, Phys. Rev C {\bf 62}, 054302 (2000).

\bibitem{Ba01}
G.C. Ball, S. Bishop, J.A. Behr, G.C. Boisvert, P. Bricault, J. Cerny, J.M. D'Auria, M. Dombsky, J.C. Hardy, V. Iacob, J.R. Leslie, T. Lindner, J.A. Macdonald, H.-B. Mak, D.M. Moltz, J. Powell, G. Savard and I.S. Towner, Phys. Rev. Lett. {\bf 86}, 1454 (2001).

\bibitem{Ba04}
P.H. Barker, I.C. Barnett, G.J. Baxter and A.P. Byrne, Phys. Rev. C {\bf 70}, 024302 (2004).

\bibitem{Ba06}
P.H. Barker and A.P. Byrne, Phys. Rev. C {\bf 73}, 064306 (2006).

\bibitem{Ba09}
P.H. Barker, K.K.H. Leung and A.P. Byrne, Phys. Rev. C {\bf 79}, 024311 (2009).

\bibitem{Ba10}
G.C. Ball, G. Boisvert, P. Bricault, R. Churchman, M. Dombsky, T. Lindner, J.A. Macdonald, E. Vandervoort, S. Bishop, J.M. D'Auria, J.C. Hardy, V.E. Iacob, J.R. Leslie and H.-B. Mak, Phys. Rev. C {\bf 82}, 045501 (2010).

\bibitem{Be68}
E. Beck and H. Daniel, Z. Phys. {\bf 216}, 229 (1968).

\bibitem{Be78}
J.A. Becker, R.A. Chalmers, B.A. Watson and D.H. Wilkinson, Nucl. Instr. Meth. {\bf 155}, 211 (1978).

\bibitem{Be85}
F.J. Bergmeister, K.P. Lieb, K. Pampus and M. Uhrmacher, Z. Phys. {\bf A320}, 693 (1985).

\bibitem{Be08}
A. Bey, B. Blank, G. Canchel, C. Dossat, J. Giovinazzo, I Matea, V.-V Elomaa, T. Eronen, U. Hager, J. Hakala,
A. Jokinen, A. Kankainen, I. Moore, H. Penttila, S. Rinta-Antila, A. Saastamoinen, T. Sonoda, J. Aysto, N. Adimi,
G. de France, J.-C. Thomas, G. Voltolini and T. Chaventre, Eur. Phys. J. A {\bf 36}, 121 (2008).

\bibitem{Bi03}
S. Bishop, R.E. Azuma, L. Buchmann, A.A. Chen, M.L. Chatterjee, J.M. D'Auria, S. Engel, D. Gigliotti, U. Greife, M. Hernanz, D. Hunter, A. Hussein, D. Hutcheon, C. Jewett, J. Jos\'{e}, J. King, S. Kubono, A.M. Laird, M. Lamey, R. Lewis, W. Liu, S. Michimasa, A. Olin, D. Ottewell, P.D. Parker, J.G. Rogers, F. Strieder and C. Wrede, Phys. Rev. Lett. {\bf 90}, 162501 (2003).

\bibitem{Bl04a}
B. Blank, G. Savard, J. Doring, A. Blazhev, G. Canchel, M. Chartier, D. Henderson, Z. Janas, R. Kirchner, I. Mukha, E. Roeckl, K. Schmidt, and J. Zylicz, Phys. Rev. C {\bf 69}, 015502 (2004).

\bibitem{Bl04b}
K. Blaum, G. Audi, D. Beck, G. Bollen, C. Guenaut, P. Delahaye, F. Herfurth, A. Kellerbauer, H.-J. Kluge, D. Lunney, D. Rodriguez, S. Schwarz, L. Schweikhard, C. Weber and C. Yazidjian, Nucl. Phys. {\bf A746}, 305c (2004).

\bibitem{Bl10}
B. Blank, A. Bey, I Matea, J. Souin, L. Audirac, M.J,G. Borge, G. Canchel, P. Delahaye, F. Delalee, C.-E. Dumonchy, R. Dominguez-Reyes, L.M. Fraile, J. Giovinazzo, Tran Trong Hui, J. Huikari, D. Lunney, F. Munoz, J.-L. Pedroza, C. Plaisir, L. Scrani, S. Sturm, O. Tengblad and F. Wenender, Eur. Phys. J. A {\bf 44}, 363 (2010).

\bibitem{Bo64}
R.O. Bondelid and J.W. Butler, Nucl. Phys. {\bf 53}, 618 (1964).

\bibitem{Br94}
S.A. Brindhaban and P.H. Barker, Phys. Rev. C {\bf 49}, 2401 (1994); reference replaces earlier conference proceedings from the same laboratory.

\bibitem{Bu61}
J.W. Butler and R.O. Bondelid, Phys. Rev. {\bf 121}, 1770 (1961).

\bibitem{Bu88}
R.H. Burch, C.A. Gagliardi and R.E. Tribble, Phys. Rev. C {\bf 38}, 1365 (1988).

\bibitem{Bu06}
J.T. Burke, P.A. Vetter, S.J. Freedman, B.K. Fujikawa and W.T. Winter, Phys. Rev. C {\bf 74}, 025501 (2006)

\bibitem{Ca05}
G. Canchel, B. Blank, M. Chartier, F. Delalee, P. Dendooven, C. Dossat, J. Giovinazzo, J. Huikari, A.S. Lalleman, M.J. Lopez Jimenez, V. Madec, J.L. Pedroza, H. Penttila and J.C. Thomas, Eur. Phys. J. {\bf A23}, 409 (2005).

\bibitem{Ch84}
N.M. Chaudri, Fizika {\bf 16}, 297 (1984).

\bibitem{Ch13}
L. Chen, J.C. Hardy, M. Bencomo, V. Horvat, N. Nica and H.I. Park, Nucl. Instr. And Meth. In Phys. Res. A {\bf 728}, 81 (2013).

\bibitem{Cl73}
G.J. Clark, J.M. Freeman, D.C. Robinson, J.S. Ryder, W.E. Burcham and G.T.A. Squier, Nucl. Phys. {\bf A215}, 429 (1973); this reference replaces
the half-life value in G.J. Clark, J.M. Freeman, D.C. Robinson, J.S. Ryder, W.E. Burcham and G.T.A. Squier, Phys. Lett {\bf 35B}, 503 (1971).

\bibitem{Da80}
C.N. Davids, in ``Atomic masses and fundamental constants 6", eds. J.A. Nolen and W. Benenson (Plenum, New York, 1980) p. 419.

\bibitem{Da85}
W.W. Daehnick and R.D. Rosa, Phys. Rev. C {\bf 31}, 1499 (1985).

\bibitem{De69}
P. De Wit and C. Van der Leun, Phys. Lett. {\bf 30B}, 639 (1969).

\bibitem{De78}
R.M. DelVecchio and W.W. Daehnick, Phys. Rev. C {\bf17}, 1809 (1978).

\bibitem{Dr75}
M.A. van Driel, H. Klijnman, G.A.P. Engelbertink, H.H. Eggebhuisen and J.A.J. Hermans, Nucl. Phys. {\bf A240}, 98 (1975).

\bibitem{Du13}
R. Dunlop, G.C. Ball, J.R. Leslie, C.E. Svensson, I.S. Towner, C. Andreoiu, S. Chagnon-Lessard, A. Chester, D.S. Cross, P. Finlay, A.B. Garnsworthy, P.E. Garrett, J. Glister, G. Hackman, B. Hadinia, K.G. Leach, E.T. Rand, K. Starosta, E.R. Tardiff, S. Triambak, S.J. Williams, J. Wong, S.W. Yates and E.F. Zganjar, Phys. Rev. C {\bf 88}, 045501 (2013).

\bibitem{En90}
P.M. Endt, Nucl. Phys. {\bf A521}, 1 (1990).

\bibitem{En98}
P.M. Endt, Nucl. Phys. {\bf A633}, 1 (1998).

\bibitem{Er06a}
T. Eronen, V. Elomaa, U. Hager, J. Hakala, A. Jokinen, A. Kankainen, I. Moore, H. Penttila, S. Rahaman, S.
Rinta-Antila, A. Saastamoinen, T. Sonoda, J. Aysto, A. Bey, B. Blank, G. Canchel, C. Dossat, J. Giovinazzo,
I. Matea and N. Adimi, Phys. Lett. B {\bf 636}, 191 (2006).

\bibitem{Er06b}
T. Eronen, V. Elomaa, U. Hager, J. Hakala, A. Jokinen, A. Kankainen, I. Moore, H. Penttila, S. Rahaman, J. Rissanen,
A. Saastamoinen, T. Sonoda, J. Aysto, J.C. Hardy and V.S. Kolhinen, Phys. Rev. Lett. {\bf 97}, 232501 (2006).

\bibitem{Er08}
T. Eronen, V.-V. Elomaa, U. Hager, J. Hakala, J.C. Hardy, A. Jokinen, A. Kankainen, I.D. Moore, H. Penttila,
S. Rahaman, S. Rinta-Antila, J. Rissanen, A. Saastamoinen, T. Sonoda, C. Weber and J Aysto, Phys. Rev. Lett.
{\bf 100}, 132502 (2008).

\bibitem{Er09a}
T. Eronen, V.-V. Elomaa, U. Hager, J. Hakala, A. Jokinen, A. Kankainen, T. Kessler, I.D. Moore, S. Rahaman, J. Rissanen, C. Weber and J Aysto, Phys. Rev. C {\bf 79}, 032802(R) (2009).

\bibitem{Er09b}
T. Eronen, V.-V. Elomaa, J. Hakala, J.C. Hardy, A. Jokinen, I.D. Moore, M. Reponen, J. Rissanen, A. Saastamoinen, C. Weber and J Aysto, Phys. Rev. Lett. {\bf 103}, 252501 (2009).

\bibitem{Er11}
T. Eronen, D. Gorelov, J. Hakala, J.C. Hardy, A. Jokinen, A. Kankainen, V.S. Kolhinen, I.D. Moore, H. Penttila, M. Reponen, J. Rissanen, A. Saastamoinen and J Aysto, Phys. Rev. C {\bf 83}, 055501 (2011).

\bibitem{Et11}
S. Ettenauer, M.C. Simon, A.T. Gallant, T. Brunner, U. Chowdhury, V.V. Simon, M. Brodeur, A. Chaudhuri, E. Man\'{e}, C. Andreoiu, G. Audi, J.R. Crespo L\'{o}pez-Urrutia, P. Delheij, G. Gwinner, A. Lapierre, D. Lunney, M.R. Pearson, R. Ringle, J. Ullrich and J. Dilling, Phys. Rev. Lett. {\bf 107}, 272501 (2011). 

\bibitem{Fa09}
T. Faestermann, R. Hertenberger, H.-F. Wirth, R. Krucken, M. Mahgoub and P. Maier-Komor, Eur. Phys. J. A {\bf 42}, 339 (2009).

\bibitem{Fi08}
P. Finlay, G.C. Ball, J.R. Leslie, C.E. Svensson, I.S. Towner, R.A.E. Austin, D. Bandyopadhyay, A. Chaffey,
R.S. Chakrawarthy, P.E. Garrett, G.F. Grinyer, G. Hackman, B. Hyland, R. Kanungo, K.G. Leach, C.M. Mattoon,
A.C. Morton, C.J. Pearson, A.A. Phillips, J.J. Ressler, F. Sarazin, H. Savajols, M.A. Schumaker and J. Wong,
Phys. Rev. C {\bf 78}, 025502 (2008); the branching-ratio result in this reference replaces the result reported
in B. Hyland {\it et al.}, Phys. Rev. Lett. {\bf 97}, 102501 (2006).

\bibitem{Fi11}
P. Finlay, S. Ettenauer, G.C. Ball, J.R. Leslie, C.E. Svensson, C. Andreoiu, R.A.E. Austin, D. Bandyopadhyay, D.S. Cross, G. Demand, M. Djongolov, P.E. Garrett, K.L. Green, G.F. Grinyer, G. Hackman, K.G. Leach, C.J. Pearson, A.A. Phillips, C.S. Sumithrarachchi, S. Triambak and S.J. Williams, Phys. Rev. Lett. {\bf 106} 032501 (2011),

\bibitem{Fi12}
P. Finlay, G.C. Ball, J.R. Leslie, C.E. Svensson, C. Andreoiu, R.A.E. Austin, D. Bandyopadhyay, D.S. Cross, G. Demand, M. Djongolov, S. Ettenauer, P.E. Garrett, K.L. Green, G.F. Grinyer, G. Hackman, K.G. Leach, C.J. Pearson, A.A. Phillips, E.T. Rand, C.S. Sumithrarachchi, S. Triambak and S.J. Williams, Phys. Rev. C {\bf 85}, 055501 (2012).

\bibitem{Fr63}
G. Frick, A. Gallmann, D.E. Alburger, D.H. Wilkinson and J.P. Coffin, Phys. Rev. {\bf 132}, 2169 (1963).

\bibitem{Fr65}
J.M. Freeman, G. Murray and W.E. Burcham, Phys.Lett. {\bf 17}, 317 (1965).

\bibitem{Fr69a}
J.M. Freeman, J.G. Jenkin, G. Murray, D.C. Robinson and W.E. Burcham, Nucl. Phys. {\bf A132}, 593 (1969); this reference replaces the
half-life values in J.M. Freeman, J.G. Jenkin, G. Murray and W.E. Burcham, Phys. Rev. Lett. {\bf 16}, 959 (1966).

\bibitem{Fr75}
J.M. Freeman, R.J. Petty, S.D. Hoath, G.T.A. Squier and W.E. Burcham, Phys. Lett {\bf 53B}, 439 (1975).

\bibitem{Fu99}
B.K. Fujikawa, S.J. Asztalos, R.M. Clark, M.A. Deleplanque-Stephens, P. Fallon, S.J. Freeman, J.P. Greene, I.-Y. Lee, L.J. Lising, A.O. Macchiavelli, R.W. MacLeod, J.C. Reich, M.A. Rowe, S.-Q. Shang, F.S. Stephens and E.G. Wasserman, Phys. Lett. {\bf B449}, 6 (1999).

\bibitem{Ga69}
A Gallmann, E. Aslanides, F. Jundt and E. Jacobs, Phys. Rev. {\bf 186}, 1160 (1969).

\bibitem{Ga01}
M. Gaelens, J. Andrzejewski, J. Camps, P. Decrock, M. Huyse, K. Kruglov, W.F. Mueller, A. Piechaczek, N. Severijns, J. Szerypo, G. Vancraeynest, P. Van Duppen and J. Wauters, Eur. Phys. J. {\bf A11}, 413 (2001).

\bibitem{Ge07}
S. George, S. Baruah, B. Blank, K. Blaum, M. Breitenfeldt, U. Hager, F. Herfurth, A. Herlert, A. Kellerbauer,
H.-J. Kluge, M. Kretzschmar, D. Lunney, R. Savreux, S. Schwarz, L. Schweikhard and C. Yazidjian, Phys. Rev. Lett.
{\bf 98}, 162501 (2007).

\bibitem{Ge08}
S. George, G. Audi, B. Blank, K. Blaum, M. Breitenfeldt, U. Hager, F. Herfurth, A. Herlert, A. Kellerbauer,
H.-J. Kluge, M. Kretzschmar, D. Lunney, R. Savreux, S. Schwarz, L. Schweikhard and C. Yazidjian, Europhysics Lett.
{\bf 82}, 50005 (2008).

\bibitem{Gi72}
H.J. Gils, D. Flothmann, R. Loehken and W. Wiesner, Nucl. Instr. and Meth. {\bf 105}, 179 (1972).

\bibitem{Go72}
D.R. Goosman and D.E. Alburger, Phys. Rev. C {\bf 5}, 1893 (1972); the branching-ratio upper limit set in this
reference is considered to replace the much higher value reported by D.R. Brown, S.M. Ferguson and D.H.
Wilkinson, Nucl. Phys. {\bf A135}, 159 (1969).

\bibitem{Gr07}
G.F. Grinyer, M.B. Smith, C. Andreoiu, A.N. Andreyev, G.C. Ball, P. Bricault, R.S. Chakrawarthy, J.J. Daoud, P.
Finlay, P.E. Garrett, G. Hackman, B. Hyland, J.R. Leslie, A.C. Morton, C.J. Pearson, A.A. Phillips, M.A. Schumaker,
C.E. Svensson, J.J. Valiente-Dobon, S.J. Williams and E.F. Zganjar, Phys. Rev. C {\bf 76}, 025503 (2007).

\bibitem{Gr08}
G.F. Grinyer, P. Finlay, C.E. Svensson, G.C. Ball, J.R. Leslie, R.A.E. Austin, D. Bandyopadhyay, A. Chaffey, R.S.
Chakrawarthy, P.E. Garrett, G. Hackman, B. Hyland, R. Kanungo, K.G. Leach, C.M. Mattoon,  A.C. Morton, C.J. Pearson,
A.A. Phillips, J.J. Ressler, F. Sarazin, H. Savajols, M.A. Schumaker and J. Wong, Phys. Rev. C {\bf 77}, 015501 (2008).

\bibitem{Gr13}
G.F. Grinyer, G.C. Ball, H. Bouzomita, S. Ettenauer, P. Finlay, A.B. Garnsworthy, P.E. Garrett, K.L. Green, G. Hackman, J.R. Leslie, C.J. Pearson, E.T. Rand, C.S. Sumithrarachchi, C.E. Svensson, J.C. Thomas, S. Triambak and S.J. Williams, Phys. Rev. C {\bf 87}, 045502 (2013).

\bibitem{Ha67}
G.I. Harris and A.K. Hyder, Phys. Rev. {\bf 157}, 958 (1967).

\bibitem{Ha68}
M. Hagen, K.H. Maier and R. Michaelsen, Phys. Lett. {\bf 26B}, 432 (1968).

\bibitem{Ha72a}
J.C. Hardy and D.E. Alburger, Phys. Lett. {\bf B42}, 341 (1972).

\bibitem{Ha74a}
J.C. Hardy, H. Schmeing, J.S. Geiger and R.L. Graham, Nucl. Phys. {\bf A223}, 157 (1974); this reference replaces results in J.C. Hardy, H. Schmeing,
J.S. Geiger, R.L. Graham and I.S. Towner, Phys. Rev. Lett. {\bf 29}, 1027 (1972).

\bibitem{Ha74b}
J.C. Hardy, H.R. Andrews, J.S. Geiger, R.L. Graham, J.A. Macdonald and H. Schmeing, Phys. Rev. Lett. {\bf 33}, 1647 (1974).

\bibitem{Ha74c}
J.C. Hardy, H. Schmeing, W. Benenson, G.M. Crawley, E. Kashy and H. Nann, Phys. Rev. C {\bf 9}, 252 (1974).

\bibitem{Ha74d}
J.C. Hardy, G.C. Ball, J.S. Geiger, R.L. Graham, J.A. Macdonald and H. Schmeing, Phys. Rev. Lett. {\bf 33}, 320 (1974); the value for the
$^{46}$V $Q_{EC}$-value from this reference was later withdrawn by J.C. Hardy and I.S. Towner, in ``Atomic masses and fundamental constants 5",
eds. J.H. Sanders and A.H. Wapstra (Plenum, New York, 1976) p 66.

\bibitem{Ha75}
J.C. Hardy, H Schmeing, J.S. Geiger and R.L. Graham, Nucl. Phys. {\bf A246}, 61 (1975); this reference replaces results in J.C. Hardy, H. Schmeing,
J.S. Geiger, R.L. Graham and I.S. Towner, Phys. Rev. Lett. {\bf 29}, 1027 (1972).

\bibitem{Ha94}
E. Hagberg, V.T. Koslowsky, J.C. Hardy, I.S. Towner, J.G. Hykawy, G. Savard and T. Shinozuka, Phys. Rev. Lett. {\bf 73}, 396 (1994);
uncertainties on the Gamow-Teller decays observed from $^{46}$V and $^{50}$Mn did not appear in this reference but have been derived
from the original data and added here.

\bibitem{Ha98}
P.D. Harty, N.S. Bowden, P.H. Barker and P.A. Amundsen, Phys. Rev. C {\bf 58}, 821 (1998).

\bibitem{Ha02}
J.C. Hardy and I.S. Towner, Phys. Rev. Lett. {\bf 88}, 252501 (2002).

\bibitem{Ha03}
J.C. Hardy, V.E. Iacob, M. Sanchez-Vega, R.G. Neilson, A Azhari, C.A. Gagliardi, V.E. Mayes, X. Tang, L. Trache and R.E. Tribble, Phys. Rev. Lett. {\bf 91}, 082501 (2003).

\bibitem{He61}
D.L. Hendrie and J.B. Gerhart, Phys. Rev. {\bf 121}, 846 (1961).

\bibitem{He81}
A.M. Hernandez and W.W. Daehnick, Phys. Rev. C {\bf 24}, 2235 (1981).

\bibitem{He82}
A.M. Hernandez and W.W. Daehnick, Phys. Rev. C {\bf 25}, 2957 (1982).

\bibitem{He00}
R.G. Helmer and C. van der Leun, Nucl. Instr. and Meth. in Phys. Res. {\bf A450}, 35 (2000).

\bibitem{He01}
F. Herfurth, J. Dilling, A. Kellerbauer, G. Audi, D. Beck, G. Bollen, H.-J. Kluge, D. Lunney, R.B. Moore,
C. Scheidenberger, S. Schwarz, G. Sikler, J. Szerypo and Isolde Collaboration, Phys. Rev. Lett. {\bf 87}, 142501 (2001).

\bibitem{He02}
F. Herfurth, A. Kellerbauer, F. Ames, G. Audi, D. Beck, K. Blaum, G. Bollen, O. Engels, H.-J. Kluge, D. Lunney,
R.B. Moore, O. Oinonen, E. Sauvan, C. Scheidenberger, S. Schwarz, G. Sikler and C. Weber, Eur. Phys. J.
{\bf A15}, 17 (2002).

\bibitem{Ho64}
I. Hofmann, Acta Phys. Aust. {\bf 18}, 309 (1964).

\bibitem{Ho74}
S.D. Hoath, R.J. Petty, J.M. Freeman, G.T.A. Squier and W.E. Burcham, Phys. Lett. {\bf 51B}, 345 (1974).

\bibitem{Hu82}
P. Hungerford and H.H. Schmidt, Nucl. Instr. and Meth. {\bf 192}, 609 (1982).

\bibitem{Hy03}
B.C. Hyman, V.E. Iacob, A. Azhari, C.A. Gagliardi, J.C. Hardy, V.E. Mayes, R.G. Neilson, M. Sanchez-Vega, X. Tang, L. Trache and R.E. Tribble, Phys. Rev. C {\bf 68}, 015501 (2003).

\bibitem{Hy05}
B. Hyland, D. Melconian, G.C. Ball, J.R. Leslie, C.E. Svensson, P. Bricault, E. Cunningham, M. Dombsky, G.F.
Grinyer, G. Hackman, K. Koopmans, F. Sarazin, M.A. Schumaker, H.C. Scraggs, M.B. Smith and P.M. Walker, J. Phys. G: Nucl. Part. Phys. {\bf 31}, S1885 (2005).

\bibitem{Ia06}
V.E. Iacob,  J.C. Hardy, J.F. Brinkley, C.A. Gagliardi, V.E. Mayes, N. Nica, M. Sanchez-Vega, G. Tabacaru,
L. Trache and R.E. Tribble,
Phys. Rev. C {\bf 74}, 055502 (2006).

\bibitem{Ia08}
V.E. Iacob, J.C. Hardy, V. Golovko, J. Goodwin, N. Nica, H.I. Park, L. Trache and R.E. Tribble, Phys. Rev. C
{\bf 77}, 045501 (2008).

\bibitem{Ia10}
V.E. Iacob, J.C. Hardy, A. Banu, L. Chen, V.V. Golovko, J. Goodwin, V. Horvat, N. Nica, H.I. Park, L. Trache and R.E. Tribble, Phys. Rev. C {\bf 82}, 035502 (2010).

\bibitem{In77}
P.D. Ingalls, J.C. Overley and H.S. Wilson, Nucl. Phys. {\bf A293}, 117 (1977).

\bibitem{Is80}
M.A. Islam, T.J. Kennett, S.A. Kerr and W.V. Prestwich, Can. J. Phys. {\bf 58}, 168 (1980).

\bibitem{Je07}
C. Jewett, C. Baktash, D. Bardayan, J. Blackmon, K. Chipps, A. Galindo-Uribarri, U. Greife, C. Gross, K. Jones, 
F. Liang, J. Livesay, R. Kozub, C. Nesaraja, D. Radford, F. Sarazin, M.S. Smith, J. Thomas, C.-H. Yu, Nucl. 
Instr. and Meth. in Phys. Res. {\bf B261}, 945 (2007).

\bibitem{Ji02}
M.J. Lopez Jimenez, B. Blank, M. Chartier, S. Czajkowski, P. Dessagne, G. de France, J. Giovinazzo, D. Karamanis,
M. Lewitowicz, V. Maslov, C. Miehe, P.H. Regan, M. Stanoiu and M. Wiescher, Phys. Rev. C {\bf 66}, 025803 (2002).

\bibitem{Ka69}
R.W. Kavanagh, Nucl. Phys {\bf A129}, 172 (1969).

\bibitem{Ke07}
A. Kellerbauer, G. Audi, D. Beck, K. Blaum, G. Bollen, C. Guenaut, F. Herfurth, A. Herlert, H.-J. Kluge, D. Lunney,
S. Schwarz, L. Schweikhard, C. Weber and C. Yazidjian, Phys. Rev. C {\bf 76}, 045504 (2007); this result for the
mass of $^{74}$Rb is the same as -- but more clearly explained than -- the result given in A. Kellerbauer, G.
Audi, D. Beck, K. Blaum, G. Bollen, B.A. Brown, P. Delahaye, C. Guenaut, F. Herfurth, H.-J. Kluge, D. Lunney,
S. Schwarz, L. Schweikhard and C. Yazidjian, Phys. Rev. Lett., {\bf 93}, 072502 (2004).

\bibitem{Ki89}
S.W. Kikstra, C. van der Leun, S. Raman, E.T. Jurney and I.S. Towner, Nucl. Phys. {\bf A496}, 429 (1989).

\bibitem{Ki91}
S.W. Kikstra, Z. Guo, C. Van der Leun, P.M. Endt, S. Raman, Walkiewicz, J.W. Starner, E.T. Jurney and I.S. Towner, Nucl. Phys. {\bf A529}, 39 (1991).

\bibitem{Ko83}
V.T. Koslowsky, E. Hagberg, J.C. Hardy, R.E. Azuma, E.T.H. Clifford, H.C. Evans, H. Schmeing, U.J. Schrewe and K.S. Sharma, Nucl. Phys. {\bf A405}, 29 (1983).

\bibitem{Ko87}
V.T. Koslowsky, J.C. Hardy, E. Hagberg, R.E. Azuma, G.C. Ball, E.T.H. Clifford, W.G. Davies, H. Schmeing, U.J.
Schrewe and K.S. Sharma, Nucl. Phys. {\bf A472}, 419 (1987); the $^{14}$O-$^{26}$Al$^m$ $Q_{EC}$-value-difference
result reported in this reference replaces an earlier value given in V.T. Koslowsky, J.C. Hardy, R.E. Azuma, G.C.
Ball, E.T.H. Clifford, W.G. Davies, E. Hagberg, H. Schmeing, U.J. Schrewe and K.S. Sharma, Phys. Lett.
{\bf 119B}, 57 (1982). 

\bibitem{Ko97a}
V.T. Koslowsky, E. Hagberg, J.C. Hardy, G. Savard, H. Schmeing, K.S. Sharma and X.J. Sun, Nucl.Instr.and Meth. {\bf A401}, 289 (1997).

\bibitem{Ko97b}
V.T. Koslowsky, E. Hagberg, J.C. Hardy, H. Schmeing and I.S. Towner, Nucl. Phys. {\bf A624}, 293 (1997).

\bibitem{Kr91}
M.A. Kroupa, S.J. Freeman, P.H. Barker and S.M. Ferguson, Nucl. Instr. And Meth. in Phys. Res. {\bf A310}, 649 (1991).

\bibitem{Ku09}
T. Kurtukian Nieto, J. Souin, T. Eronen, L. Audirac, J. Aysto, B. Blank, V.-V. Elomaa, J. Giovinazzo, U. Hager, J. Hakala, A. Jokinen, A. Kankainen, P. Karvonen, T. Kessler, I.D. Moore, H. Penttila, S. Rahaman, M. Reponen, S. Rinta-Antila, J. Rissanen, A. Saastamoinen, T. Sonoda and C. Weber, Phys. Rev. C {\bf 80}, 035502 (2009).

\bibitem{Kw10}
A.A. Kwiatkowski, B.R. Barquest, G. Bollen, C.M. Campbell, R. Ferrer, A.E. Gehring, D.L. Lincoln, D.J. Morrissey, G.K. Pang, J. Savory and S. Schwarz, Phys. Rev. C {\bf 81}, 058501 (2010).

\bibitem{Kw13}
A.A. Kwiatkowski, A. Chaudhuri, U. Chowdhury, A.T. Gallant, T.D. Macdonald, B.E. Schultz, M.C. Simon and J. Dilling, Ann. Phys. (Berlin) {\bf 525}, 529 (2013).

\bibitem{La13}
A.T. Laffoley, C.E. Svensson, C. Andreoiu, R.A.E. Austin, G.C. Ball, B. Blank, H. Bouzomita, D.S. Cross, A. Diaz Varela, R. Dunlop, P. Finlay, A.B. Garnsworthy, P.E. Garrett, J. Giovinazzo, G.F. Grinyer, G. Hackman, B. Hadinia, D.S. Jamieson, S. Ketelhut, K.G. Leach, J.R. Leslie, E. Tardiff, J.C. Thomas and C. Unsworth, Phys. Rev. C {\bf 88}, 015501 (2013).

\bibitem{Le08}
K.G. Leach, C.E. Svensson, G.C. Ball, J.R. Leslie, R.A.E. Austin, D. Bandyopadhyay, C. Barton, E. Bassiachvilli,
S. Ettenauer, P. Finlay, P.E. Garrett, G.F. Grinyer, G. Hackman, D. Melconian, A.C. Morton, S. Mythili, O. Newman,
C.J. Pearson, M.R. Pearson, A.A. Phillips, H. Savajols, M.A. Schumaker and J. Wong, Phys. Rev. Lett. {\bf 100},
192504 (2008).

\bibitem{Li94}
S. Lin, S.A. Brindhaban and P.H. Barker, Phys. Rev. C {\bf 49}, 3098 (1994).

\bibitem{Ma94}
P.V. Magnus, E.G. Adelberger and A. Garcia, Phys. Rev. C {\bf 49}, R1755 (1994).

\bibitem{Ma08}
I. Matea, J. Souin, J. Aysto, B. Blank, P. Delahaye, V.-V. Elomaa, T. Eronen, J. Giovinazzo, U. Hager, J. Hakala,
J. Huikari, A. Jokinen, A. Kankainen, I.D. Moore, J.-L. Pedroza, S. Rahaman, J. Rissanen, J. Ronkainen, A.
Saastamoinen, T. Sonoda and C. Weber, Eur. Phys. J. A {\bf 37}, 151 (2008); erratum {\bf 38}, 247 (2008).

\bibitem{Mc67}
W.R. McMurray, P. Van der Merwe and I.J. Van Heerden, Nucl. Phys. {\bf A92}, 401 (1967).

\bibitem{Mi67}
R.G. Miller and R.W. Kavanagh, Nucl. Phys. {\bf A94}, 261 (1967).

\bibitem{Mo71}
C.E. Moss, C. Detraz and C.S. Zaidins, Nucl. Phys. {\bf A174}, 408 (1971).

\bibitem{Mu04}
M. Mukherjee, A. Kellerbauer, D. Beck, K. Blaum, G. Bollen, F. Carrel, P. Delahaye, J. Dilling, S. George, C. Guenaut, F. Herfurth, A. Herlert, H.-J. Kluge, U. Koster, D. Lunney, S. Schwarz, L. Schweikhard and C. Yazidjian, Phys. Rev. Lett. {\bf 93}, 150801 (2004).

\bibitem{Na91}
Y. Nagai, K. Kunihiro, T. Toriyama, S. Harada, Y. Torii, A. Yoshida, T. Nomura, J. Tanaka and T. Shinozuka, Phys. Rev. C {\bf 43}, R9 (1991).

\bibitem{No74}
J.A. Nolen, G. Hamilton, E. Kashy and I.D. Proctor, Nucl. Instr. and Meth. {\bf 115}, 189 (1974).

\bibitem{Oi01}
M.Oinonen {\it et al}, Phys. Lett. {\bf B511}, 145 (2001).

\bibitem{Pa72}
R.A. Padock, Phys. Rev. C {\bf 5}, 485 (1972).

\bibitem{Pa05}
A. Parikh, J.A. Caggiano, C. Deibel, J.P. Greene, R. Lewis, P.D. Parker and C. Wrede, Phys. Rev. C {\bf 71}, 055804 (2005).

\bibitem{Pa11}
H.I. Park, J.C. Hardy, V.E. Iacob, A. Banu, L. Chen, V.V. Golovko, J. Goodwin, V. Horvat, N. Nica, E. Simmons, L. Trache and R.E. Tribble, Phys. Rev. C {\bf 84}, 065502 (2011).

\bibitem{Pa12}
H.I. Park, J.C. Hardy, V.E. Iacob, L. Chen, J. Goodwin, N. Nica, E. Simmons, L. Trache and R.E. Tribble, Phys. Rev. C {\bf 85}, 035501 (2012).

\bibitem{Pa14}
H.I. Park, J.C. Hardy, V.E. Iacob, M. Bencomo, L. Chen, V. Horvat, N. Nica, B.T. Roeder, E. Simmons, R.E. Tribble and I.S. Towner , Phys. Rev. Lett. {\bf 112}, 102502 (2014).

\bibitem{Pi03}
A. Piechaczek, E.F. Zganjar, G.C. Ball, P. Bricault, J.M. D'Auria, J.C. Hardy, D.F. Hodgson, V. Iacob, P Klages, W.D. Kulp, J.R. Leslie, M. Lipoglavsek, J.A. Macdonald, H.-B. Mak, D.M. Moltz, G. Savard, J. von Schwarzenberg, C.E. Svensson, I.S. Towner and J.L. Wood, Phys. Rev. C {\bf 67}, 051305(R) (2003); the branching-raio results from this measurement are considered to replace the contradictory upper limit set in an earlier less-precise measurement\cite{Oi01}.

\bibitem{Pr67}
F.W. Prosser, G.U. Din and D.D. Tolbert, Phys. Rev. {\bf 157}, 779 (1967).

\bibitem{Pr90}
W.V. Prestwich and T.J. Kennett, Can. J. Phys. {\bf 68}, 261 (1990); erratum {\bf 68}, 1352 (1990).

\bibitem{Ra83}
S. Raman, E.T. Jurney, D.A. Outlaw and I.S. Towner, Phys. Rev. C {\bf 27}, 1188 (1983).

\bibitem{Re85}
J.P.L. Reinecke, F.B. Waanders, P. Oberholtzer, P.J.C. Janse van Rensburg, J.A. Cilliers, J.J.A. Smit, M.A. Meyer and P.M. Endt,
Nucl. Phys. {\bf A435}, 333 (1985).

\bibitem{Ri07}
R. Ringle, T. Sun, G. Bollen, D. Davies, M. Facina, J. Huikari, E. Kwan, D.J. Morrissey, A. Prinke, J. Savory,
P. Schury, S. Schwarz and C.S. Sumithrarachchi, Phys. Rev. C {\bf 75}, 055503 (2007); this result is the same as that
appearing in G. Bollen, D. Davies, M. Facina, J. Huikari, E. Kwan, P.A. Lofy, D.J. Morrissey, A. Prinke, R. Ringle,
J. Savory, P. Schury, S. Schwarz, C. Sumithrarachchi, T. Sun and L. Weissman, Phys. Rev. Lett. {\bf 96}, 152501 (2006).

\bibitem{Ro70}
M.L. Roush, L.A. West and J.B. Marion, Nucl. Phys. {\bf A147}, 235 (1970).

\bibitem{Ro72}
D.C. Robinson, J.M. Freeman and T.T. Thwaites, Nucl. Phys. {\bf A181}, 645 (1972); this reference replaces the $^{10}$C branching ratio
from J.M. Freeman, J.G. Jenkin and G. Murray, Nucl. Phys. {\bf A124}, 393 (1969).

\bibitem{Ro74}
D.C. Robinson and P.H. Barker, Nucl. Phys. {\bf A225}, 109 (1974).

\bibitem{Ro75}
C. Rolfs, W.S. Rodney, S. Durrance and H. Winkler, Nucl. Phys. {\bf A240}, 221 (1975).

\bibitem{Ro06}
D. Rodriguez, G. Audi, J. Aysto, D. Beck, K. Blaum, G. Bollen, F. Herfurth, A. Jokinen, A. Kellerbauer, H.-J.
Kluge, V.S. Kohlinen, M. Oinonen, E. Sauvan and S. Schwarz, Nucl. Phys. {\bf A769}, 1 (2006); this result for
the mass of $^{74}$Kr is the same as -- but more clearly explained than -- the result given in A. Kellerbauer,
G. Audi, D. Beck, K. Blaum, G. Bollen, B.A. Brown, P. Delahaye, C. Guenaut, F. Herfurth, H.-J. Kluge, D. Lunney,
S. Schwarz, L. Schweikhard and C. Yazidjian, Phys. Rev. Lett., {\bf 93}, 072502 (2004).

\bibitem{Ry73}
J.S. Ryder, G.J. Clark, J.E. Draper, J.M. Freeman, W.E. Burcham and G.T.A. Squier, Phys. Lett {\bf 43B}, 30 (1973).

\bibitem{Ry91}
A Rytz, At. Data and Nucl. Data Tables {\bf 47}, 205 (1991).

\bibitem{Sa80}
A.M. Sandorfi, C.J. Lister, D.E. Alburger and E.K. Warburton, Phys. Rev. C {\bf 22}, 2213 (1980).

\bibitem{Sa95}
G. Savard, A. Galindo-Uribarri, E. Hagberg, J.C. Hardy, V.T. Koslowsky, D.C. Radford and I.S. Towner, Phys. Rev. Lett. {\bf 74}, 1521 (1995).

\bibitem{Sa04}
G. Savard, J.A. Clark, F. Buchinger, J.E. Crawford, S. Gulick, J.C. Hardy, A.A. Hecht, V.E. Iacob, J.K.P. Lee, A.F. Levand,  B.F. Lundgren, N.D. Scielzo, K.S. Sharma, I. Tanihata, I.S. Towner, W. Trimble, J.C. Wang, Y. Wang and Z. Zhou, Phys. Rev. C {\bf 70}, 042501(R) (2004).

\bibitem{Sa05}
G. Savard, F. Buchinger, J.A. Clark, J.E. Crawford, S. Gulick, J.C. Hardy, A.A. Hecht,  J.K.P. Lee, A.F. Levand,
N.D. Scielzo, H. Sharma, K.S. Sharma, I. Tanihata, A.C.C. Villari and Y. Wang, Phys. Rev. Lett. {\bf 95}, 102501 (2005).

\bibitem{Sa09}
J. Savory, P. Schury, C. Bachelet, M. Block, G. Bollen, M. Facina, C.M. Folden III, C. Gu\'{e}naut, E. Kwan, A.A. Kwiatkowski, D.J. Morrissey, G.K. Pang, A. Prinke, R. Ringle, H. Schatz, S. Schwarz and C.S. Sumithrarachchi, Phys. Rev. Lett. {\bf 102}, 132501 (2009)

\bibitem{Sc07}
P. Schury, C. Bachelet, M. Block, G. Bollen, D.A. Davies, M. Facina, C.M. Folden III, C. Gu\'{e}naut, J. Huikari,
E. Kwan, A. Kwiatkowski, D.J. Morrissey, R. Ringle, G.K. Pang, A. Prinke, J. Savory, H. Schatz, S. Schwarz,
C.S. Sumithrarachchi and T.Sun, Phys. Rev. C {\bf 75}, 055801 (2007).

\bibitem{Sc11}
R. J. Scott, G.J. O'Keefe, M.N. Thompson and R.P. Rassool, Phys. Rev. C, {\bf 84}, 024611 (2011)

\bibitem{Se73}
J.C. Sens, A. Pape and R. Armbruster, Nucl. Phys. {\bf A199}, 241 (1973).

\bibitem{Se05}
D. Seweryniak, P.J. Woods, M.P. Carpenter, T. Davinson, R.V.F. Janssens, D.G. Jenkins, T. Lauritsen, C.J. Lister, 
C. Ruiz, J. Shergur, S. Sinha and A. Woehr, Phys. Rev. Lett. {\bf 94}, 032501 (2005).

\bibitem{Sh55}
R. Sherr, J.B. Gerhart, H. Horie and W.F. Hornyak, Phys. Rev. {\bf 100}, 945 (1955).

\bibitem{Si66}
G.S. Sidhu and J.B. Gerhart, Phys. Rev. {\bf 148}, 1024 (1963).

\bibitem{Si72}
J. Singh, Indian J. Pure Appl. Phys. {\bf 10}, 289 (1972).

\bibitem{So11}
J. Souin, T. Eronen, P. Ascher, L. Audirac, J. Aysto, B. Blank, V.-V. Elomaa, J. Giovinazzo, J. Hakala, A. Jokinen, V.S. Kolhinen, P. Karvonen, I.D. Moore, S. Rahaman, J. Rissanen, A. Saastamoinen and J.C. Thomas, Eur. Phys. J. A {\bf 47}, 40 (2011)

\bibitem{Sq75}
G.T.A. Squier, W.E. Burcham, J.M. Freeman, R.J. Petty, S.D. Hoath and J.S. Ryder, Nucl. Phys {\bf A242}, 62 (1975).

\bibitem{Ta12}
V.T. Takau, M.N. Thompson, R.J. Scott, R.P. Rassool and G.J. O'Keefe, Rad. Phys. And Chem. {\bf 81}, 1669 (2012)

\bibitem{Ti95}
D.R. Tilley, H.R. Weller, C.M. Cheves and R.M. Chasteler, Nucl. Phys. {\bf A595}, 1 (1995).

\bibitem{To03}
N.R. Tolich, P.H. Barker, P.D. Harty and P.A. Amundsen, Phys. Rev. C {\bf 67}, 035503 (2003).

\bibitem{To05}
I.S. Towner and J.C. Hardy, Phys. Rev. C {\bf 72}, 055501 (2005).

\bibitem{Vo77}
H. Vonach, P. Glaessel, E. Huenges, P. Maier-Komor, H. Roesler, H.J. Scheerer, H. Paul and D. Semrad, Nucl. Phys. {\bf A278}, 189 (1977).

\bibitem{Wa83}
F.B. Waanders, J.P.L. Reinecke, H.N. Jacobs, J.J.A. Smit, M.A. Meyer and P.M. Endt, Nucl. Phys. {\bf A411}, 81 (1983).

\bibitem{Wa92}
T.A. Walkiewicz, S. Raman, E.T. Jurney, J.W. Starner and J.E. Lynn, Phys. Rev. C {\bf 45}, 1597 (1992).

\bibitem{Wa12}
M. Wang, G. Audi, A.H. Wapstra, F.G. Kondev, M. MacCormick, X. Xu and B. Pfeiffer, Chinese Physics C {\bf 36}, 1603 (2012)

\bibitem{We68}
H. Wenninger, J. Stiewe and H. Leutz, Nucl. Phys. {\bf A109}, 561 (1968).

\bibitem{Wh77}
R.E. White and H. Naylor, Nucl. Phys. {\bf A276}, 333 (1977).

\bibitem{Wh81}
R.E. White, H. Naylor, P.H. Barker, D.M.J. Lovelock and R.M. Smythe, Phys. Lett. {\bf 105B}, 116 (1981).

\bibitem{Wh85}
R.E. White, P.H. Barker and D.M.J. Lovelock, Metrologea {\bf 21}, 193 (1985).

\bibitem{Wi76}
D.H. Wilkinson and D.E. Alburger, Phys. Rev. C {\bf 13}, 2517 (1976).

\bibitem{Wi78}
D.H. Wilkinson, A. Gallmann and D.E. Alburger, Phys. Rev. C {\bf 18}, 401 (1978).

\bibitem{Wi80}
H.S. Wilson, R.W. Kavanagh and F.M. Mann, Phys. Rev. C {\bf 22}, 1696 (1980).

\bibitem{Zi87}
F. Zijderhand, R.C. Markus and C. van der Leun, Nucl. Phys. {\bf A466}, 280 (1987).

\bibitem{PDG14}
K.A. Olive \etal [Particle Data Group], Chin. Phys. C {\bf 38}, 090001 (2014).

\bibitem{MS06}
W.J. Marciano and A. Sirlin, \prl {\bf 96}, 032002 (2006).

\bibitem{Si67}
A. Sirlin, \pr C {\bf 164}, 1767 (1967).

\bibitem{SZ86}
A. Sirlin and R. Zucchini, \prl {\bf 57}, 1994 (1986).

\bibitem{JR87}
W. Jaus and G. Rasche, \pr D {\bf 35}, 3420 (1987).

\bibitem{Si87}
A. Sirlin, \pr D {\bf 35}, 3423 (1987).

\bibitem{CWS04}
A. Czarnecki, W.J. Marciano, and A. Sirlin, \pr D {\bf 70}, 093006 (2004). 

\bibitem{TH08}
I.S. Towner and J.C. Hardy, \pr C {\bf 77}, 025501 (2008).

\bibitem{MA14}
M. MacCormick and G. Audi, \np A {\bf 925}, 61 (2014);
corrigendum, \np A {\bf 925}, 296 (2014).
 
\bibitem{AME12}
G. Audi, F.G. Kondev, M. Wang, B. Pfeiffer, X. Sun, J. Blachot and
M. MacCormick, Chinese Physics C {\bf 36}, 1157 (2012).

\bibitem{deV87}
H. De Vries, C.W. De Jager and C. De Vries,
At. Data and Nucl. Data Tables {\bf 36}, 495 (1987).

\bibitem{TH02}
I.S. Towner and J.C. Hardy, \pr C {\bf 66}, 035501 (2002).

\bibitem{An04}
I. Angeli, At. Data and Nucl. Data Tables {\bf 87}, 185 (2004).

\bibitem{Ma11}
E. Man\'{e}, A. Voss, J.A. Behr, J. Billowes, T. Brunner, F. Buchinger,
J.E. Crawford, J. Dilling, S. Ettenauer, C.D.P. Levy, O. Shelbaya
and M.R. Pearson, \prl {\bf 107}, 212502 (2011).

\bibitem{OB89}
W.E. Ormand and B.A. Brown, \prl {\bf 62}, 866 (1989).

\bibitem{Le11}
J. Le Bloas, L. Bonneau, P. Quentin and J. Bartel,
Int. J. of Modern Phys. E {\bf 20}, 274 (2011).

\bibitem{SGS86}
H. Sagawa, N.V. Giai and T. Suzuki, \pr C {\bf 53}, 2163 (1986).

\bibitem{LGM09}
H. Liang, N.V. Giai and J. Meng, \pr C {\bf 79}, 064316 (2009).

\bibitem{LYC11}
Z.X. Li, J.M. Yao and H. Chen, Sci China Phys Mech Astron {\bf 54},
1131 (2011).

\bibitem{Au09}
N. Auerbach, \pr C {\bf 79}, 035502 (2009).

\bibitem{Ro13}
V. Rodin, \pr C {\bf 88}, 064318 (2013).

\bibitem{Sa12}
W. Satula, J. Dobaczewski, W. Nazarewicz and T.R. Werner, \pr C
{\bf 86}, 054316 (2012).

\bibitem{Be75}
M. Beiner, H. Flocard, N. van Giai and P. Quentin, \np A {\bf 238},
29 (1975).

\bibitem{TH10}
I.S. Towner and J.C. Hardy, \pr C {\bf 82}, 065501 (2010).

\bibitem{TH95}
I.S. Towner and J.C. Hardy, in {\it Symmetries and Fundamental Interactions in Nuclei},
eds. W.C. Haxton and E.M. Henley, (World-Scientific, Singapore, 1995) pp. 183-249.

\bibitem{TH99}
I.S. Towner and J.C. Hardy, in {\it Proceedings of the V International WEIN Symposium: Physics
Beyond the Standard Model, Santa Fe, NM 1998}, eds. P. Herzeg, C.M. Hoffman and H.V.
Klapdor-Kleingrothaus, (World-Scientific, Singapore, 1999) pp. 338-359.


\bibitem{Fl10}
M. Antonelli \etal [FlaviaNet Working Group on Kaon Decays], Eur. Phys.
J. C {\bf 69}, 399 (2010).

\bibitem{Mo14}
M. Moulson, Proceedings of the 8$^{th}$
International Workshop on the CKM Unitarity Triangle (CKM 14), Vienna,
2014, and private communication.

\bibitem{FLAG14}
S. Aoki \etal [Flavour Lattice Averaging Group (FLAG)],
arXiv:1310.8555v2 (2014).

\bibitem{Ba14}
A. Bazavov \etal [FermiLab Lattice and MILC collaborations], \prl {\bf 112},
112001 (2014).

\bibitem{Ba14a}
A. Bazavov \etal [FermiLab Lattice and MILC collaborations], 
arXiv:1407.3772 (2014).

\bibitem{BM13}
E. Blucher and W.J. Marciano, {\it $V_{ud}$, $V_{us}$, the Cabibbo angle,
and CKM unitarity}, mini-review for Particle Data Group \cite{PDG14} (2013).

\bibitem{JTW57}
J.D. Jackson, S.B. Treiman and H.W. Wyld Jr., \pr {\bf 106}, 517 (1957).

\bibitem{Go05}
A. Gorelov, D. Melconian, W.P. Alford, D. Ashery, G. Ball, J.A. Behr,
P.G. Bricault, J.M. D'Auria, J. Deutsch, J. Dilling, M. Dombsky,
P. Dube, J. Fingler, U. Giesen, F. Gl\"{u}ck, S. Gu, O. Ha\"{u}sser,
K.P. Jackson, B.K. Jennings, M.R. Pearson, T.J. Stocki,
T.B. Swanson and M. Trinczek, \prl {\bf 94}, 142501 (2005).

\bibitem{NG13}
O. Naviliat-Cuncic and M. Gonz\'{a}lez-Alonso, Ann. Phys. (Berlin)
{\bf 525}, 600 (2013).

\bibitem{BB82}
H. Behrens and W. B\"{u}hring, {\it Electron Radial Wave Functions and 
Nuclear Beta-decay} (Clarendon Press, Oxford, 1982).



\end{thebibliography}
\end{document}